\documentclass[11pt]{article}%
\usepackage{amsmath,color}
\usepackage{times,graphics,float,subfig}
\usepackage{enumerate}
\usepackage{amsfonts,amsthm,amscd}
\usepackage{amssymb}
\usepackage{graphicx}
\usepackage{rotating}
\usepackage{acronym}
\usepackage{amsmath}
\usepackage{amsfonts}%
\newtheorem{theorem}{Theorem}

\newtheorem{definition}[theorem]{Definition}

\newtheorem{remark}[theorem]{Remark}

\usepackage{hyperref}

\newcommand{\eps}{\varepsilon}
\DeclareMathOperator{\rank}{rank}
\DeclareMathOperator{\kmax}{kmax}
\begin{document}

\title{Topological Data Analysis for Object Data}
\author{Vic Patrangenaru\\Florida State University, Florida, U.S.A.
\and Peter Bubenik\\University of Florida, Florida, U.S.A.
\and Robert L. Paige\\Missouri S \& T , Missouri, U.S.A.
\and Daniel Osborne\\Florida A\&M University, Florida, U.S.A.}
\maketitle


\begin{abstract}
Statistical analysis on object data presents many challenges. Basic summaries such as means and variances are difficult to compute. We apply ideas from topology to study object data. We present a framework for using persistence landscapes to vectorize object data and perform statistical analysis. We apply to this pipeline to some biological images that were previously shown to be challenging to study using shape theory. Surprisingly, the most persistent features are shown to be ``topological noise'' and the statistical analysis depends on the less persistent features which we refer to as the ``geometric signal''. We also describe the first steps to a new approach to using topology for object data analysis, which applies topology to distributions on object spaces.
\end{abstract}

\noindent\textbf{Keywords:} topological data analysis; persistence landscapes; object spaces; extrinsic object data analysis; .

\section{Introduction}

Object data may be considered to be sampled from some underlying object space, which may be a manifold or stratified space.
Topology produces \emph{homology invariants} that lend themselves to the investigation of holes or voids in this underlying structure.
Topological data analysis (TDA) uses distance (i.e. \emph{metric}) data to provide a multiscale summary of these topological features.

In the remainder of the introduction, we summarize: our framework for using topological data analysis for object data; our results; and make comparisons to related work. Mathematical terms that will be define in Sections \ref{sec:tda} and \ref{sec:oda} are in italics, and non-technical terms that are explained in Section~\ref{sec:leafs} are in double quotes.

\subsection{TDA framework for object data}
\label{sec:tda-framework}

The methods of topological data analysis are quite flexible, and there are many possible ways to apply them to object data. We will use \emph{persistent homology}. The crucial step is encoding the object data by an increasing sequence of spaces that contain enough of the structure so that the subsequent statistical analysis will be successful. We will represent the object data by a finite sample of points. From the pairwise distances between these points we will construct an increasing family of \emph{simplicial complexes}, called \emph{Vietoris-Rips complexes}. We will calculate their persistent homology and convert this data to vectors using \emph{death vectors} and \emph{persistence landscapes}. These vectors will constitute our topological summary of the object data and we will the input to our statistical analysis.

\subsection{Results}
\label{sec:results}

We applied the pipeline described above to a collection of images of leaves (see Figure~\ref{fig:sample-leaf-images}). Looking at these images, we see a number of regions bounded by the veins, the midrib (the large central vein), and the boundary of the leaf. These regions are approximately rectangular shaped and their boundaries are topologically equivalent (homeomorphic or homotopy equivalent) to circles. Furthermore, the union of these boundaries (the veins and the boundary of the leaf) is topologically equivalent (has the same homotopy-type as) a collection of circles attached at a common point (a wedge or bouquet of circles). We expected that our analysis would be insensitive to ``geometric noise'' and detect this ``topological signal'' (homotopy-type).

In fact, the analysis was successful for the opposite reason.
The point samples we obtained from the leaves were of low quality (see Figure~\ref{fig:sample-leaf-points}). From these points it was impossible to see all of the rectangular regions of the leaves, and in addition, outlier points (see the bottom of Figure~\ref{fig:sample-leaf-points}) created ``topological noise''. However, the arrangement of sampled points contained enough geometric information on the leaves so that our statistical analysis was successful by using this ``geometric signal''.

\subsection{Related work}
\label{sec:related-work}

Our approach and results are closely related to work by
Bendich et al. (2016)~\cite{BeMaMiPiSk:2016}, who also applied TDA to object data. In their case they considered brain artery structures extracted from magnetic resonance images and applied persistence homology, which was encoded in vectors by the order statistic on the most persistent points in the persistence diagram. Also, closely related is work by Kovacev-Nikolic et al. (2016)~\cite{Ko-NiBuNiHe:2015} who applied persistent homology and persistence landscapes to protein structure data.

Our analysis differs by considering lower quality data (see Figure~\ref{fig:sample-leaf-points}), using a sophisticated feature vector (the persistence landscape from which the persistence diagram can be reconstructed), and in highlighting the distinction between ``topological noise'' and ``geometric signal''.

\subsubsection*{Outline of the paper}

In Section~\ref{sec:tda} we summarize parts of topological data analysis (TDA) and introduce our framework for applying TDA to object data.
In Section~\ref{sec:leafs} we apply our framework to a particular set of object data. In Section~\ref{sec:oda} we introduce a new approach that applies differential or algebraic topology methods to data analysis of distributions on object spaces.

\section{Objects TDA}
\label{sec:tda}

Topological data analysis (TDA) summarizes the topological and geometric
structure of data by applying tools from algebraic topology to certain
geometric structures built from the data at hand.

\subsection{Simplicial complexes}

\label{sec:sc}

The basic building block is a \emph{simplex}, which we will now define.
A $0$-simplex is a single point or vertex, a $1$-simplex is the line segment
or edge determined by $2$ distinct vertices, $2$-simplex is the solid triangle
determined by $3$ vertices, that do not lie on a line, and so on.
More formally, a $p$-simplex is 
the convex hull of 
points $x_0,x_1,\ldots,x_p \in \mathbb{R}^d$ such that the vectors $x_{1}-x_{0},\ldots,x_{p}-x_{0}$ are linearly independent.


In data-analytic applications, one treats a data point cloud, $\mathcal{X}$,
as a noisy sampling of a metric space $\mathcal{M}$. In topological data analysis, one obtains
summaries of the topology and geometry of $\mathcal{M}$ by defining
a parametric family of nested simplicial complexes which is built on top of
$\mathcal{X}$ and considering its topology. This family is known as a
\emph{filtered simplicial complex} or simply a \emph{filtration}.

There are a number of different complexes which are used in topological data analysis.
For example, we may consider a coarse-graining of $\mathcal{X}$ by taking the
union of closed $\varepsilon$-neighborhoods;%
\[
\mathcal{X}_{\varepsilon}= \bigcup_{x \in\mathcal{X}} B_{\varepsilon}(x),
\text{ where } B_{\varepsilon}(x) \left\{  y \in\mathcal{M} :d\left(  x,
y\right)  \leq\varepsilon\right\}  .
\]
The \v{C}ech complex is mainly of theoretical interest and is defined as follows.

\begin{definition}
The \v{C}ech complex, $\mathcal{C}_{\varepsilon}$, generated from
$\mathcal{X}$
is the simplicial complex
which has a $p$-simplex whenever the
closed $\varepsilon$-neighborhoods of subset of $p$ data cloud points have a
common intersection
(this is also called the nerve of cover $\mathcal{X}_{\varepsilon}$).
\end{definition}

\begin{figure}[h]
\begin{center}
\reflectbox{\includegraphics[scale = 0.45]{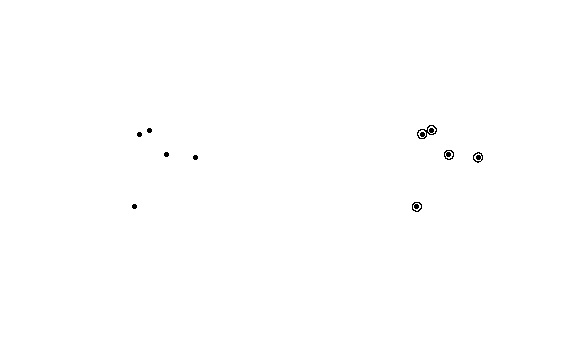}}
\reflectbox{\includegraphics[scale = 0.45]{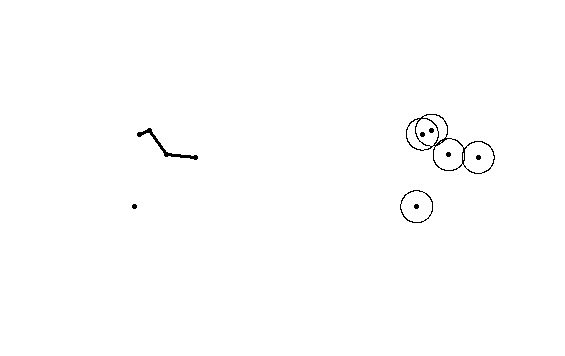}}\newline%
\reflectbox{\includegraphics[scale = 0.45]{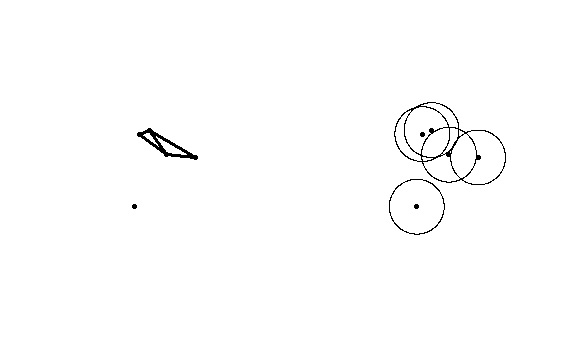}}
\reflectbox{\includegraphics[scale = 0.45]{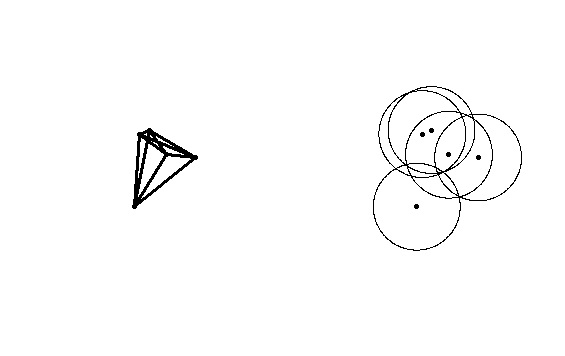}}
\end{center}
\caption{{\small \v{C}ech complexes of a point cloud for various radii.}}%
\label{fig:epsilon-disks}%
\end{figure}

Figure~\ref{fig:epsilon-disks} shows for a certain finite set of points $\mathcal{X}$ in the plane, the disks composing $\mathcal{X}_{\varepsilon}$ and the corresponding \v{C}ech complexes, $\mathcal{C}_{\eps}$, for disks of
radii $\varepsilon=0.053,0.184,0.316$ and $0.5$.

The Nerve theorem states that if $\mathcal{M}=\mathbb{R}^{d}$ then the
homotopy types of $\mathcal{C}_{\varepsilon}$ and $\mathcal{X}_{\varepsilon}$
are the same. This means that \v{C}ech complex, $\mathcal{C}_{\varepsilon}$,
is a topologically faithful simplicial model for the topology of
$\mathcal{X}_{\varepsilon}$, a point cloud fattened by balls. Unfortunately,
it is often expensive to compute and to store the \v{C}ech complex since to
determine the $p$-simplices one has to compute all subsets of size $p$ of
which there are a total of $2^{p}$. The Vietoris-Rips complex, which we now define,
is a more computationally efficient alternative to the \v{C}ech complex.

\begin{definition}
The Vietoris-Rips (VR) complex, $\mathcal{R}_{\varepsilon}$, generated from
open cover $\mathcal{X}_{\varepsilon}$ is the simplicial complex which has a
$p$-simplex any time that the closed $\varepsilon$-neighborhoods for subset of
$p$ points all have pairwise nonempty intersections.
\end{definition}

All Vietoris-Rips complexes $\mathcal{R}_{\varepsilon}$ (for all $\eps$) can be computed for $n$
data cloud points once one has computed all $\binom{n}{2}$ pairwise distances.

In general, there is no single proximity parameter $\varepsilon$ that yields a
Vietoris-Rips complex $\mathcal{R}_{\varepsilon}$ which best describes the topological
and geometric structure from which that data point cloud was sampled. Instead
one considers all possible values of $\varepsilon$ and one determines which
topological features persist as $\varepsilon$ increases.

\subsection{Persistent homology}

\label{sec:ph}

Persistent homology completely describes how homology persists as one steps
through the filtration. For example, consider a filtration of Vietoris-Rips
complexes%
\[
\mathcal{R}_{\varepsilon_{0}}\subset\mathcal{R}_{\varepsilon_{1}}\subset
\cdots\subset\mathcal{R}_{\varepsilon_{m}},%
\]
for $\eps_0 < \eps_1 < \cdots < \eps_m$.
One is interested in topological features that persist as the proximity parameter
$\varepsilon$ ranges from $\varepsilon_{0}$ to $\varepsilon_{m}$. For a given
value of $\varepsilon$, the number of $p$-dimensional holes of the Vietoris-Rips
complex $\mathcal{R}_{\varepsilon}$ is determined as the dimension of the vector space given by the $p$th
homology group $H_{p}(\mathcal{R}_{\varepsilon})$, where coefficients are
taken to be in some fixed field, typically $\mathbb{Z}/2$. Let
\[
\beta_{p}\left(  \mathcal{R}_{\varepsilon}\right)  =\dim\left[  H_{p}\left(
\mathcal{R}_{\varepsilon}\right)  \right]
\]
which is known as the \emph{$p$th Betti number}. For instance, $\beta_{0}\left(
\mathcal{R}_{\varepsilon}\right)  $ is the number of connected components or
clusters of the point cloud data set while $\beta_{1}\left(  \mathcal{R}%
_{\varepsilon}\right)  $ the number of holes or tunnels in the Vietoris-Rips complex
$\mathcal{R}_{\varepsilon}$.
However, even knowing the Betti numbers at all values of $\varepsilon$, one has no information on whether or not the corresponding topological features persist from one value of $\varepsilon$ to the next.
Persistent homology remedies this defect by encoding not only the Betti numbers, but the \emph{persistent Betti numbers}, given by
\[
  \beta_i^j = \rank \left( H_p(\mathcal{R}_{\varepsilon_i}) \to H_p(\mathcal{R}_{\varepsilon_j}) \right)
\]
where $H_p(\mathcal{R}_{\varepsilon_i}) \to H_p(\mathcal{R}_{\varepsilon_j})$ is the linear map induced by the inclusions $\mathcal{R}_{\varepsilon_i} \subset \mathcal{R}_{\varepsilon_j}$. The image of this linear map is called a \emph{persistent homology group}.

\begin{figure}
  \centering
  \includegraphics[width=60mm]{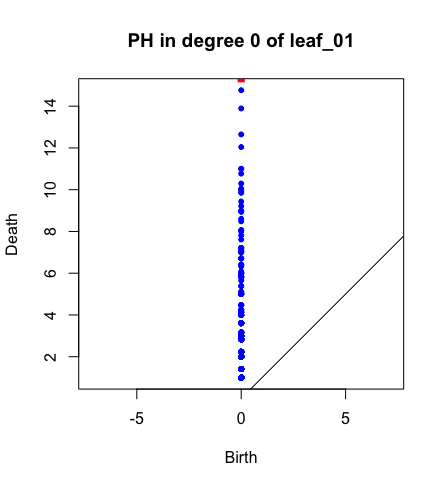} \quad
  \includegraphics[width=60mm]{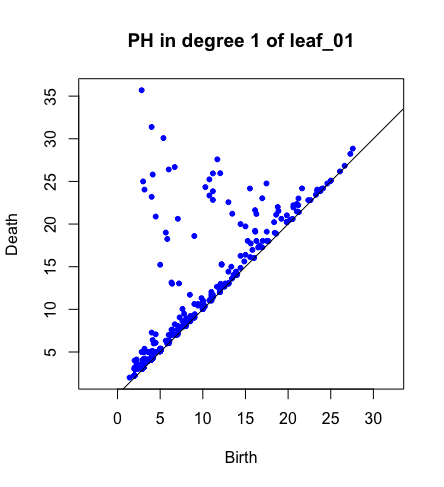}
  \caption{Persistence diagrams for homology in degree 0 (left) and degree 1 (right) of a Vietoris-Rips complex.}
  \label{fig:pd}
\end{figure}

The \emph{persistence diagram} gives a complete summary of persistent homology as a collection of points $\{(b,d)\}$, where each $(\varepsilon_i,\varepsilon_j)$ represents a homology class that is born at $\varepsilon_i$ and dies at $\varepsilon_j$. To be precise, the multiplicity of the point $(\eps_i,\eps_j)$ in the persistence diagram is given by
\[
  \mu_i^j = \beta_{i-1}^j - \beta_i^j + \beta_i^{j-1} -\beta_{i-1}^{j-1}.
\]
See~\cite{CoEdHa:2007} for more details.
Two persistence diagrams are given in Figure~\ref{fig:pd}.
For a point $(b,d)$ in the persistence diagram, the quantity $d-b$ is called its \emph{persistence}.

It is sometimes said that proximity parameter $\varepsilon$ ranges from
$\varepsilon_{0}$ to $\varepsilon_{m}$ short-lived topological features are
assumed to represent topological (statistical) noise while the features which
persist over a wide range proximity parameter $\varepsilon$ values represent a
topological signal.
However, we will show that it can be the case that long-lived features
represent noise and that short-lived features represent a geometric signal.

\begin{remark}
\emph{Persistent homology of the \v Cech complex and the Vietoris-Rips complex.}
  Note that while, in general, the homotopy types of $\mathcal{R}_{\varepsilon}$
and $\mathcal{X}_{\varepsilon}$ are not the same, it is true that for all
$\varepsilon$%
\[
\mathcal{C}_{\varepsilon}\subset\mathcal{R}_{\varepsilon}\subset
\mathcal{C}_{2\varepsilon}%
\]
Which means that if \v{C}ech complexes $\mathcal{C}_{\varepsilon}$ and
$\mathcal{C}_{2\varepsilon}$ are effective in detecting persistent topological
and geometric features then $\mathcal{R}_{\varepsilon}$ will also be effective.
\end{remark}

\subsection{Persistence landscapes and statistical inference}
\label{sec:pl}

In order to facilitate statistical inference we wish to give a complete (i.e. invertible) unique (i.e. injective) encoding of the persistence diagram as a vector.

\begin{figure}
  \centering
  \includegraphics[width=60mm]{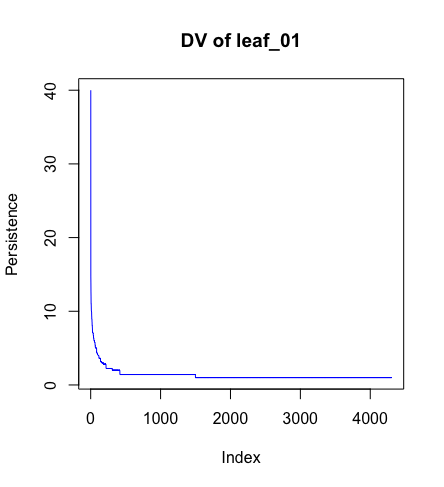} \quad
  \includegraphics[width=60mm]{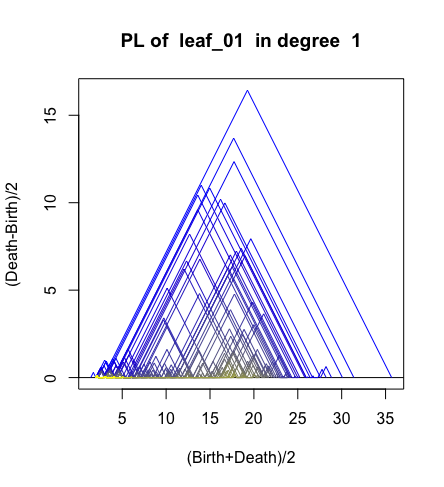}
  \caption{The death vector and persistence landscape corresponding to the persistence diagrams in Figure~\ref{fig:pd}.}
  \label{fig:dv-pl}
\end{figure}

For the Vietoris-Rips complex, since all vertices appear at filtration value $0$, all of the points in the persistence diagram for homology in degree 0 have birth coordinate 0 (see the left side of Figure~\ref{fig:pd}).
Thus,
all of the information is included in the death times (the times when connected
components merge). As such, we encode the persistence diagram using the corresponding order
statistic. We call this the \emph{death vector}. See the left hand side of Figure~\ref{fig:dv-pl}.

For more general persistence diagrams, such as for homology in degree 1 for the Vietoris-Rips complex (see the right hand figure in Figure~\ref{fig:pd}), we use the persistence landscape~\cite{Bu:2015}, which we now describe.

For each point $(b,d)$ in the persistence diagram, consider the following function
\[
  f_{(b,d)}(t) =
  \begin{cases}
    t-b, \quad \text{if } b \leq t < \frac{b+d}{2},\\
    d-t, \quad \text{if } \frac{b+d}{2} \leq t < d,\\
    0, \quad \text{otherwise}.
  \end{cases}
\]
Then for $k \geq 1$, the \emph{$k$th persistence landscape function} of the persistence diagram $\mathcal{D}$ is given by
\begin{equation*}
  \lambda_k(t) = \kmax_{(b_i,d_i) \in \mathcal{D}} f_{(b_i,d_i)}(t),
\end{equation*}
where $\kmax$ denotes the $k$th largest element. The \emph{persistence landscape} consists of the sequence of functions $\{\lambda_1,\lambda_2,\lambda_3,\ldots\}$. Notice that by definition,
for all $t \in \mathbb{R}$, $\lambda_1(t) \geq \lambda_2(t) \geq \lambda_3(t) \geq \ldots$.
That is, the persistence landscape is a decreasing sequence of functions.

It remains to turn this sequence of functions into a vector. This is done be evaluating the functions on a grid. Specifically, we evaluate the persistence landscape functions $\lambda_1, \ldots, \lambda_K$ for some sufficiently large $K$ at the values $a,a+\delta,a+2\delta,a+3\delta,\ldots,a+m\delta$ for appropriate choices of $a$, $\delta$ and $m$. The resulting values are concatenated to obtain a vector in $\mathbb{R}^{K(m+1)}$.

\section{Objects TDA Example}
\label{sec:leafs}

\begin{figure}[h]
\centering
\includegraphics[width=50mm]{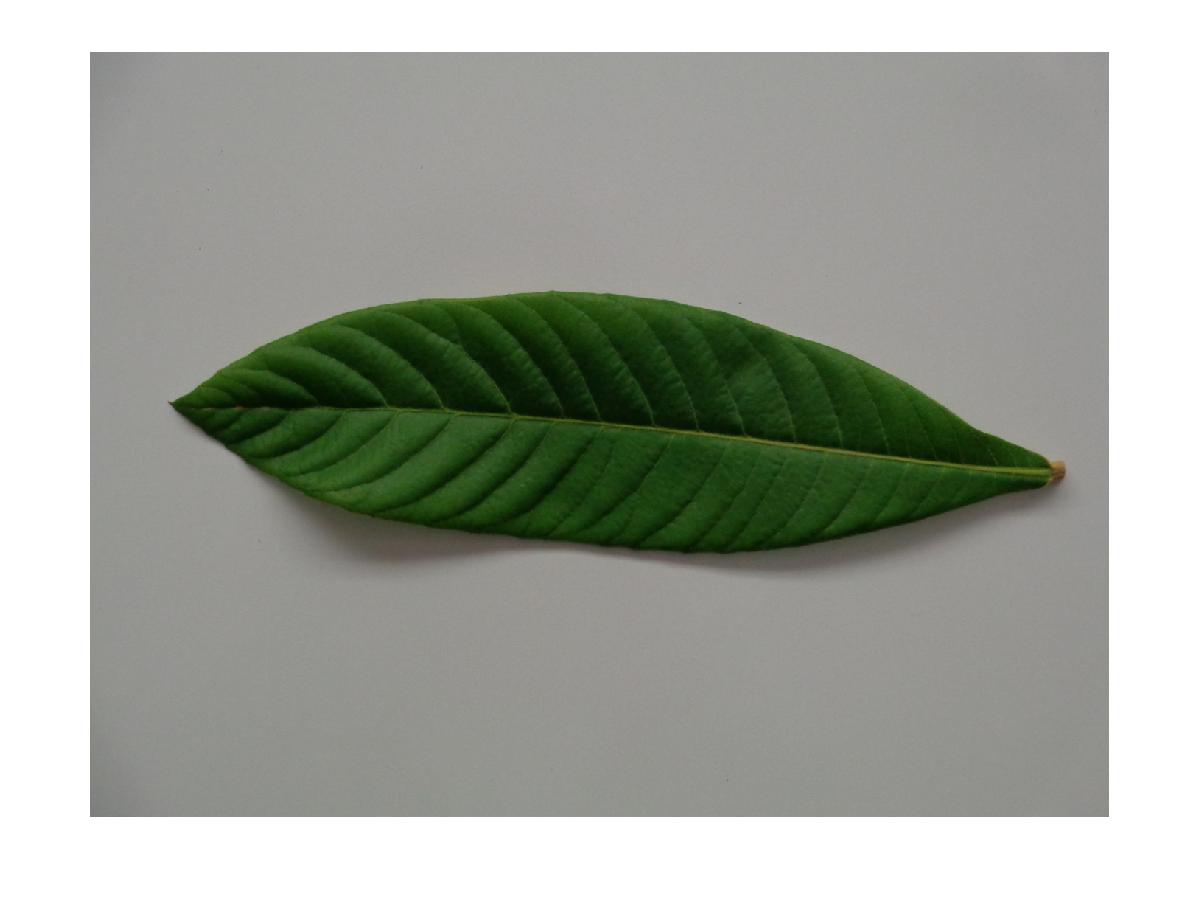} \quad
\includegraphics[width=50mm]{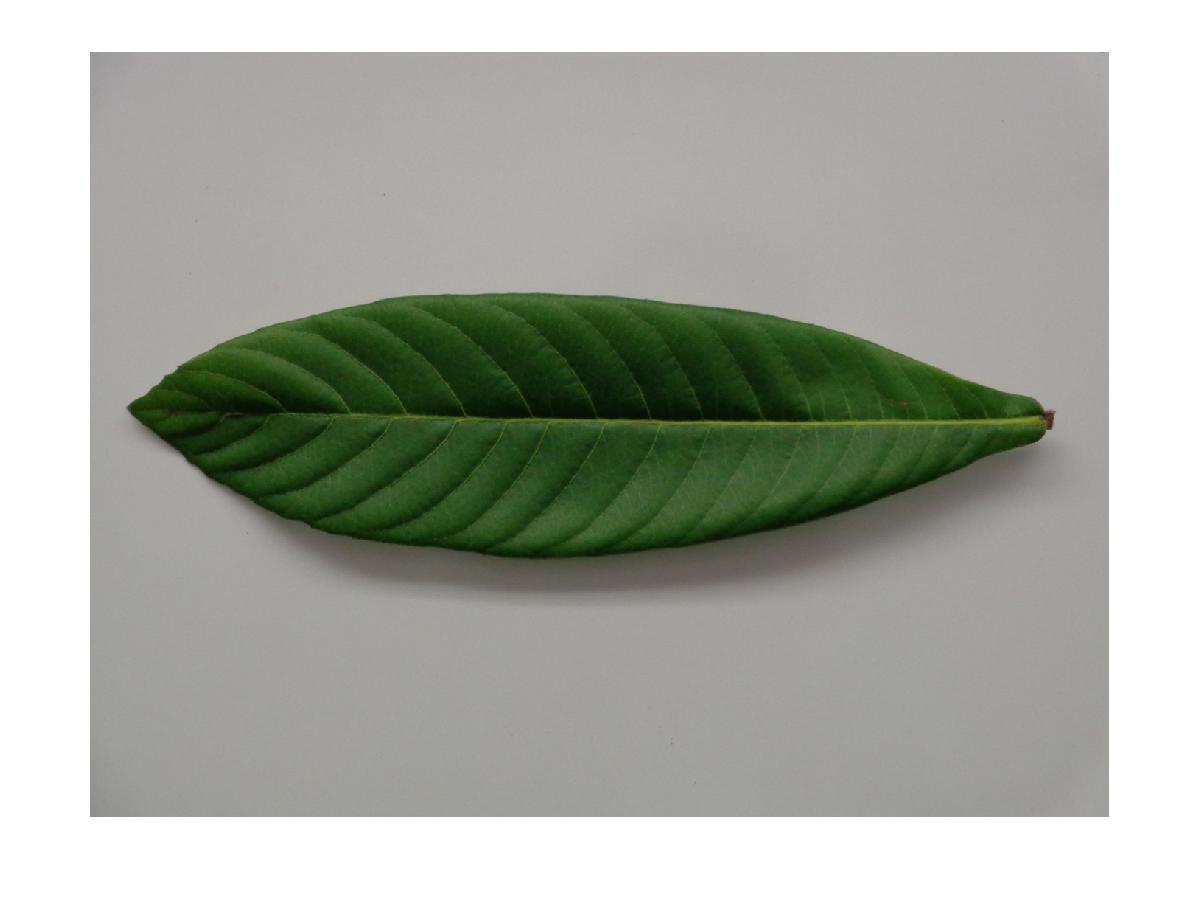}  \caption{Sample
original images of leaf A (left) and leaf B (right).}%
\label{fig:sample-leaf-images}%
\end{figure}

Here we analyze the leaf data from Patrangenaru et al. (2016)~\cite{PaPaYaQiLe:2016} (see www.stat.fsu.edu/$\sim$vic/Original-figures). This image data set
consists of two leaves, call them leaf A and leaf B, from the same
tree. Twenty pictures were taken of each leaf from different
perspectives, to yield a total of 40 pictures which are shown below in Figure
\ref{fig:ori_figure}. Two larger images are shown in Figure~\ref{fig:sample-leaf-images}.

\begin{figure}[h]
\begin{center}
\includegraphics[scale=0.035]{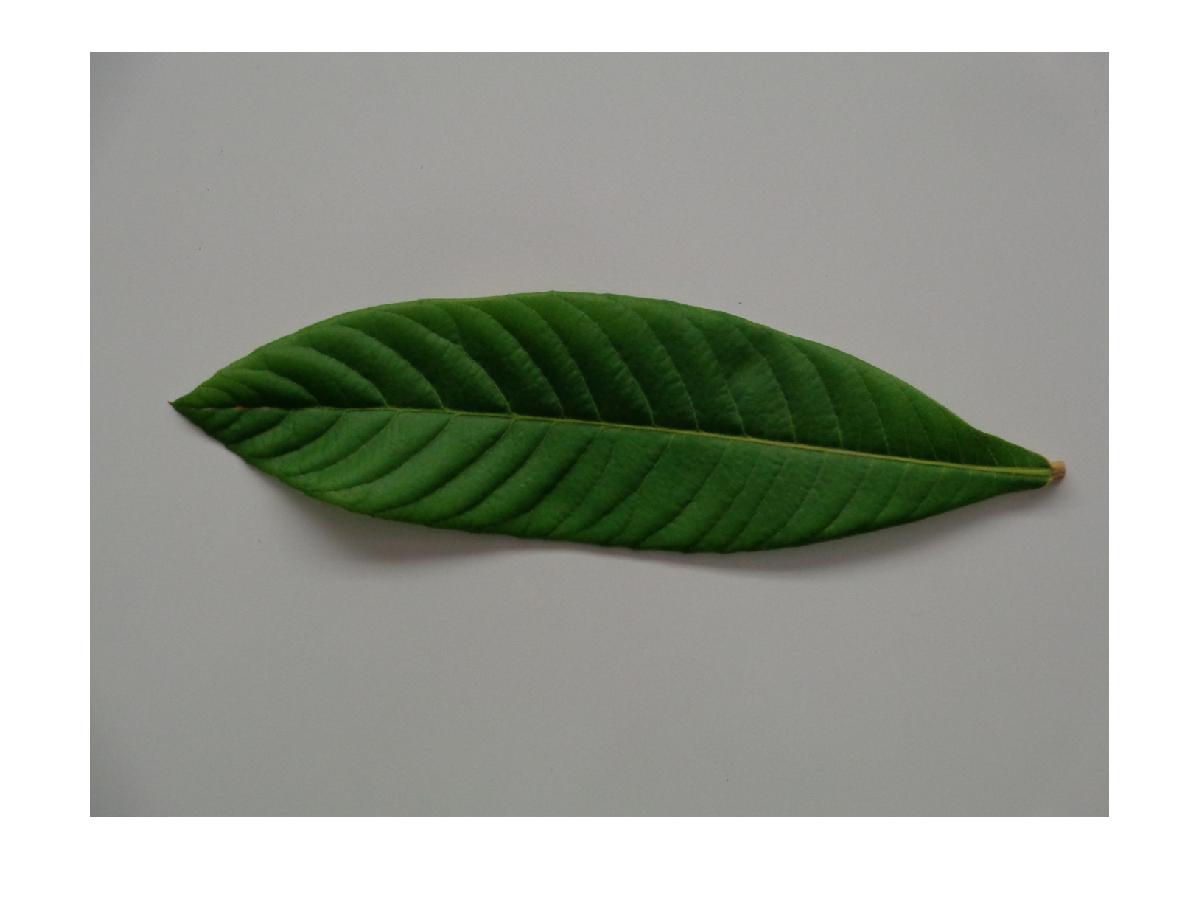}
\includegraphics[scale=0.035]{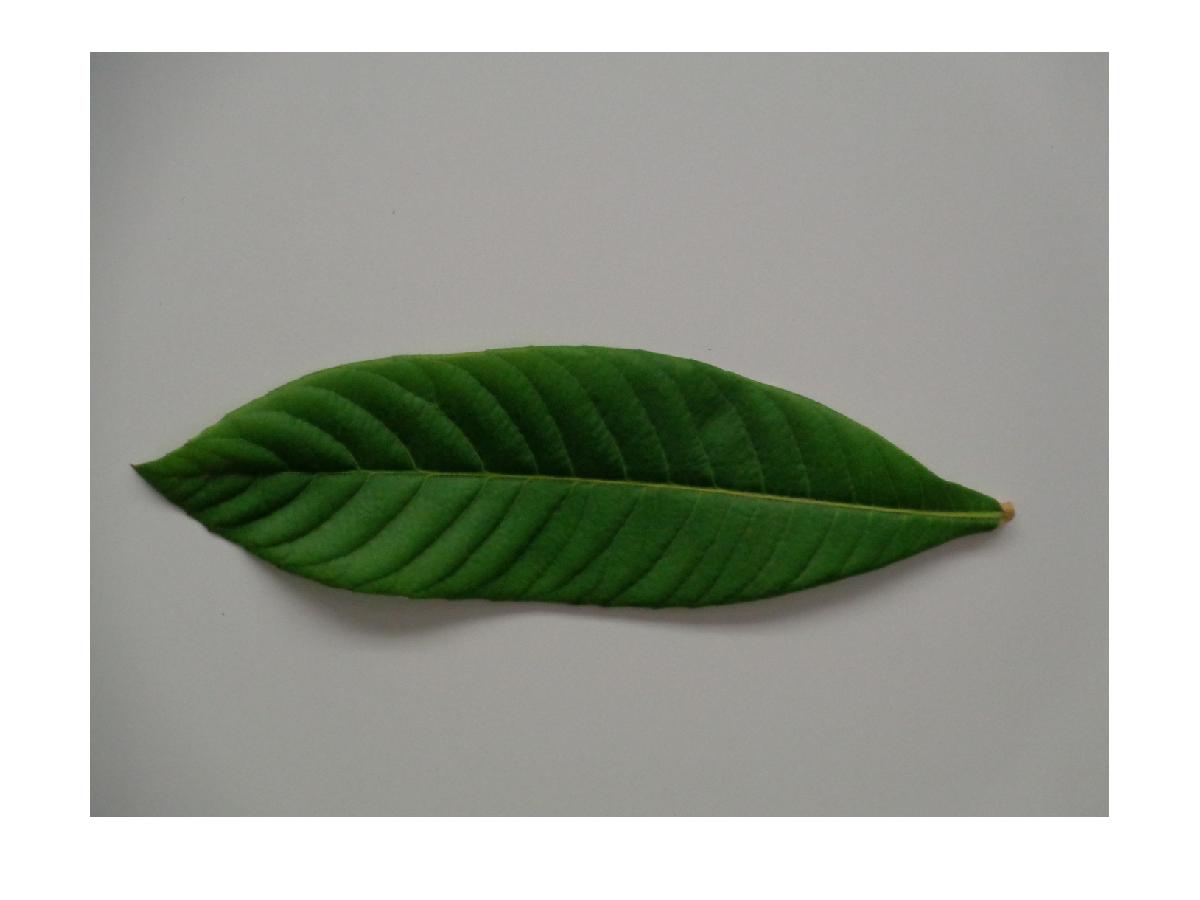}
\includegraphics[scale=0.035]{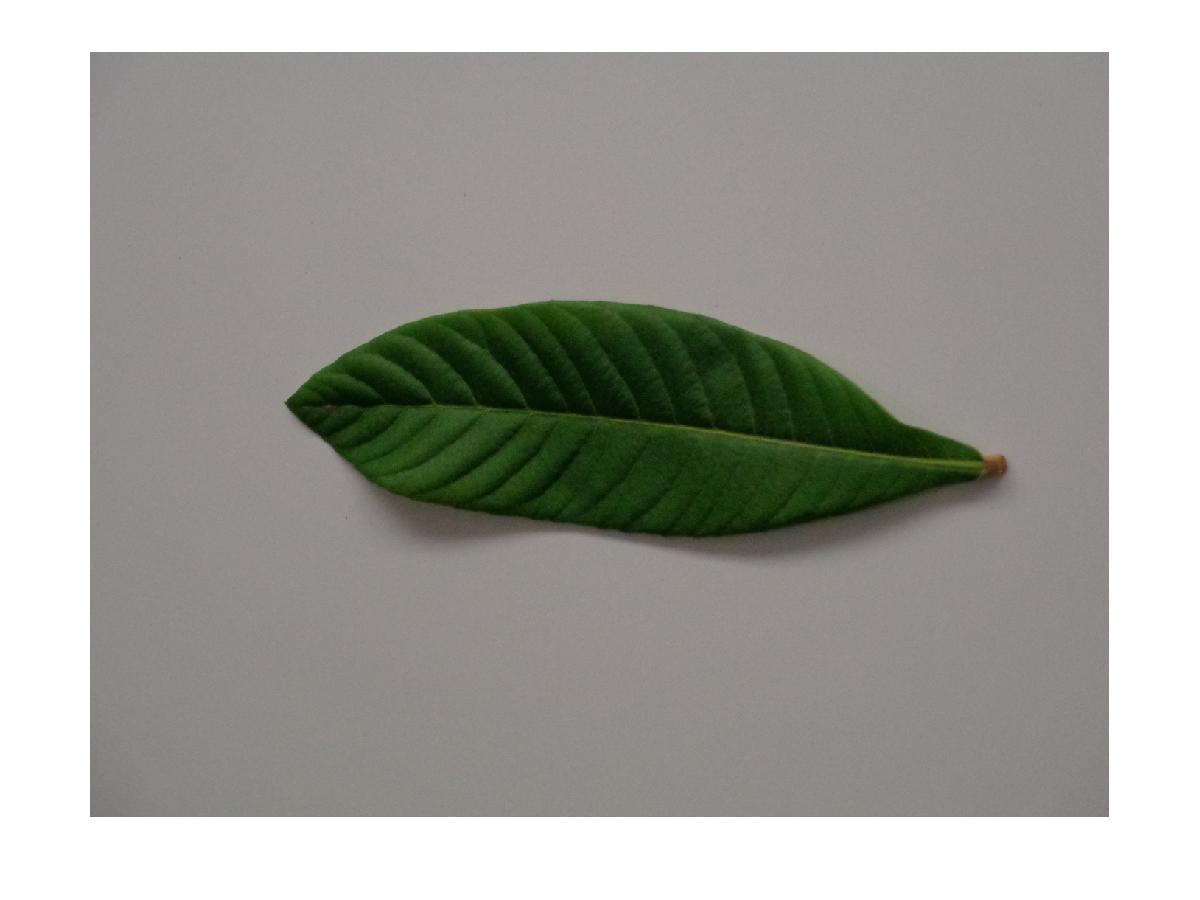}
\includegraphics[scale=0.035]{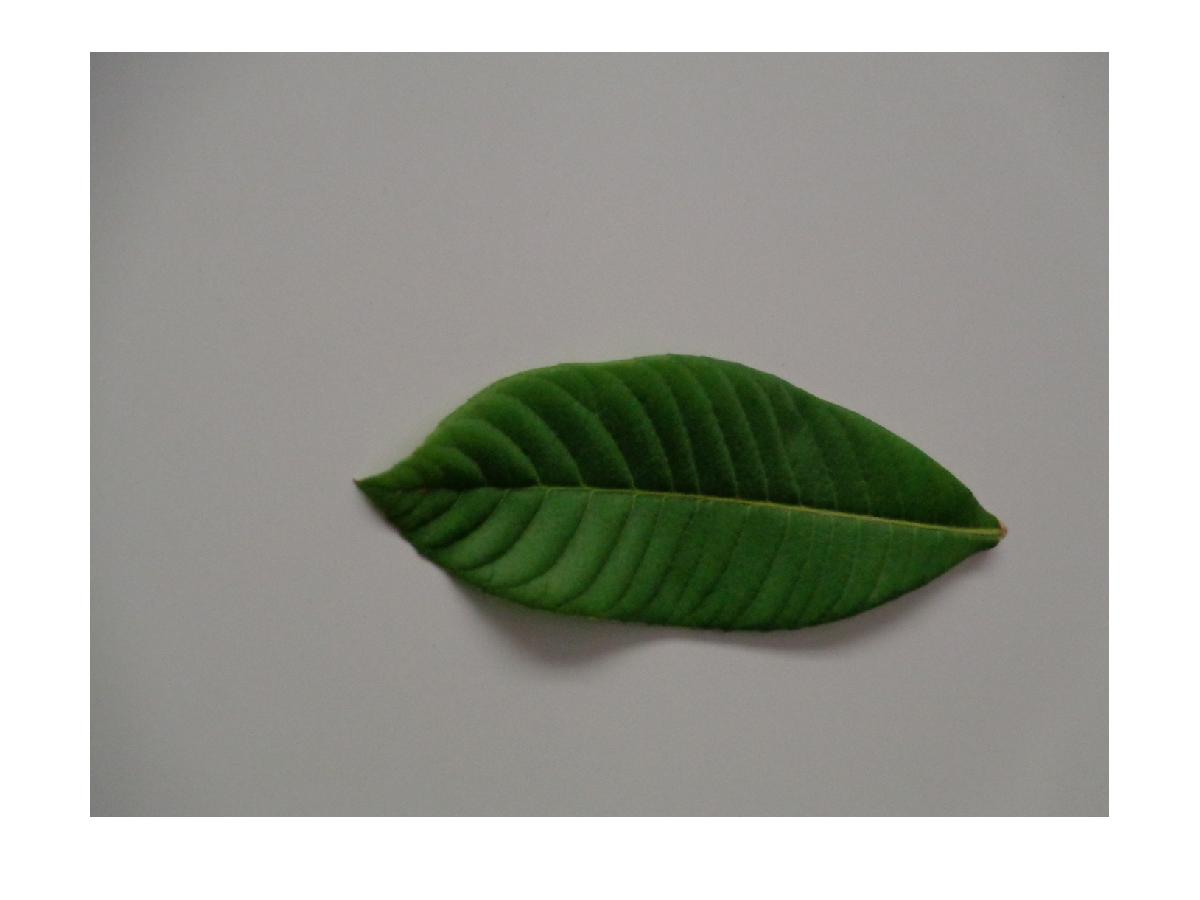}
\includegraphics[scale=0.035]{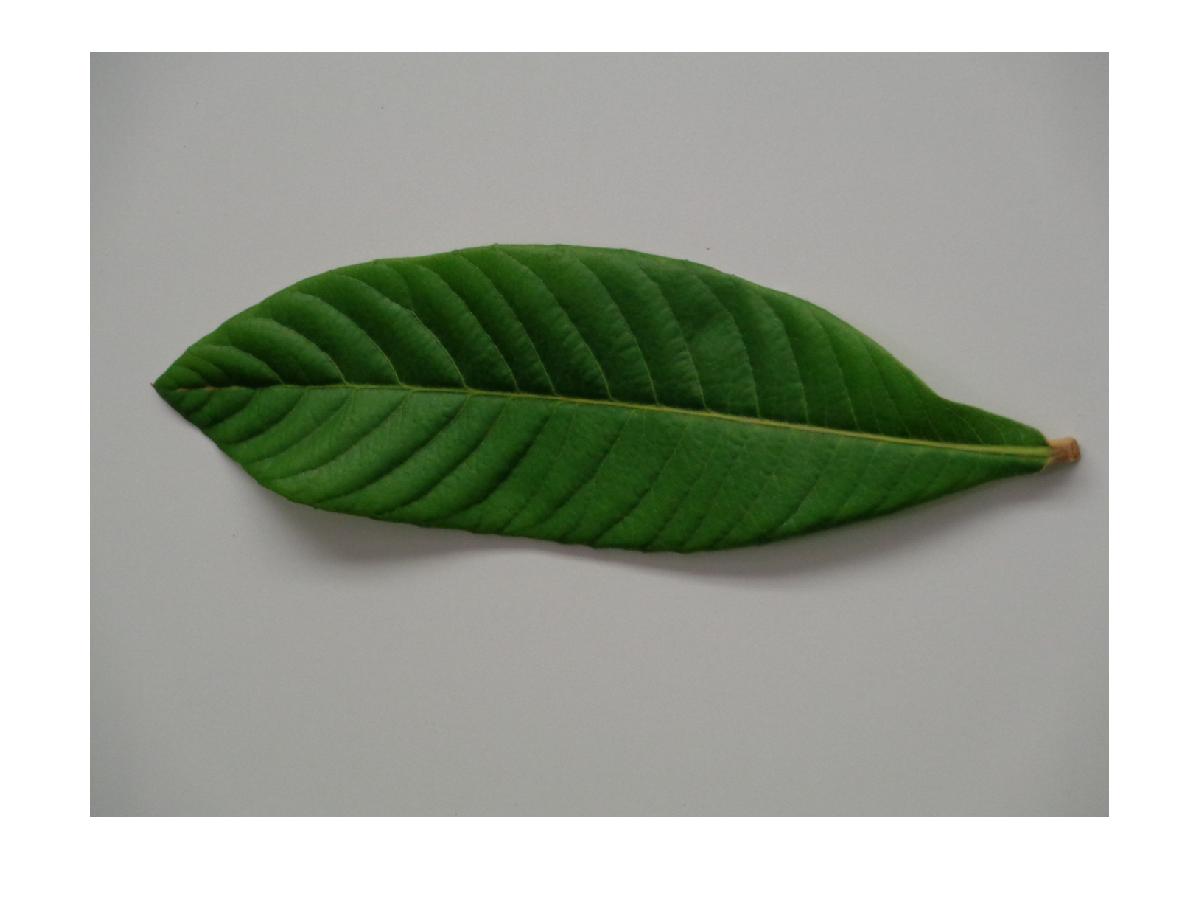}
\includegraphics[scale=0.035]{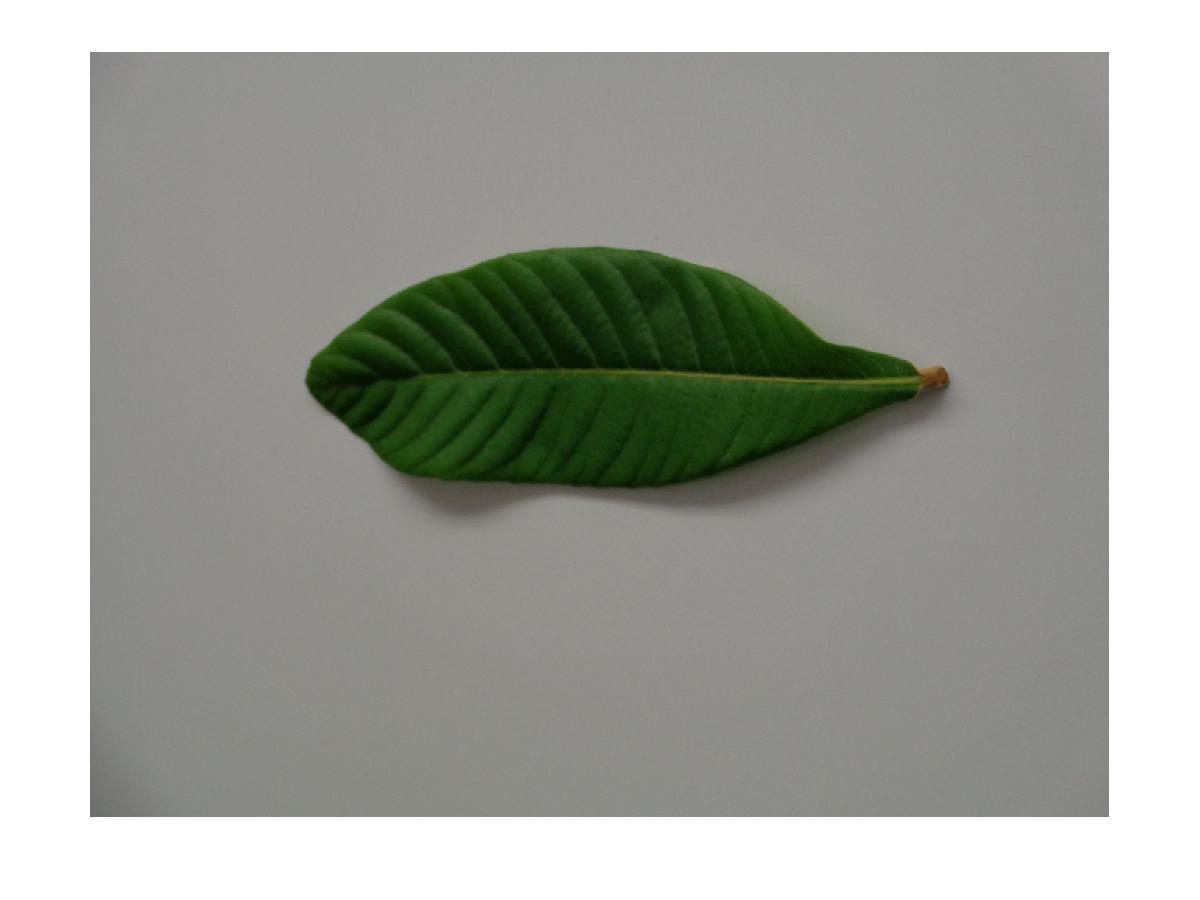}
\includegraphics[scale=0.035]{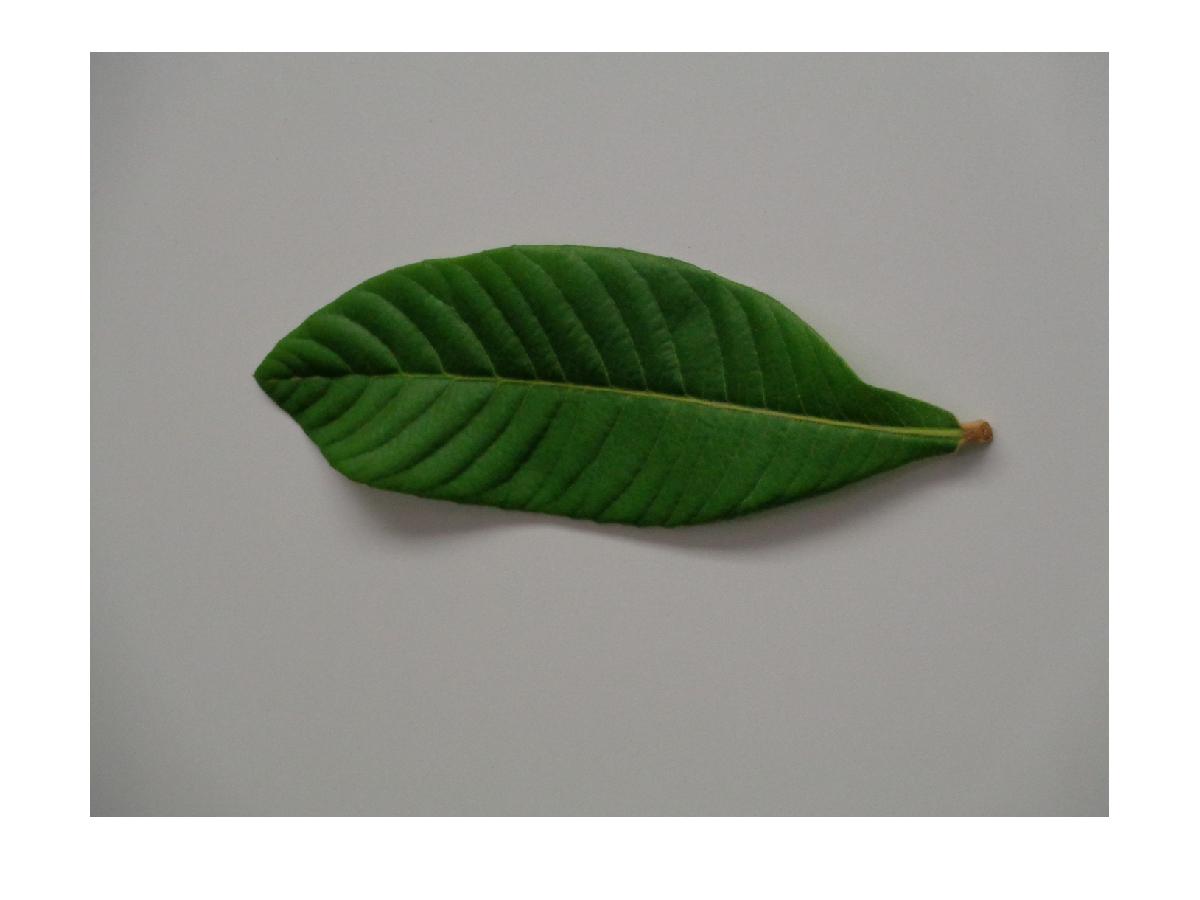}
\includegraphics[scale=0.035]{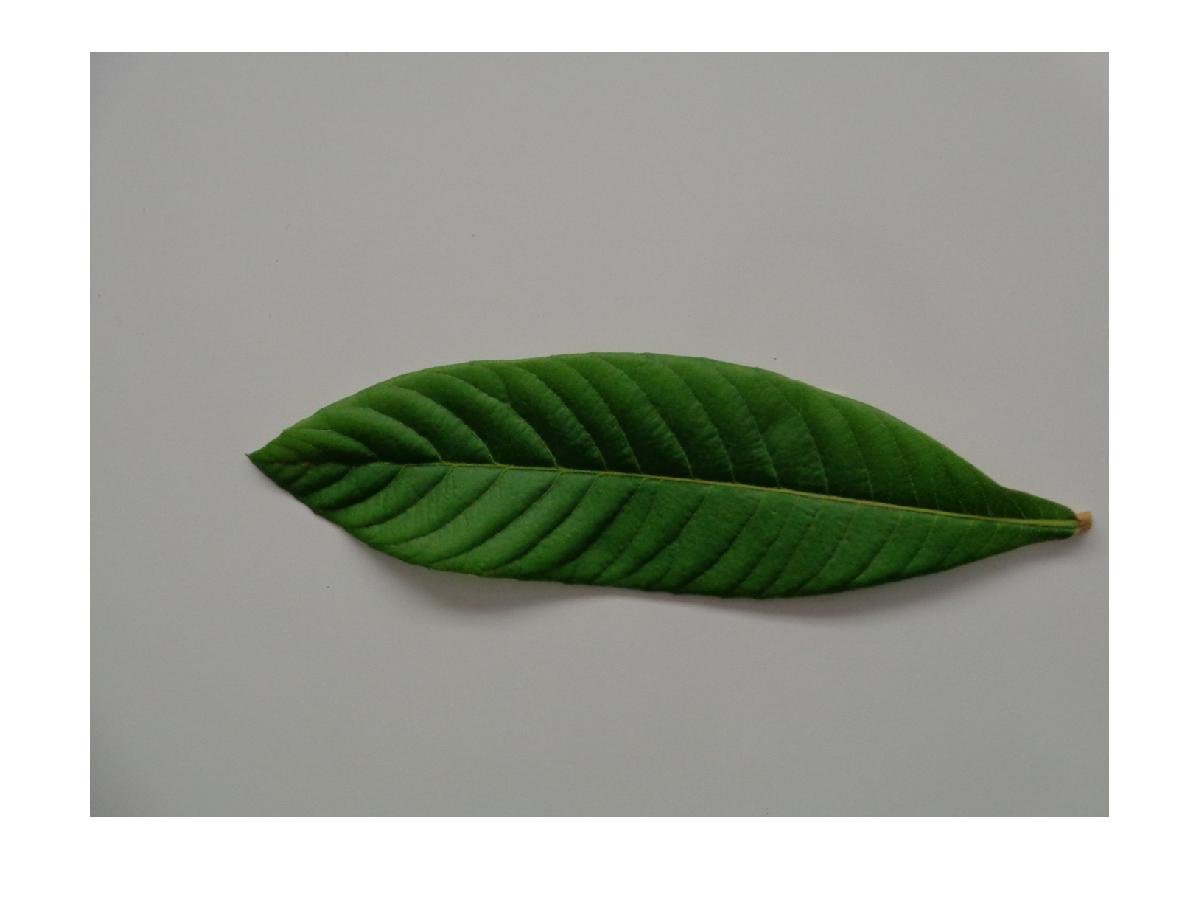}
\includegraphics[scale=0.035]{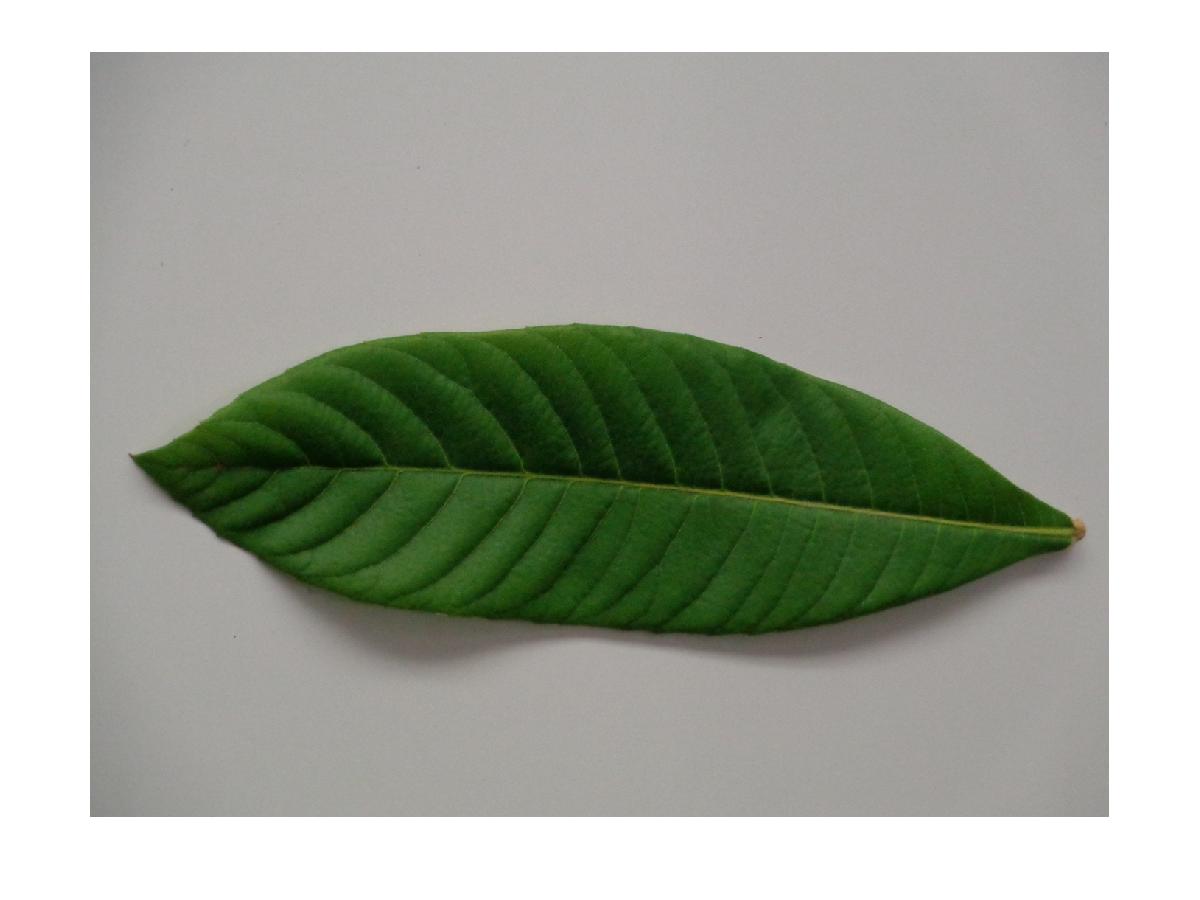}
\includegraphics[scale=0.035]{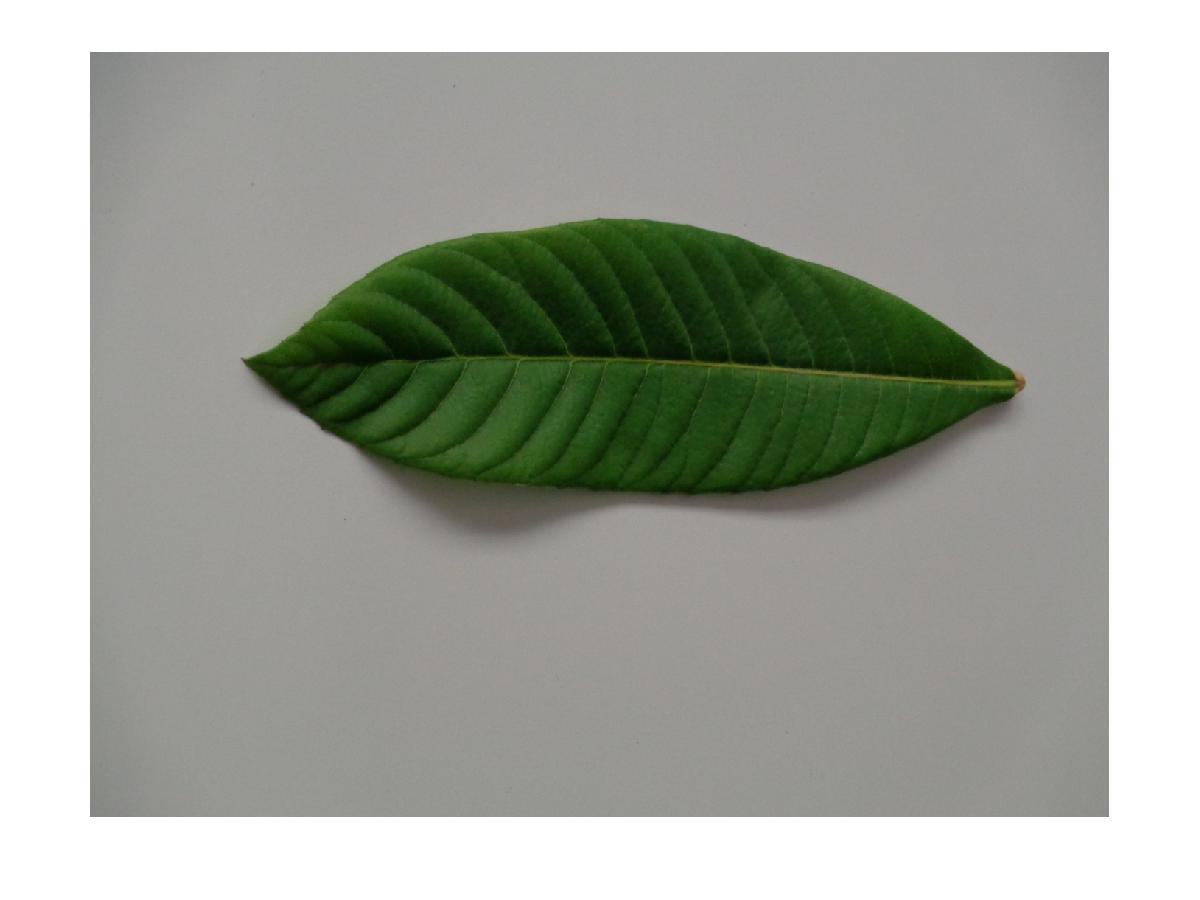}%
\\
\includegraphics[scale=0.035]{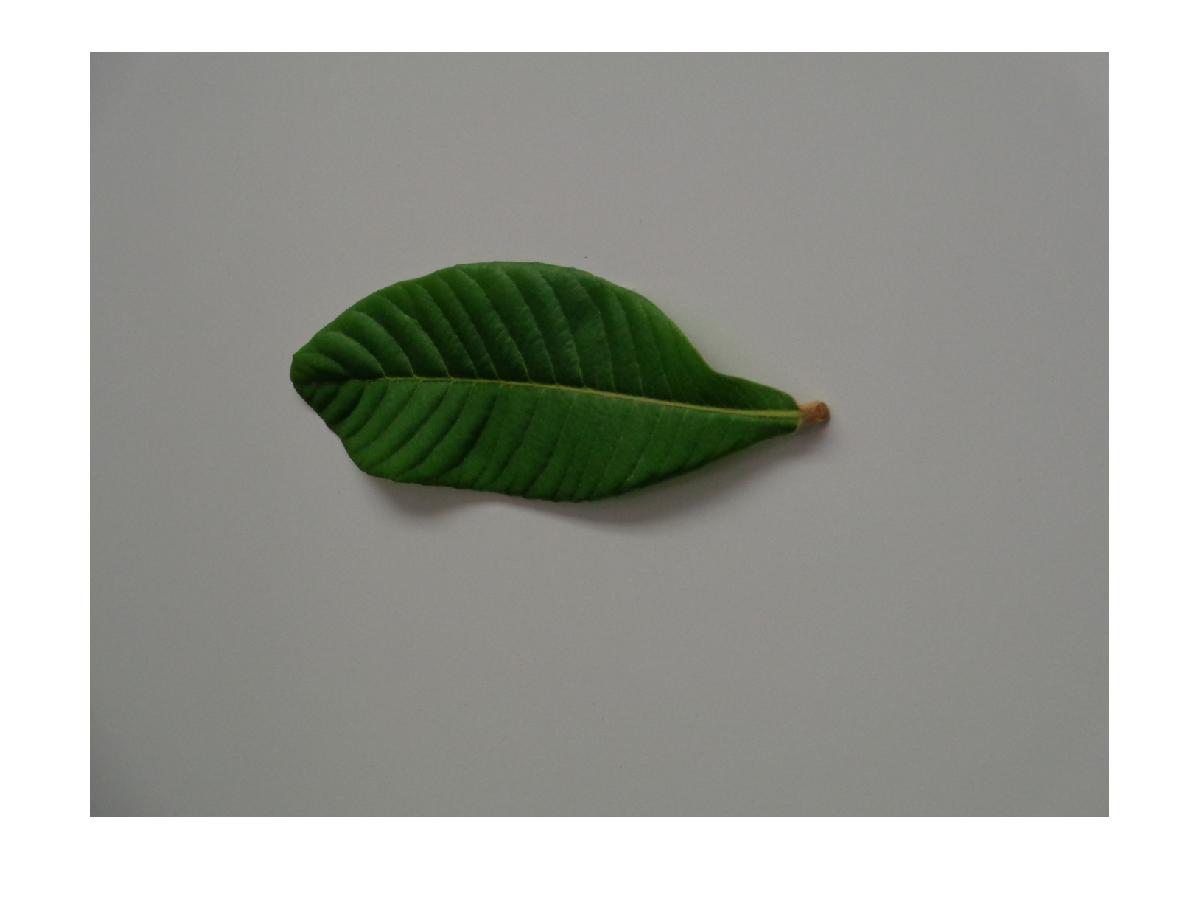}
\includegraphics[scale=0.035]{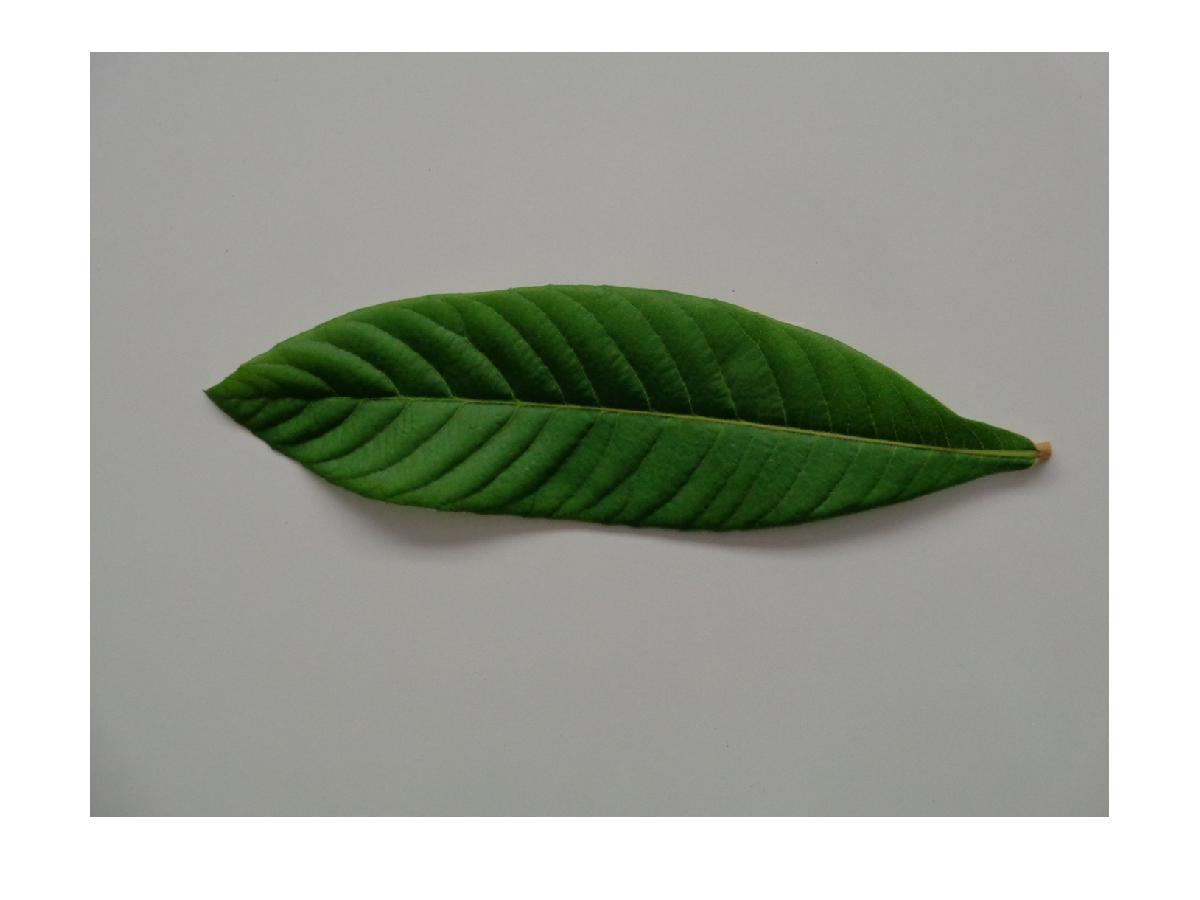}
\includegraphics[scale=0.035]{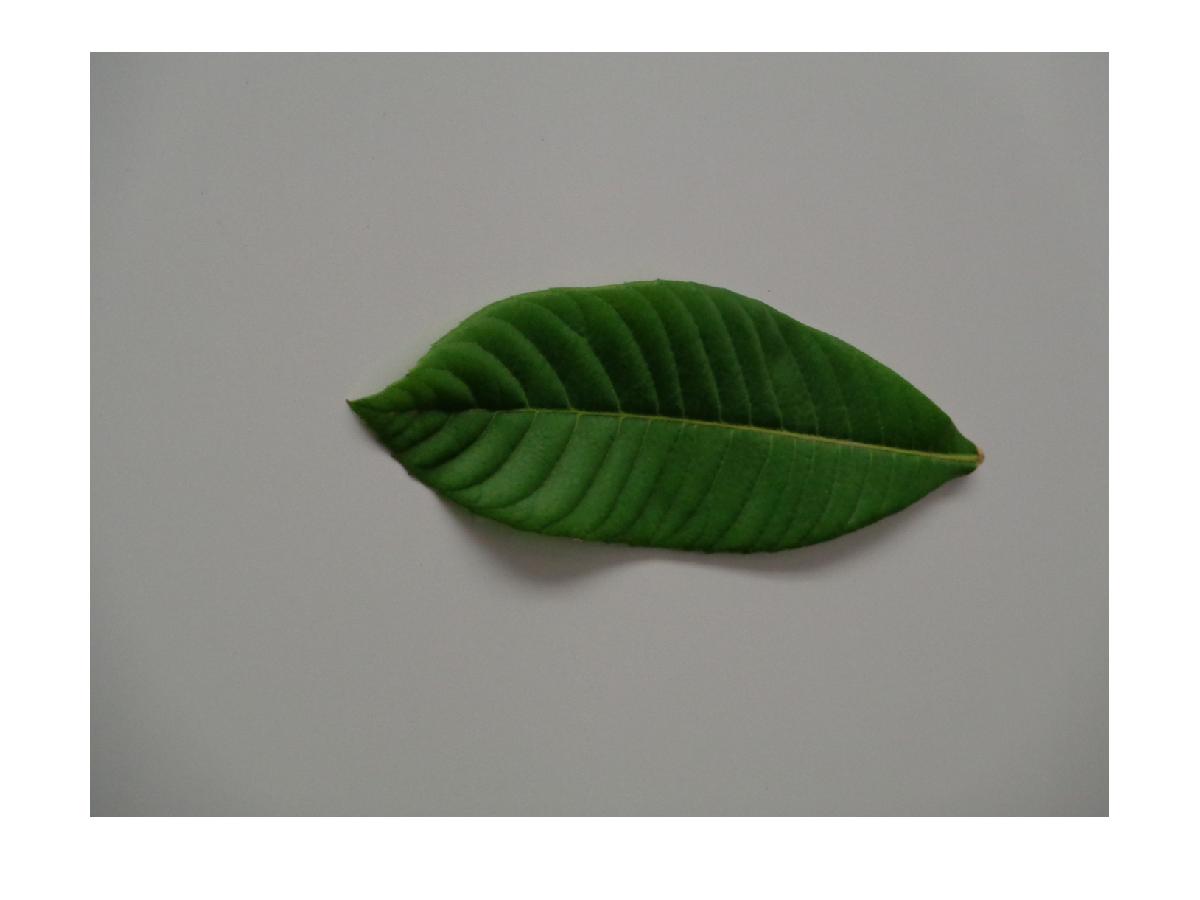}
\includegraphics[scale=0.035]{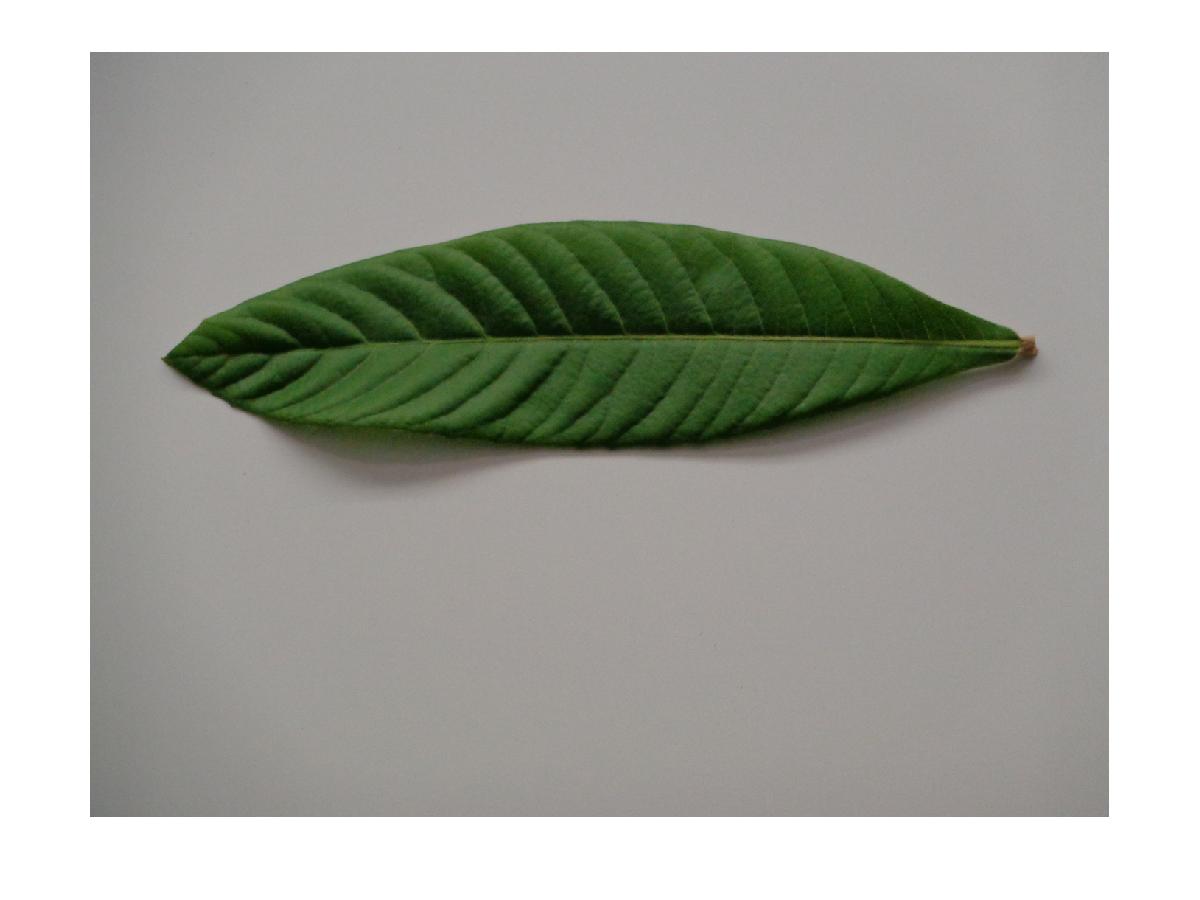}
\includegraphics[scale=0.035]{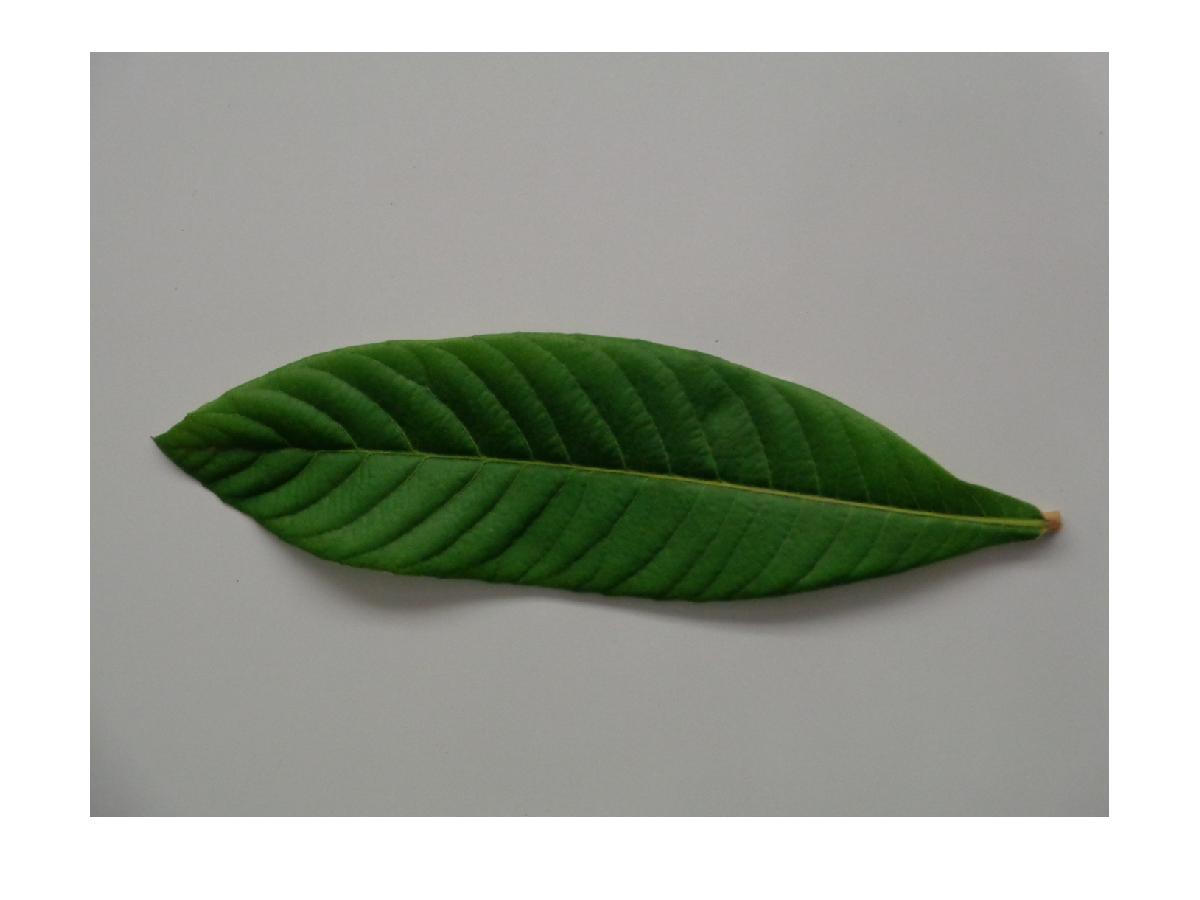}
\includegraphics[scale=0.035]{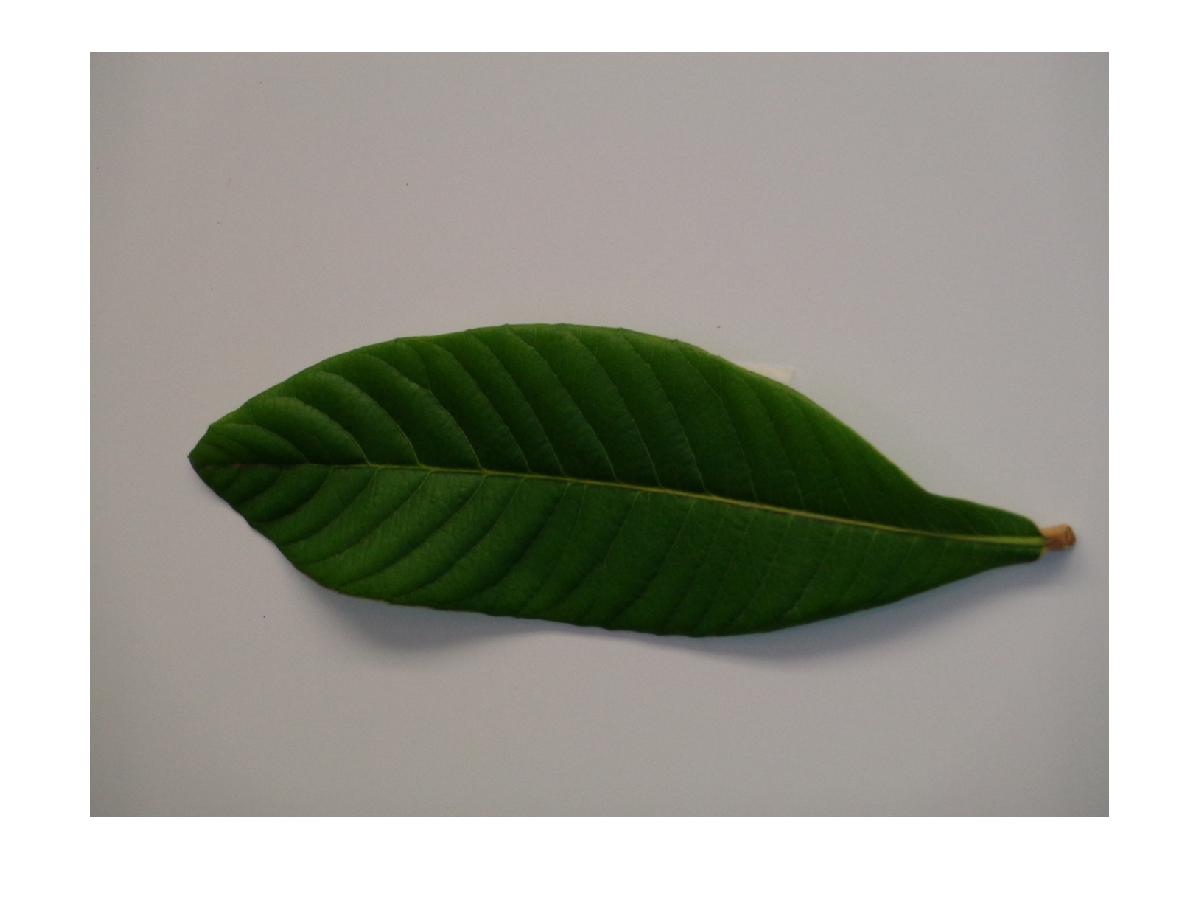}
\includegraphics[scale=0.035]{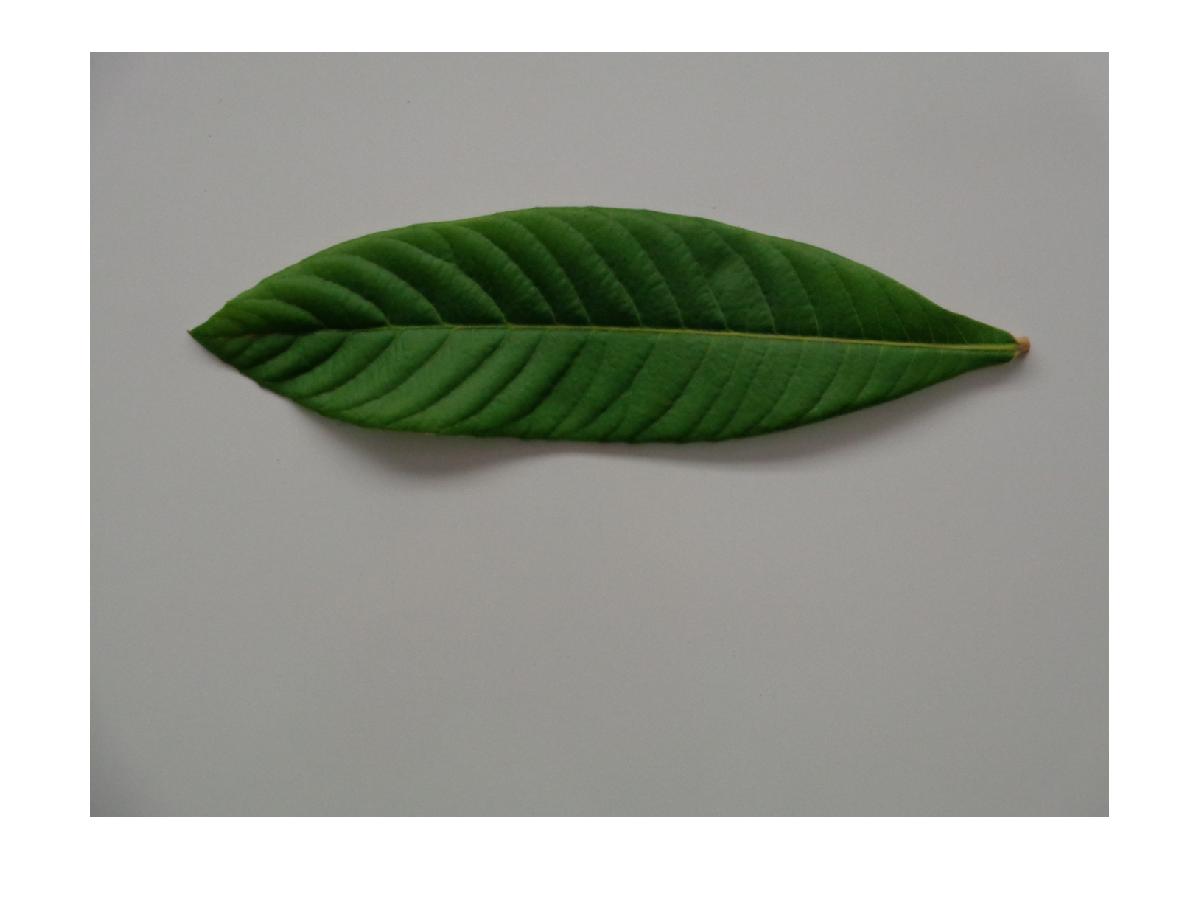}
\includegraphics[scale=0.035]{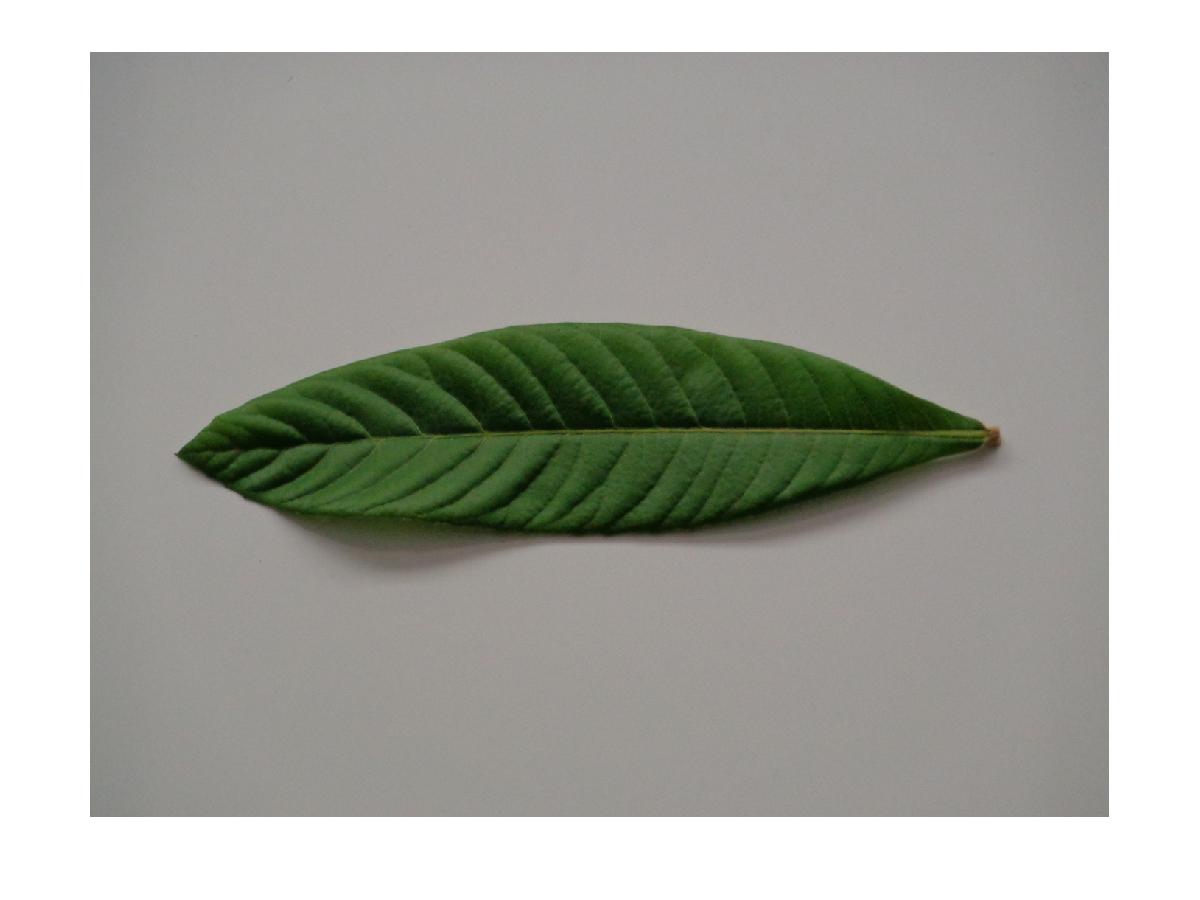}
\includegraphics[scale=0.035]{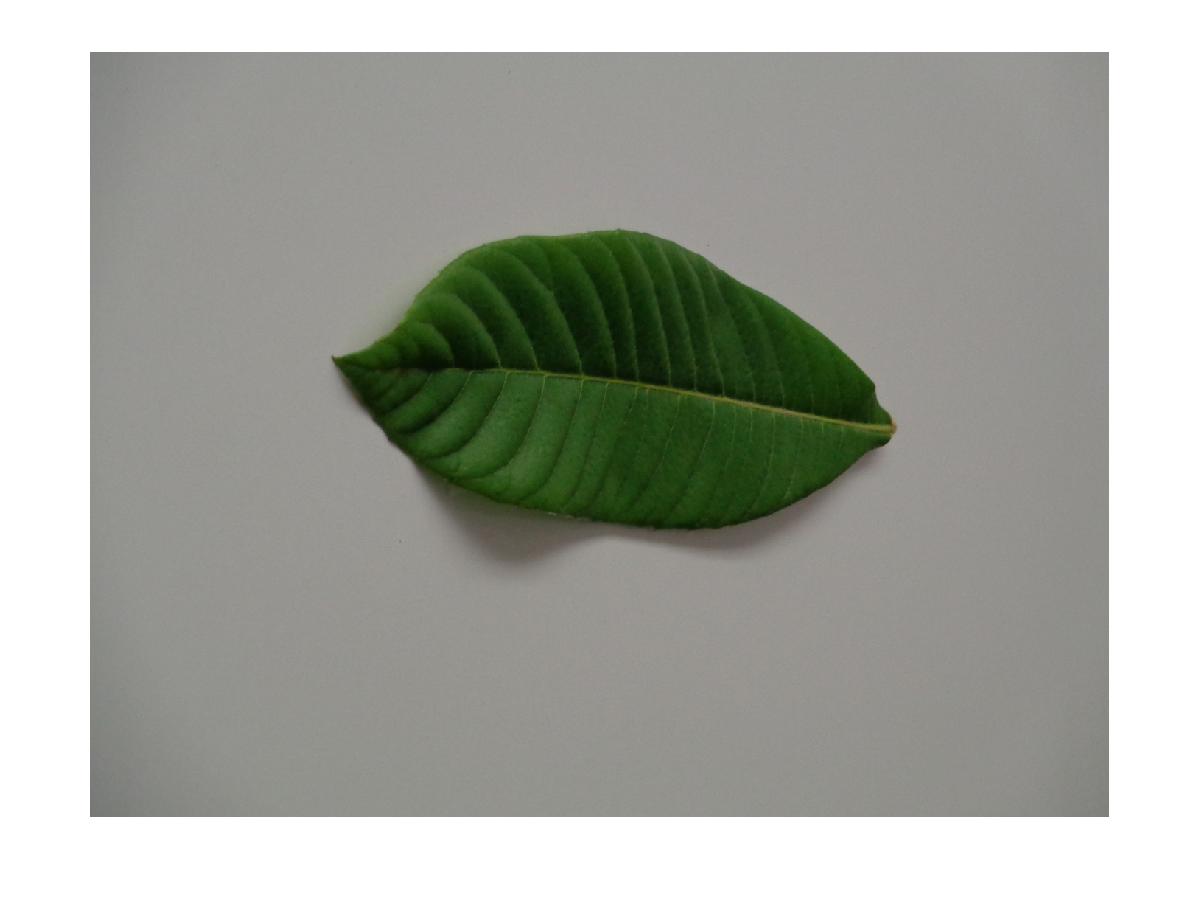}
\includegraphics[scale=0.035]{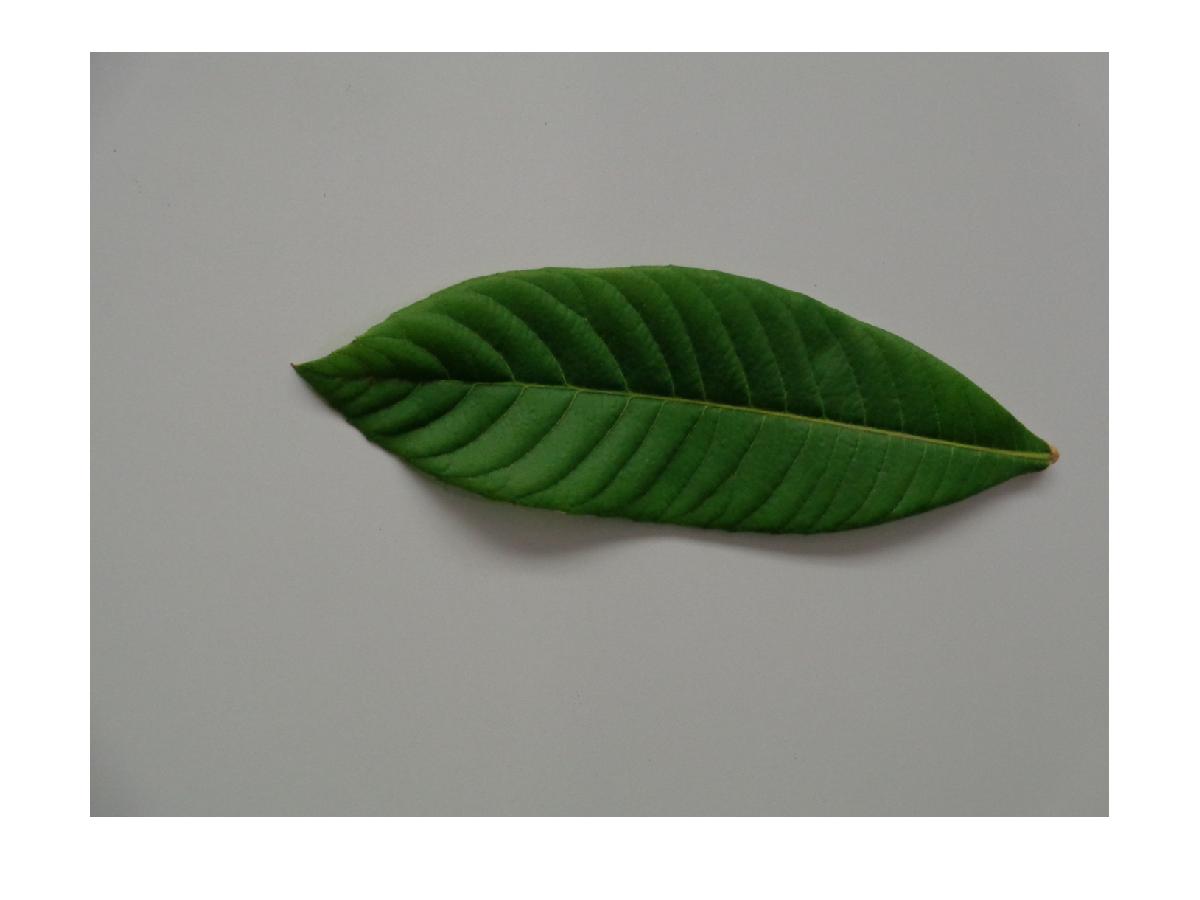}%
\\
\includegraphics[scale=0.035]{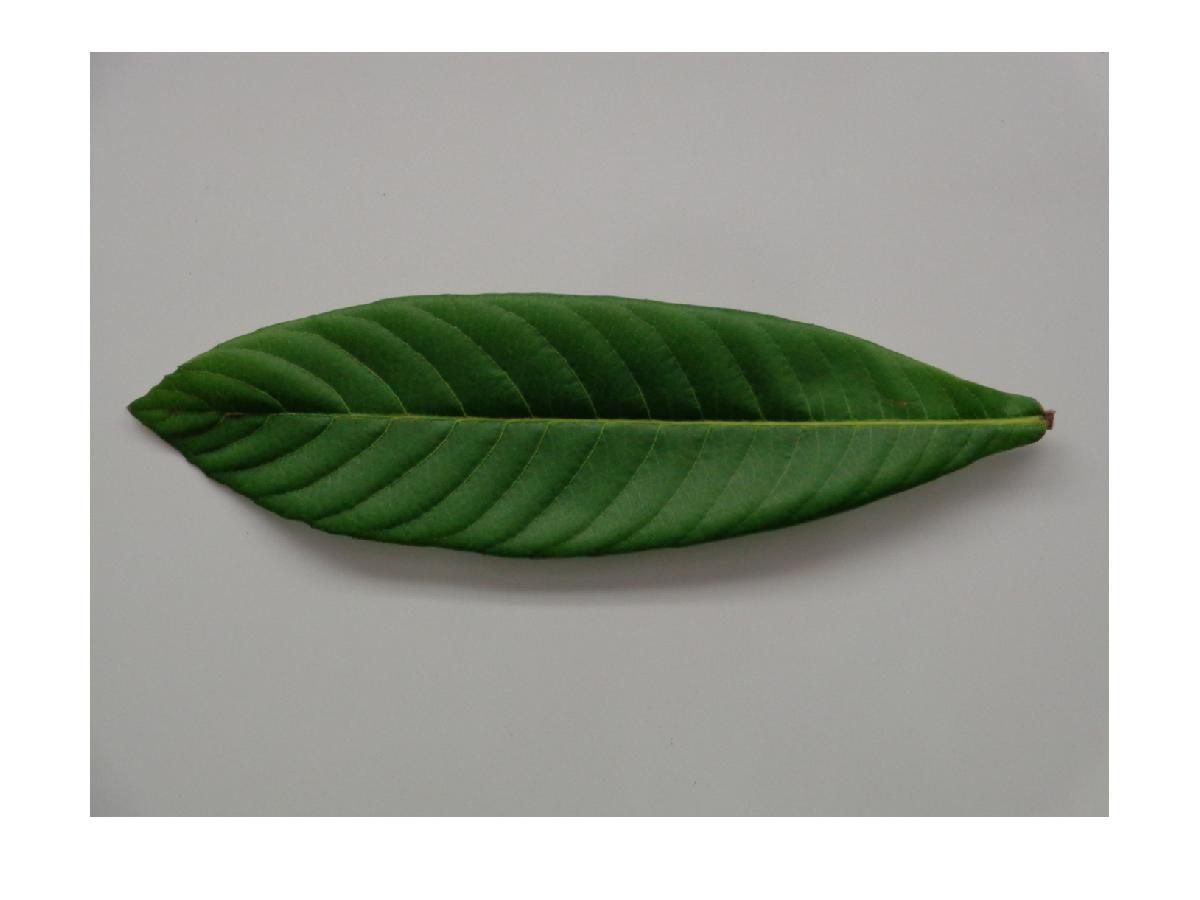}
\includegraphics[scale=0.035]{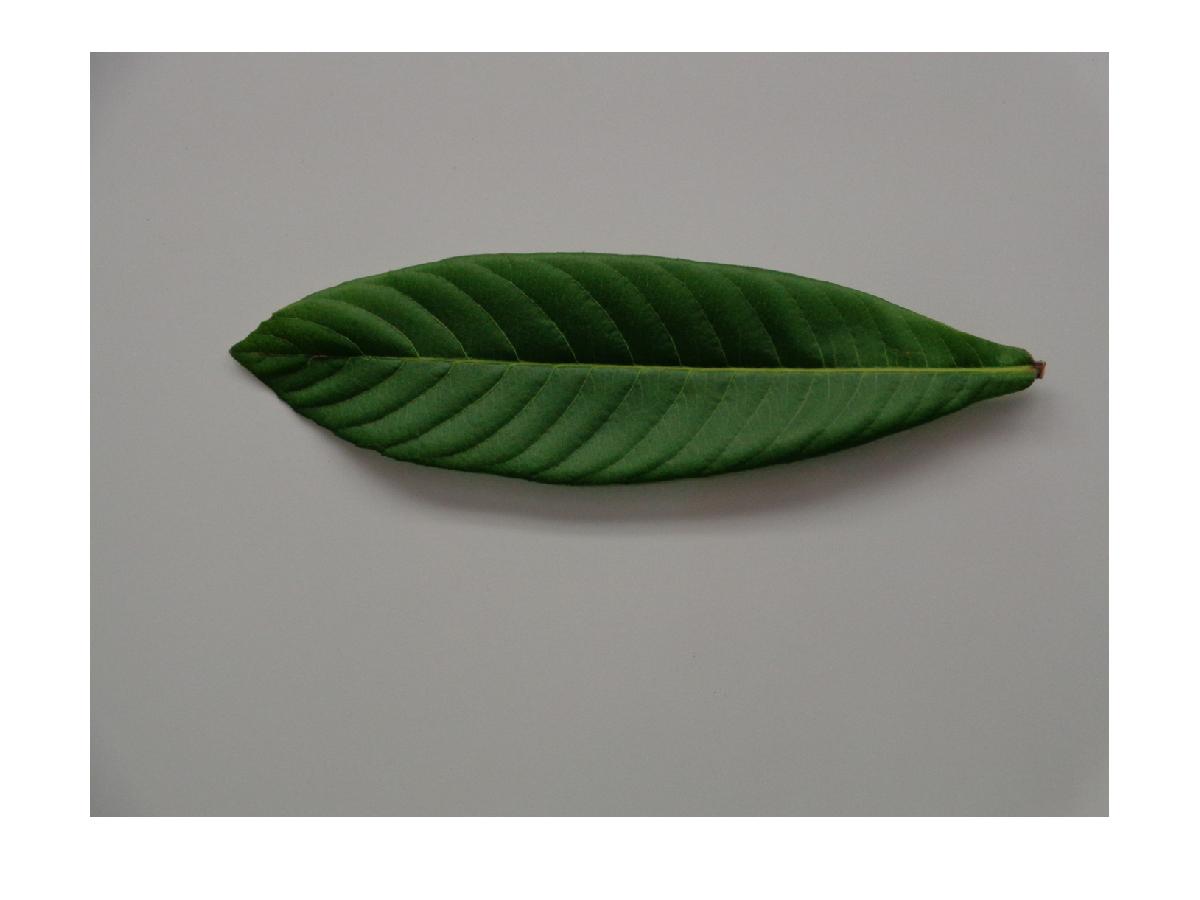}
\includegraphics[scale=0.035]{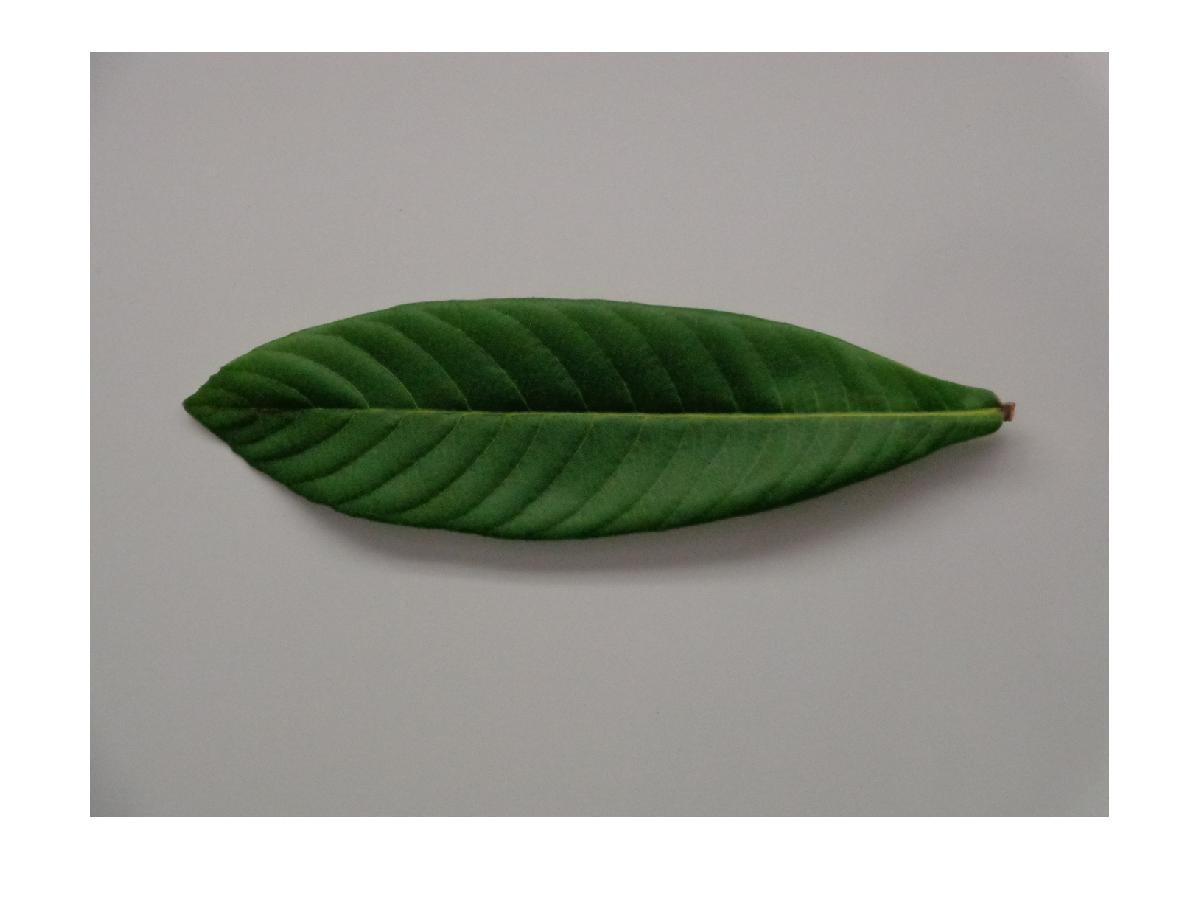}
\includegraphics[scale=0.035]{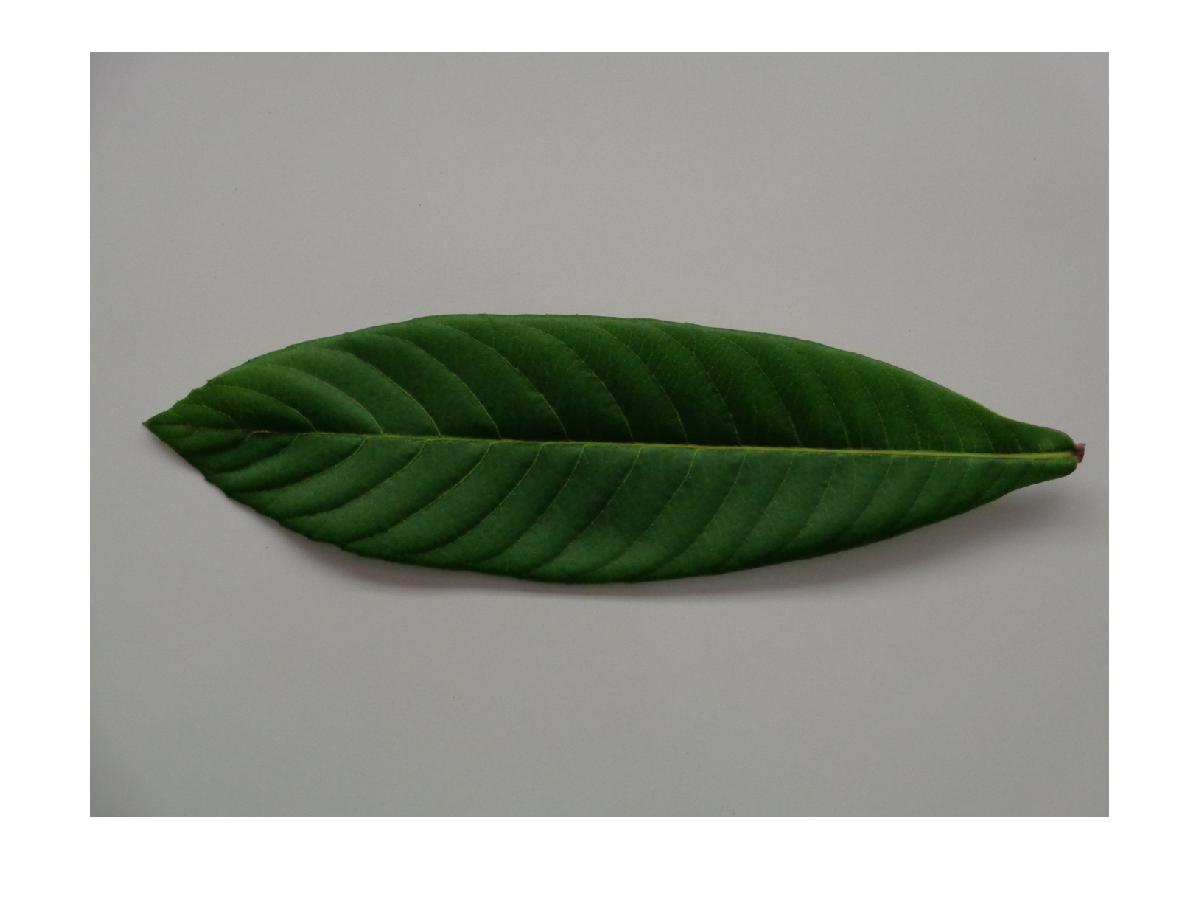}
\includegraphics[scale=0.035]{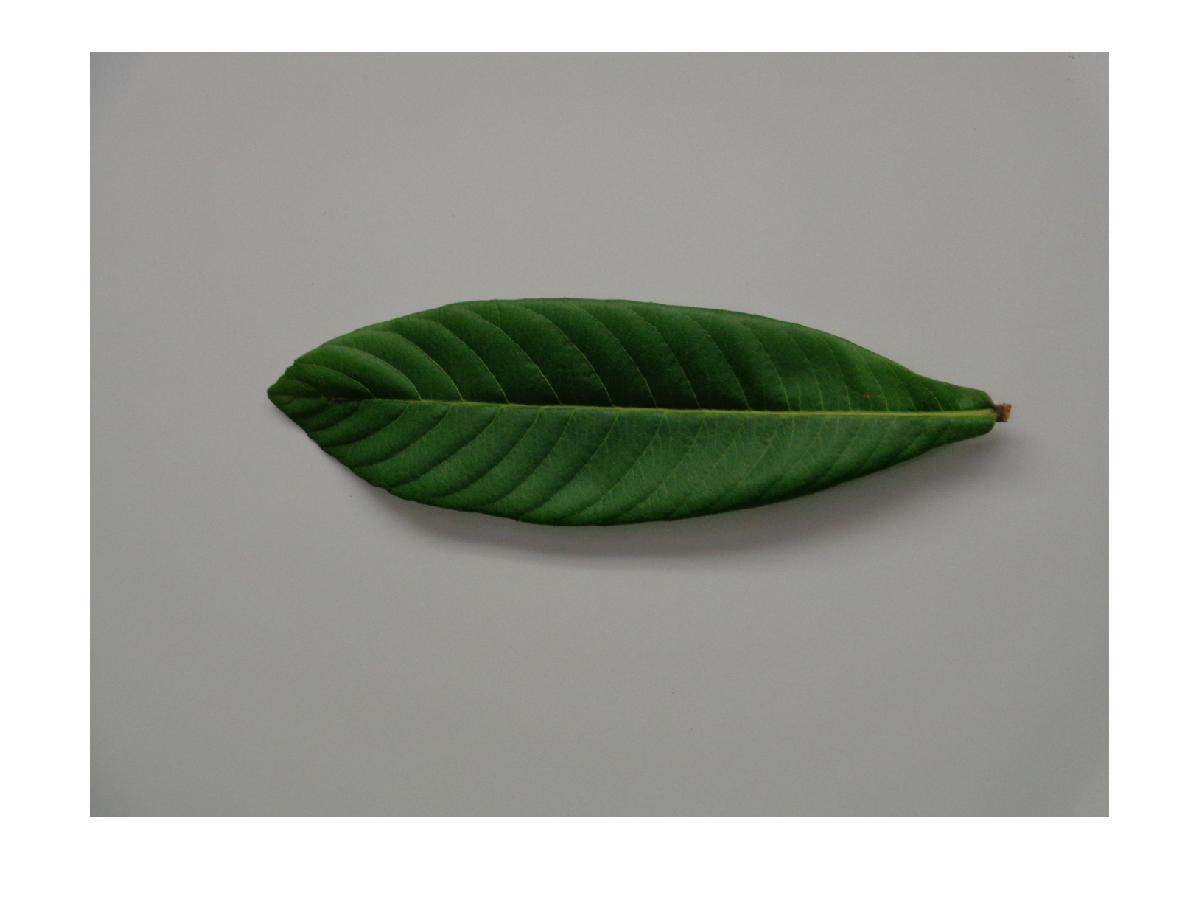}
\includegraphics[scale=0.035]{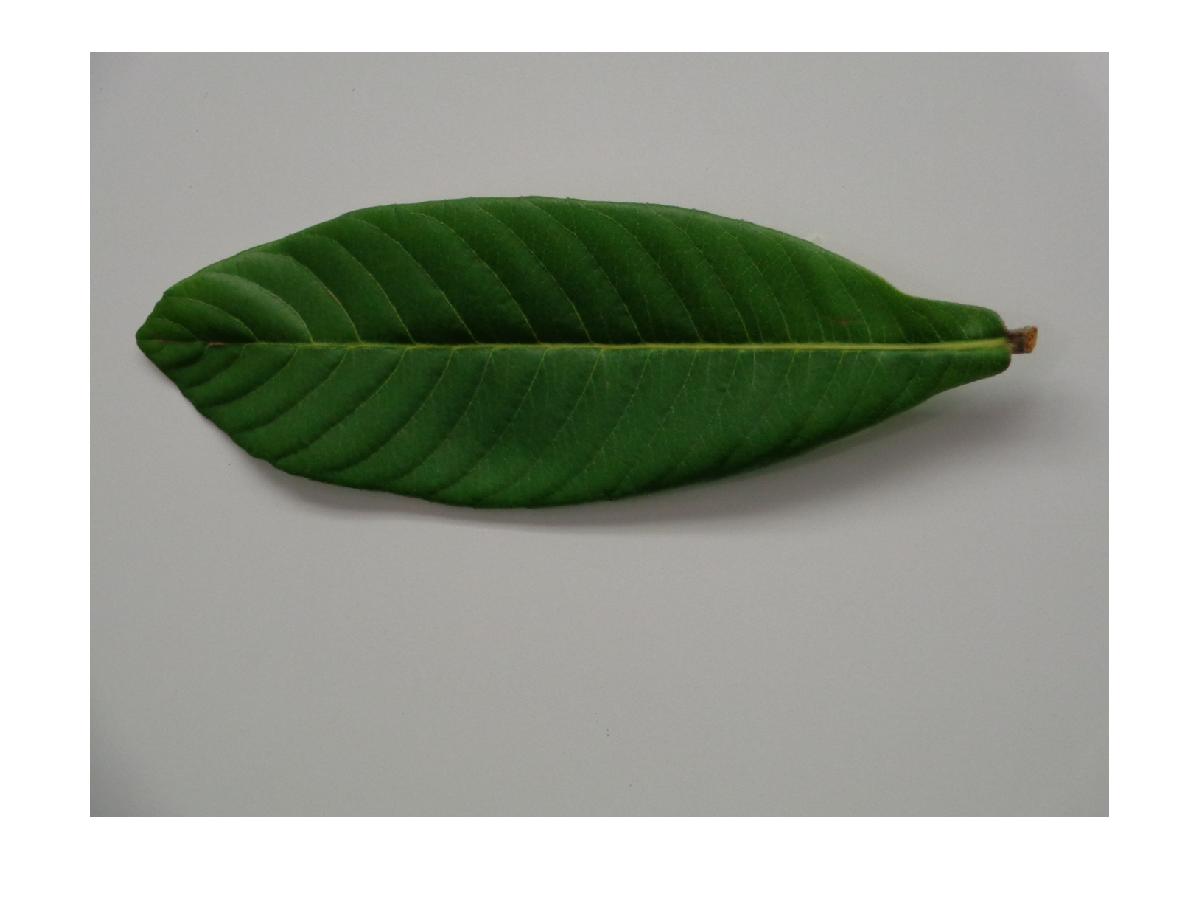}
\includegraphics[scale=0.035]{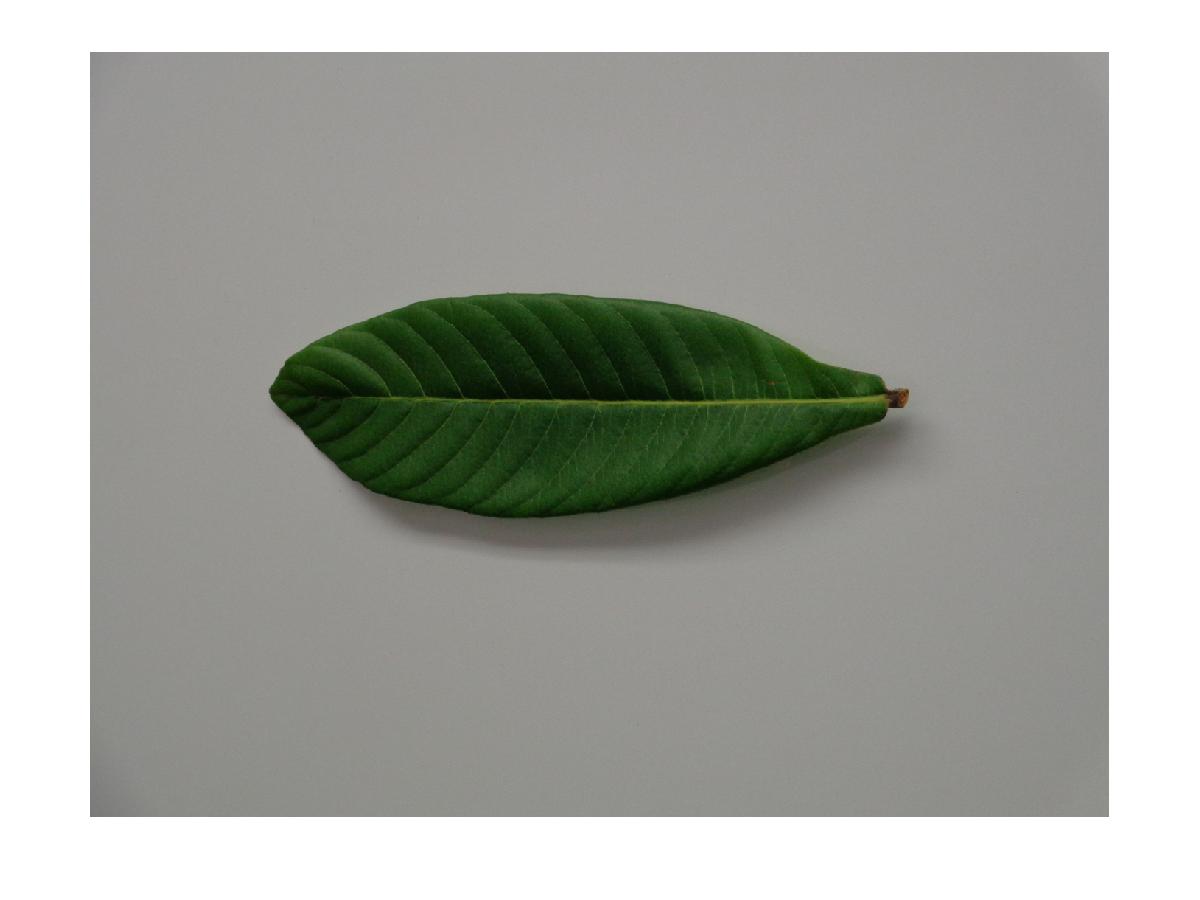}
\includegraphics[scale=0.035]{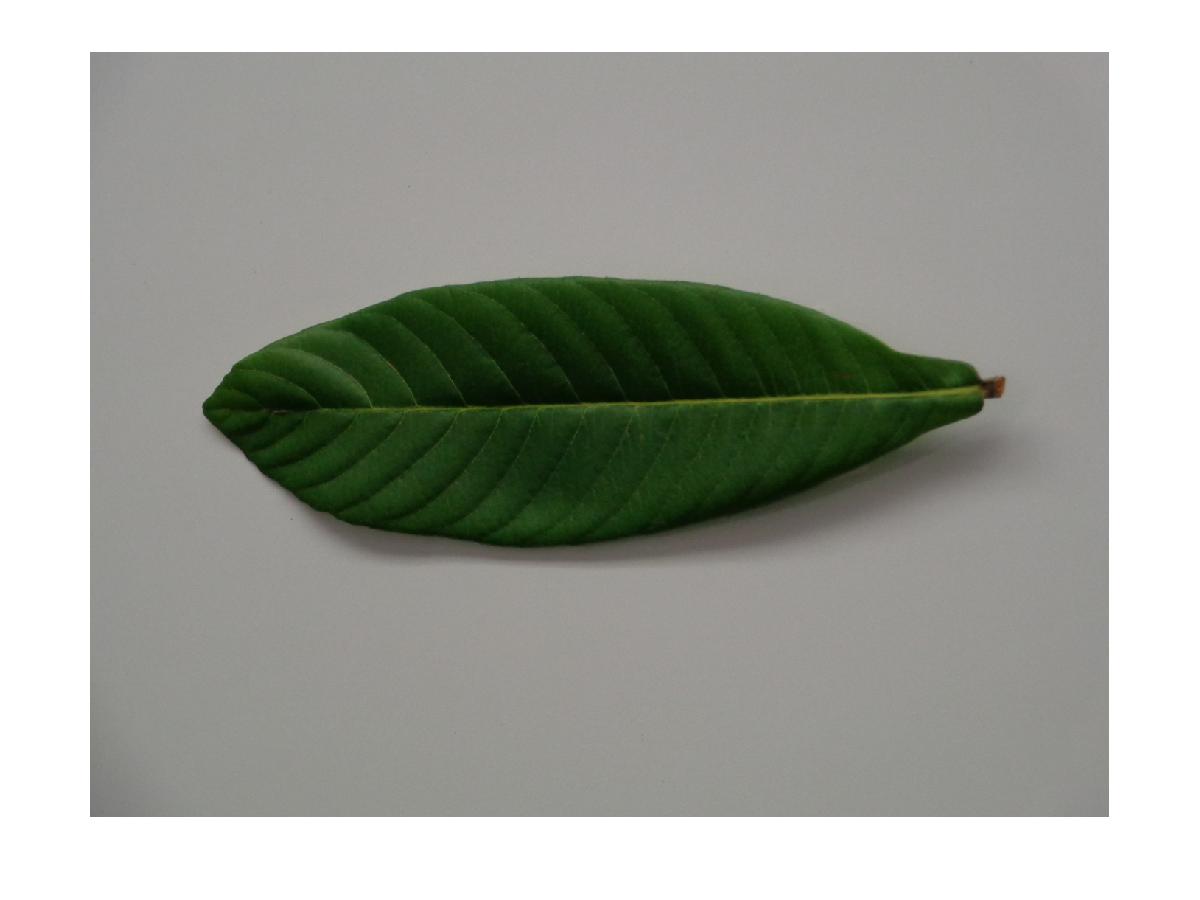}
\includegraphics[scale=0.035]{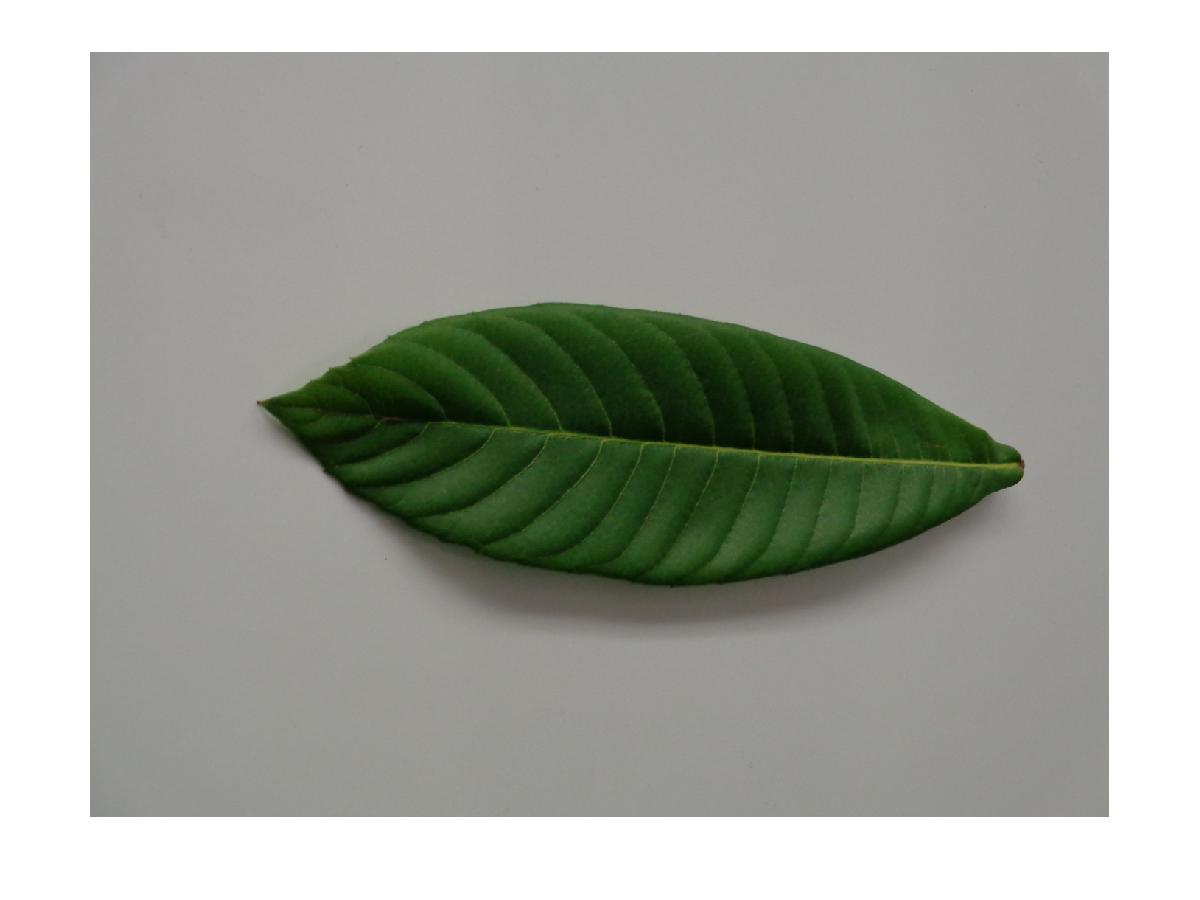}
\includegraphics[scale=0.035]{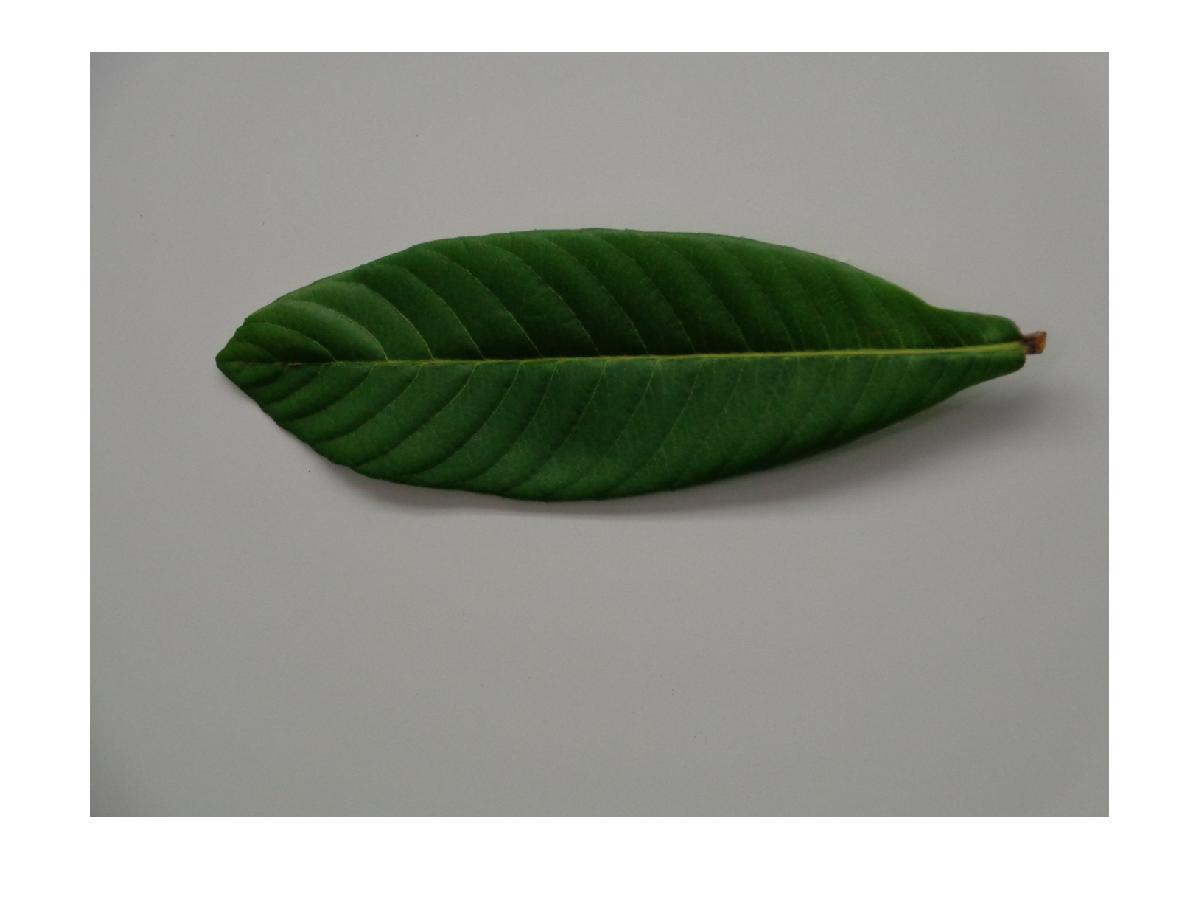}%
\\
\includegraphics[scale=0.035]{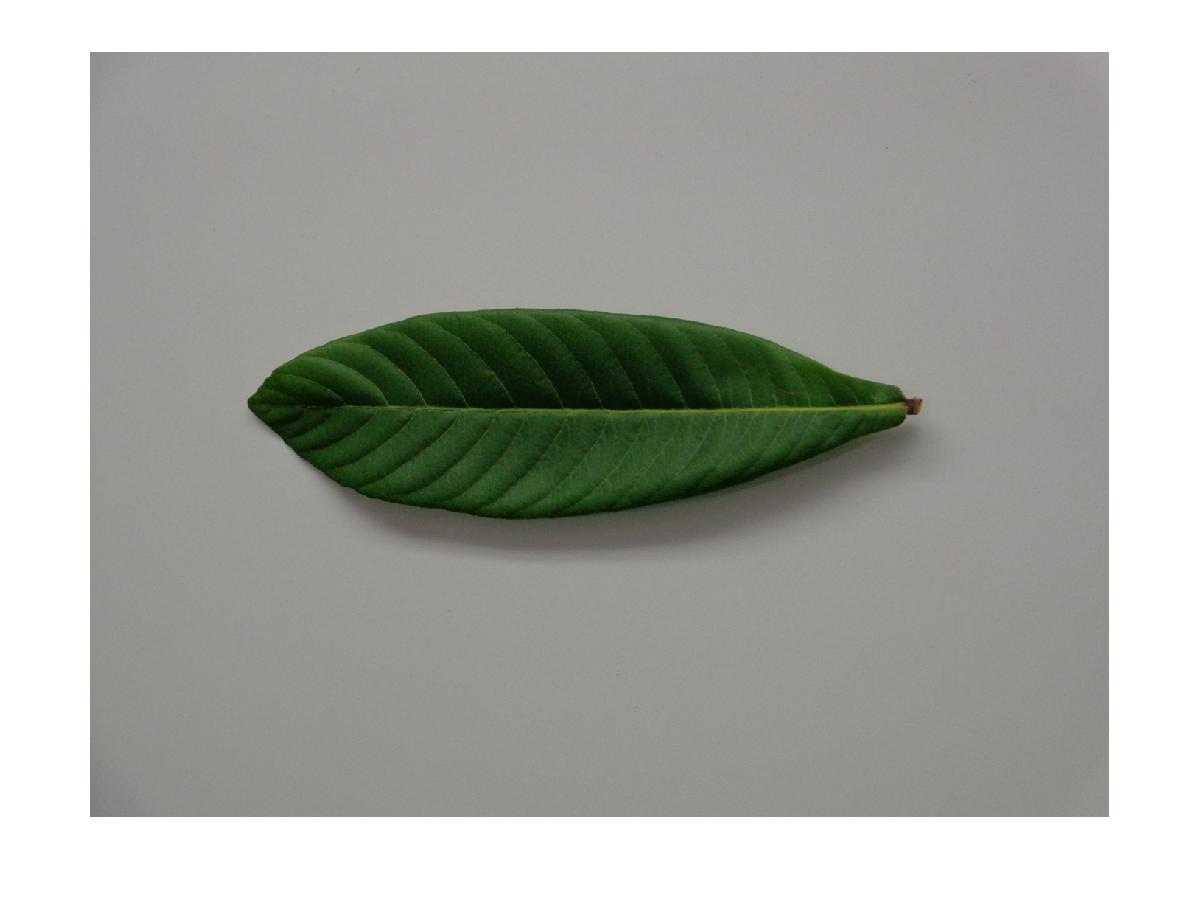}
\includegraphics[scale=0.035]{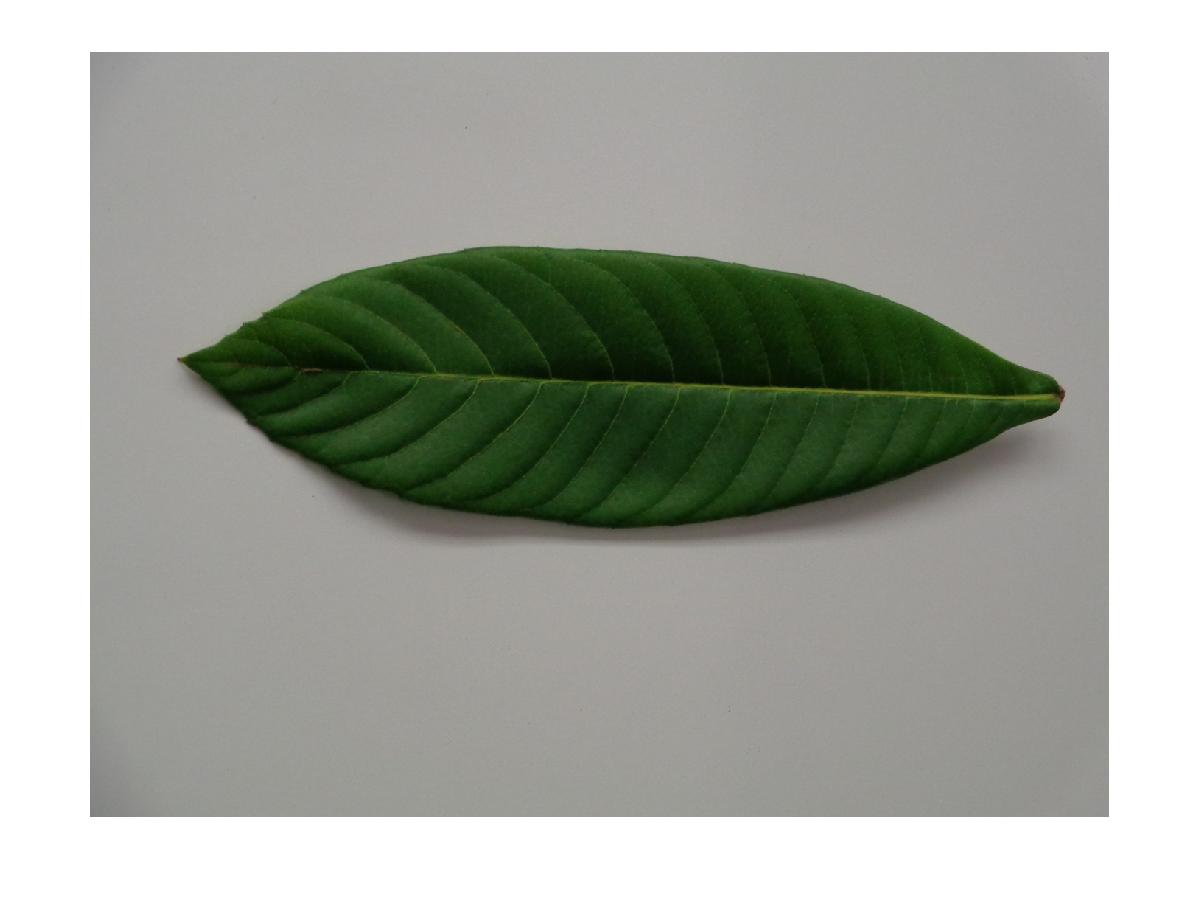}
\includegraphics[scale=0.035]{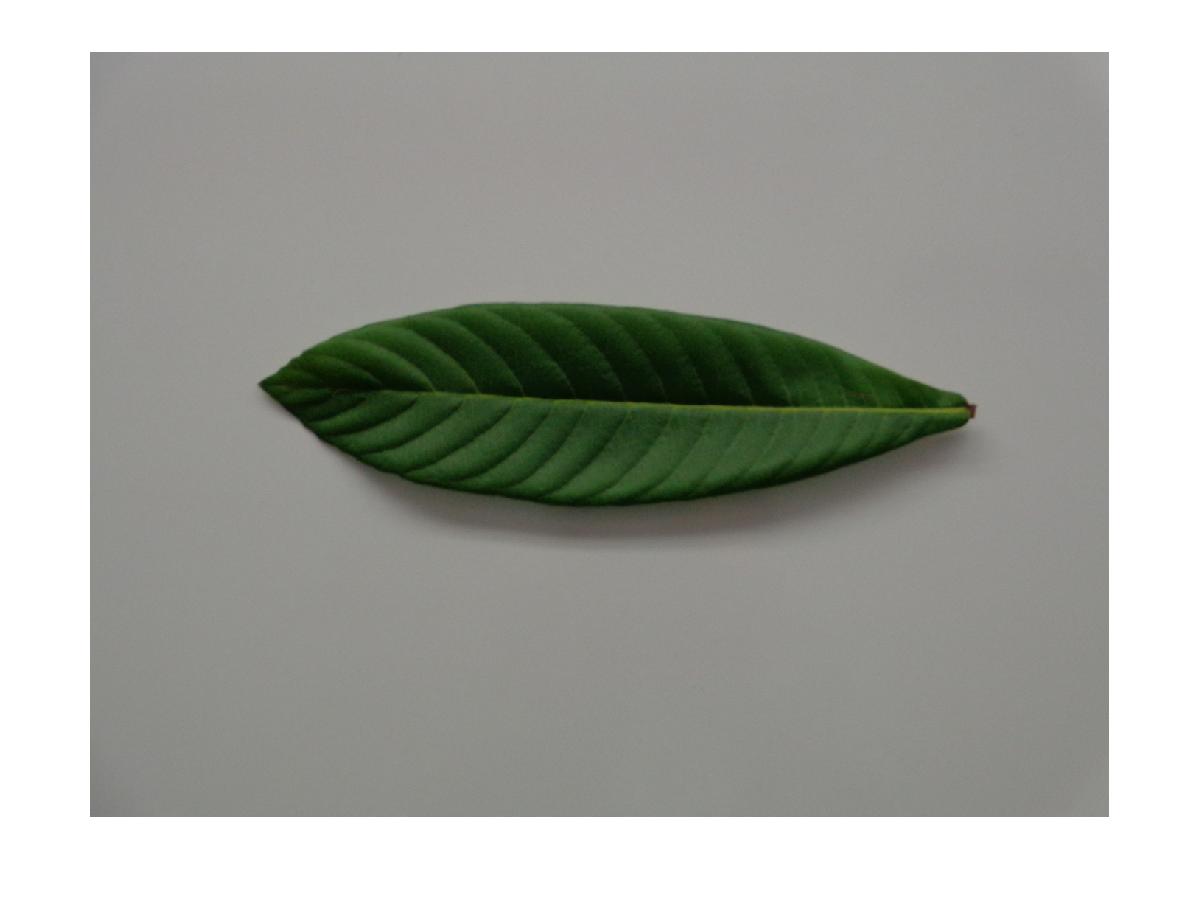}
\includegraphics[scale=0.035]{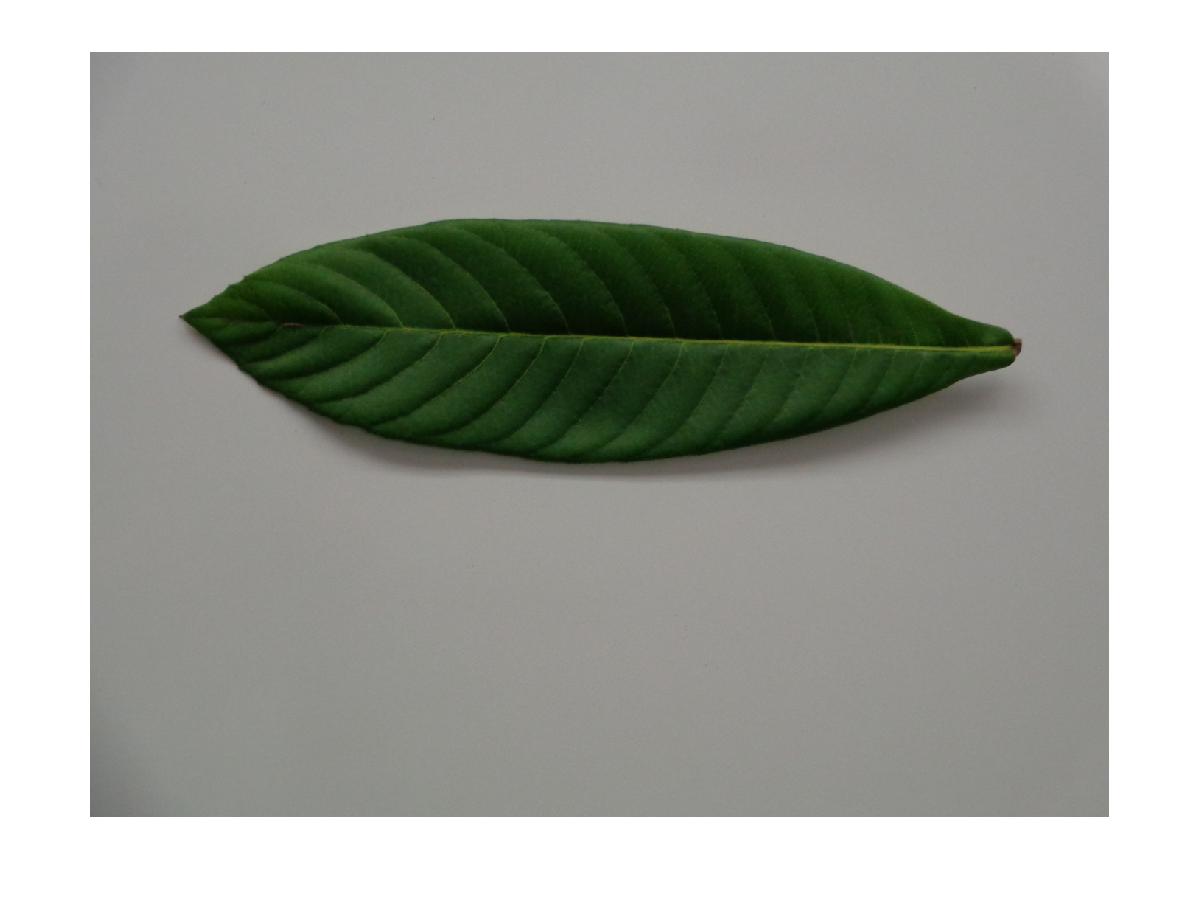}
\includegraphics[scale=0.035]{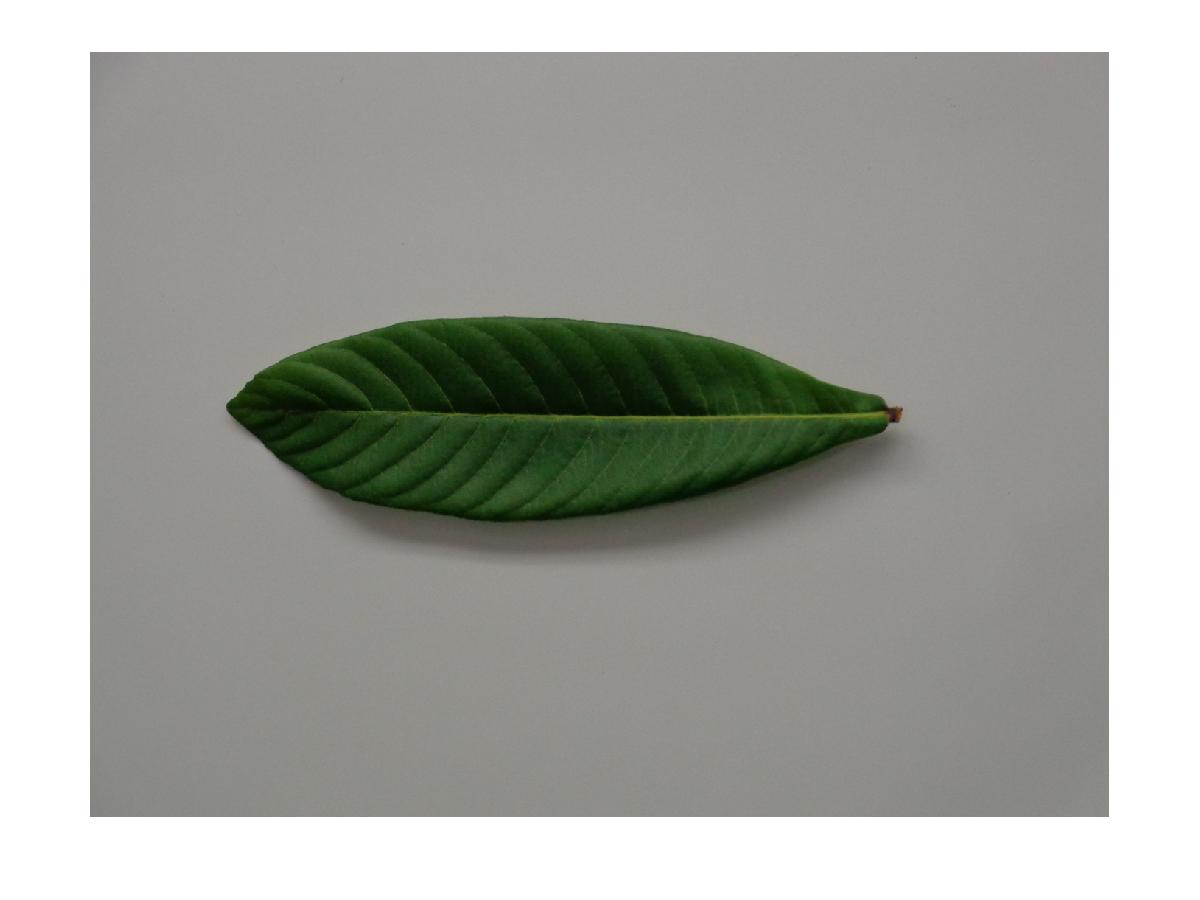}
\includegraphics[scale=0.035]{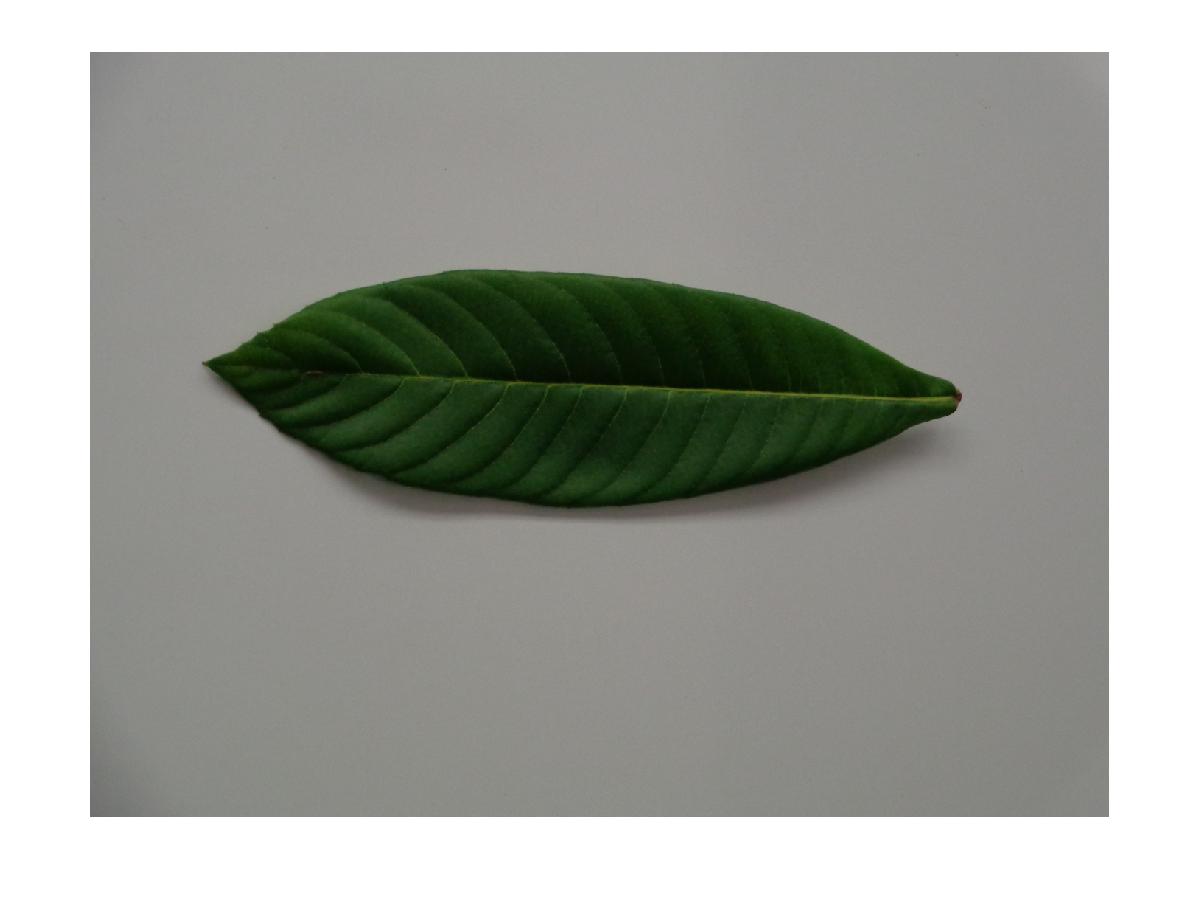}
\includegraphics[scale=0.035]{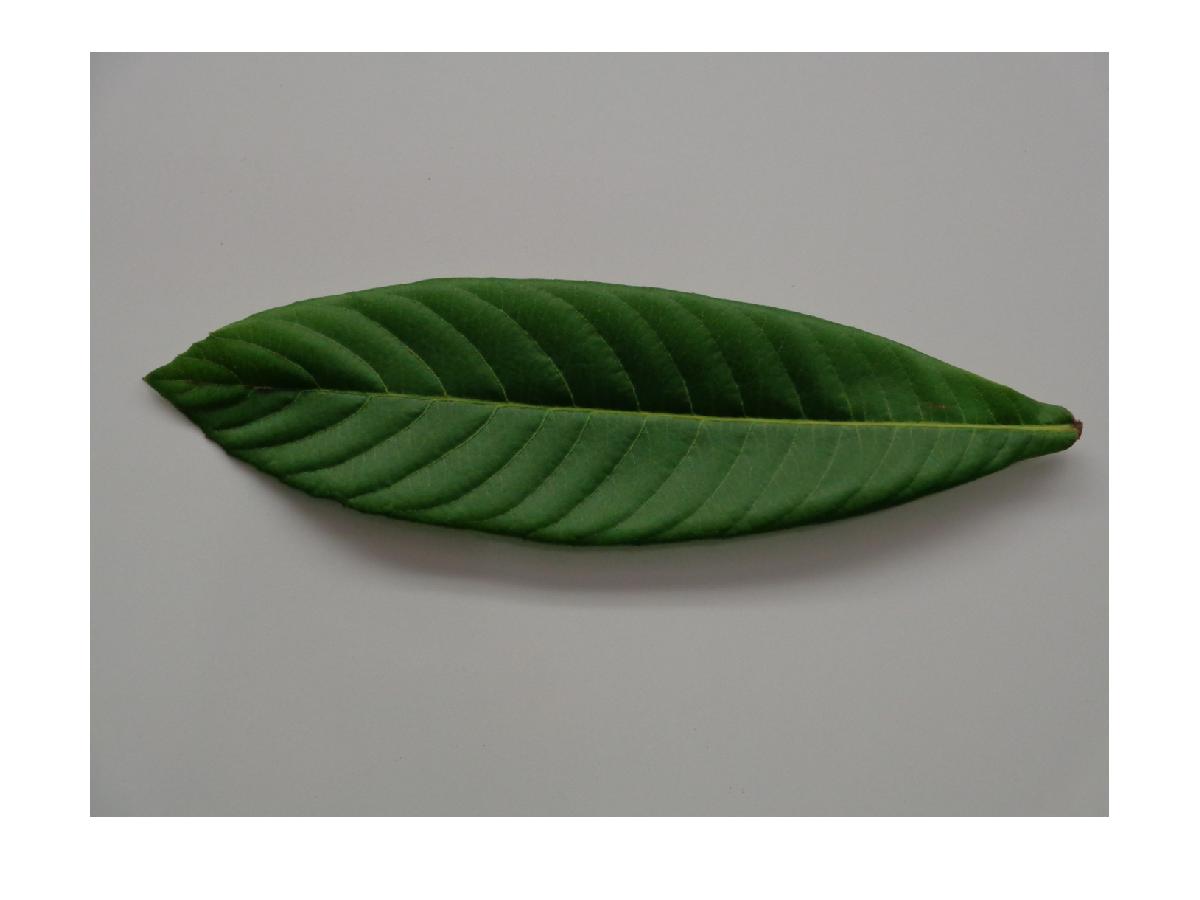}
\includegraphics[scale=0.035]{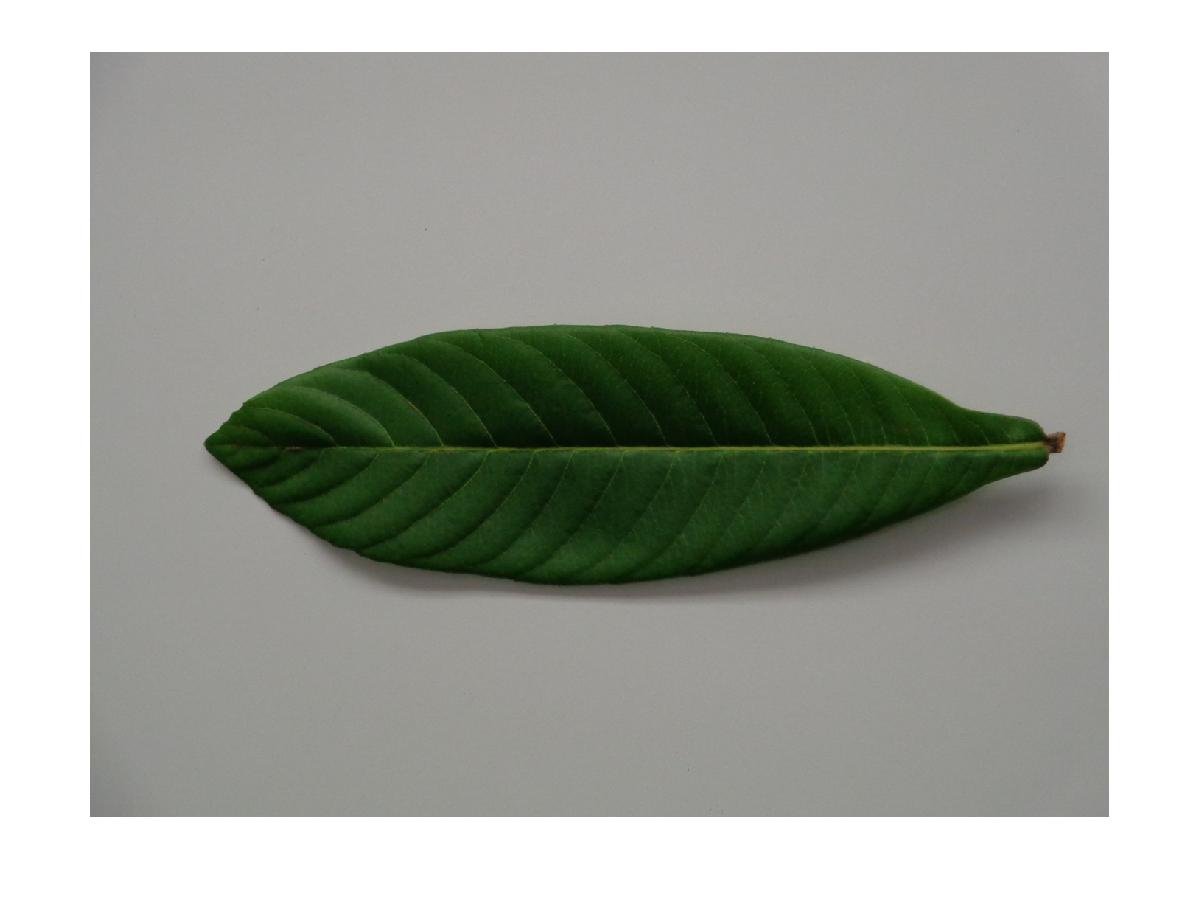}
\includegraphics[scale=0.035]{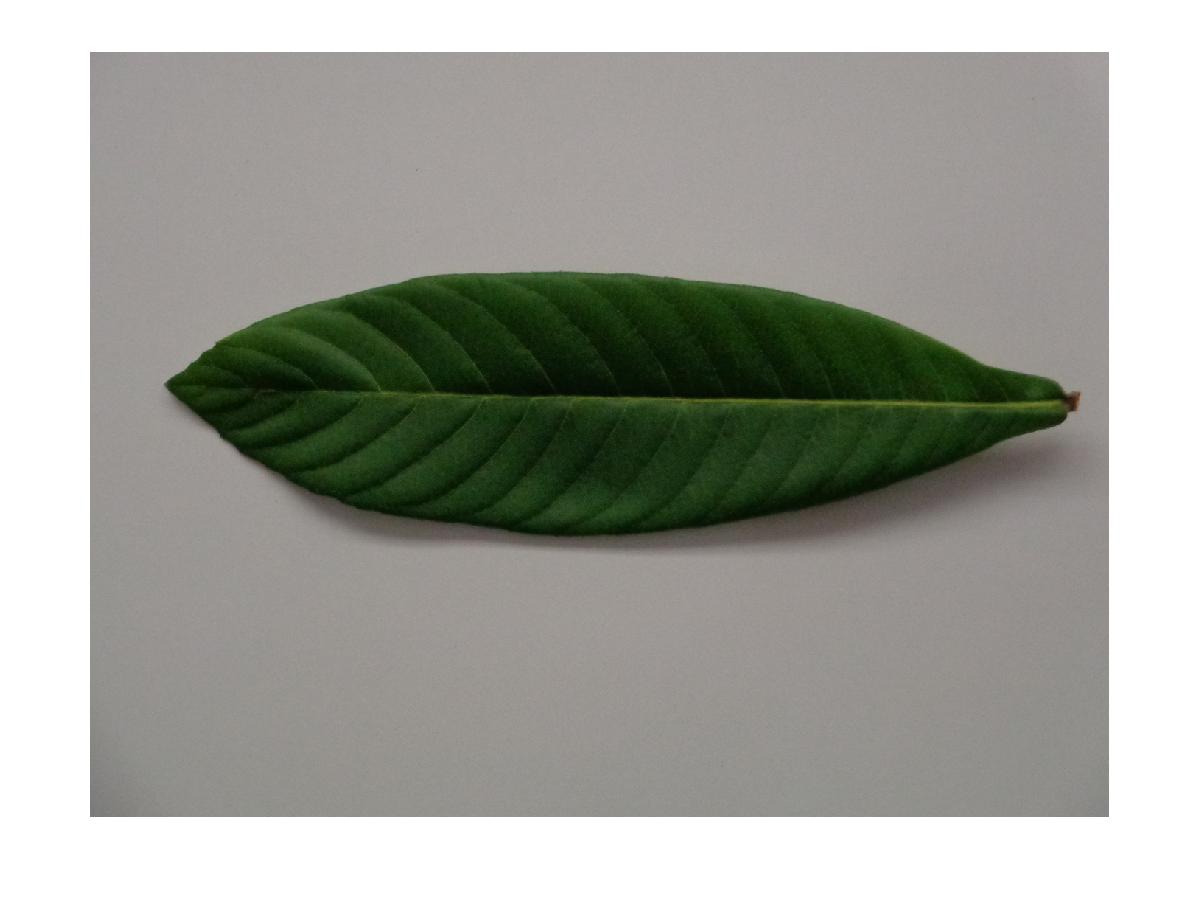}
\includegraphics[scale=0.035]{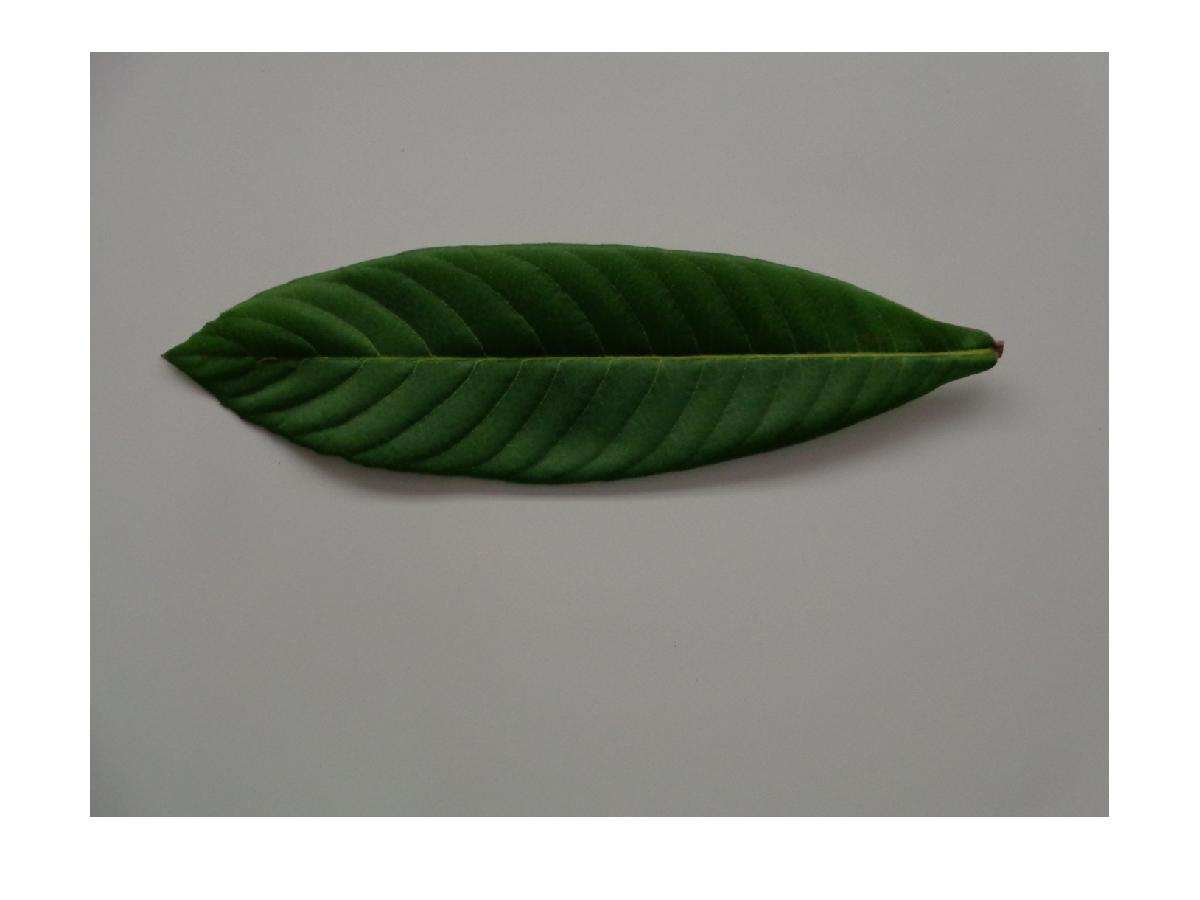}
\end{center}
\caption{Original images of leaf A (top 20 figures) and leaf B (bottom 20
figures).}%
\label{fig:ori_figure}%
\end{figure}

\begin{figure}[h]
\centering
\includegraphics[scale=0.03]{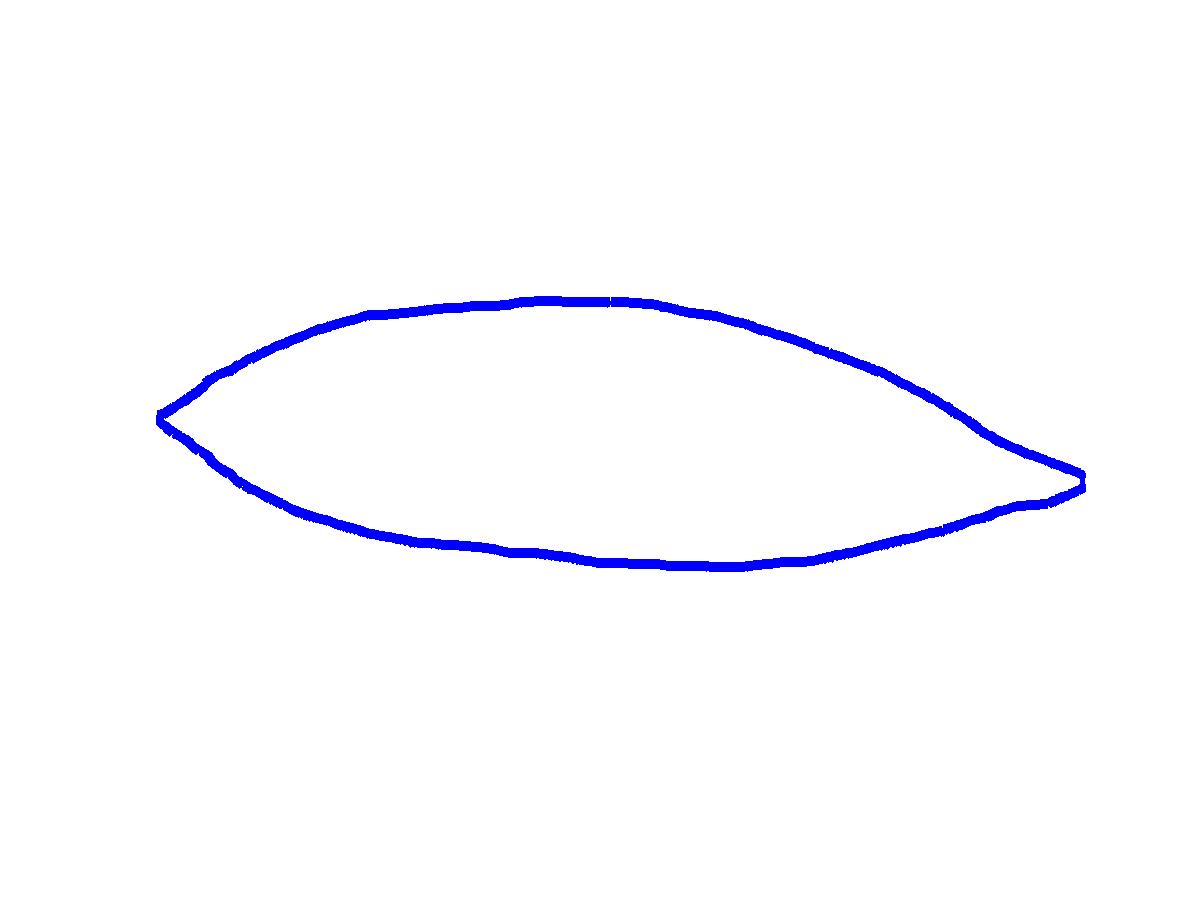}
\includegraphics[scale=0.03]{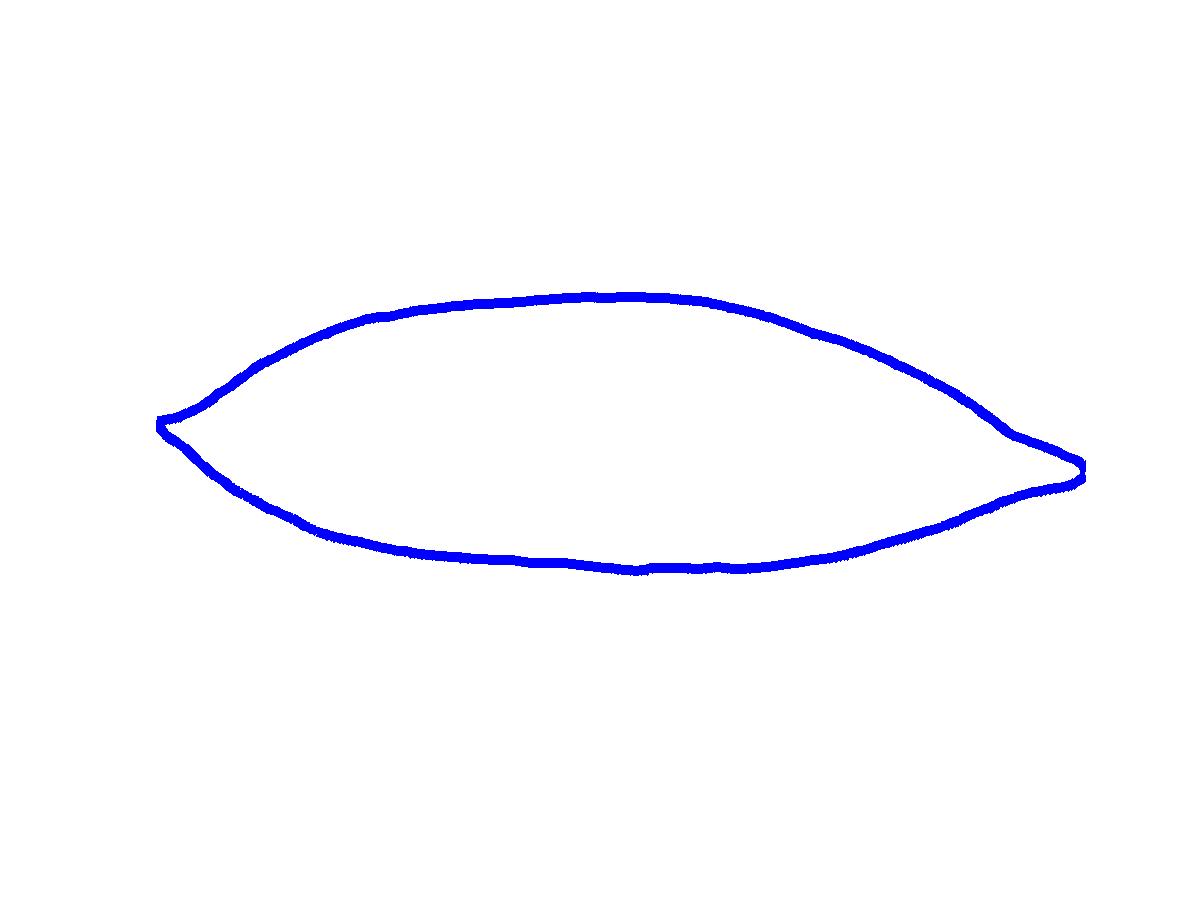}
\includegraphics[scale=0.03]{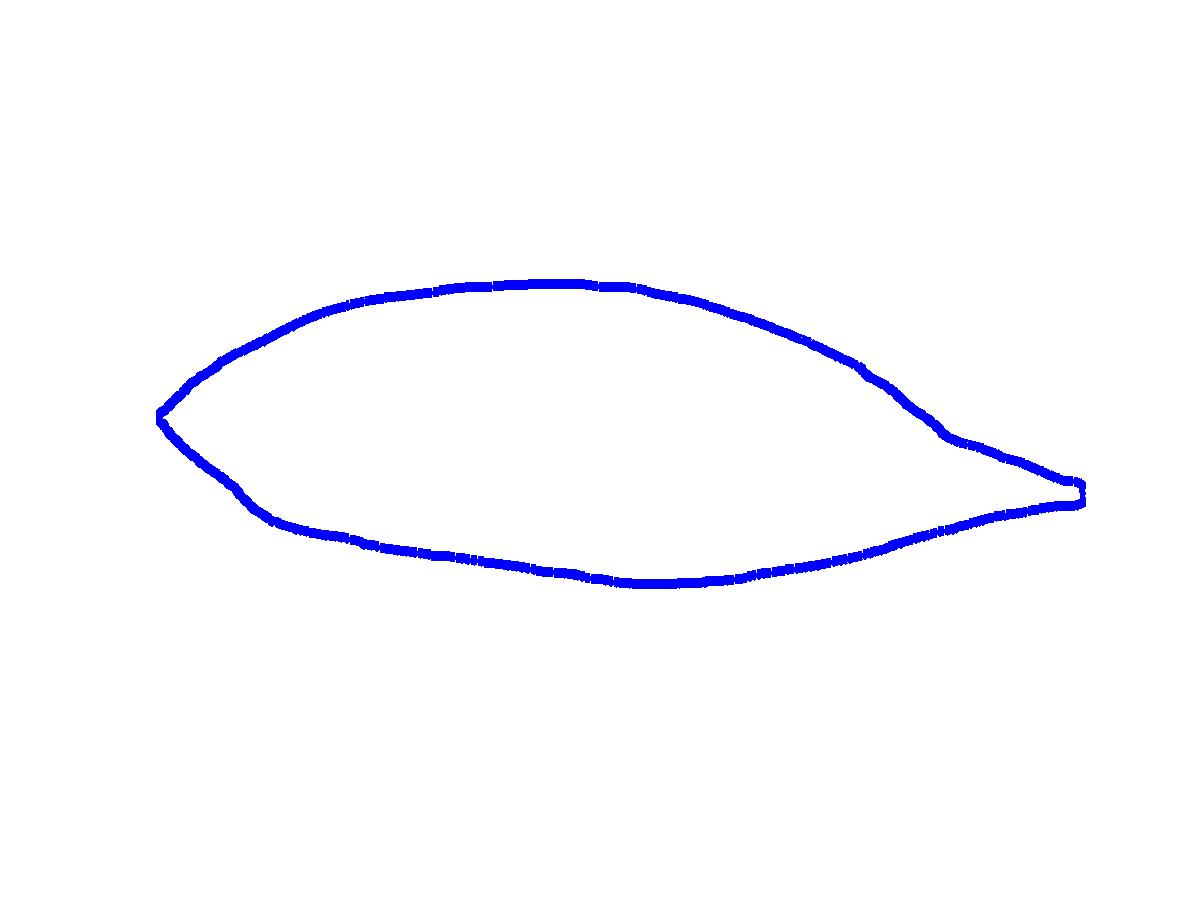}
\includegraphics[scale=0.03]{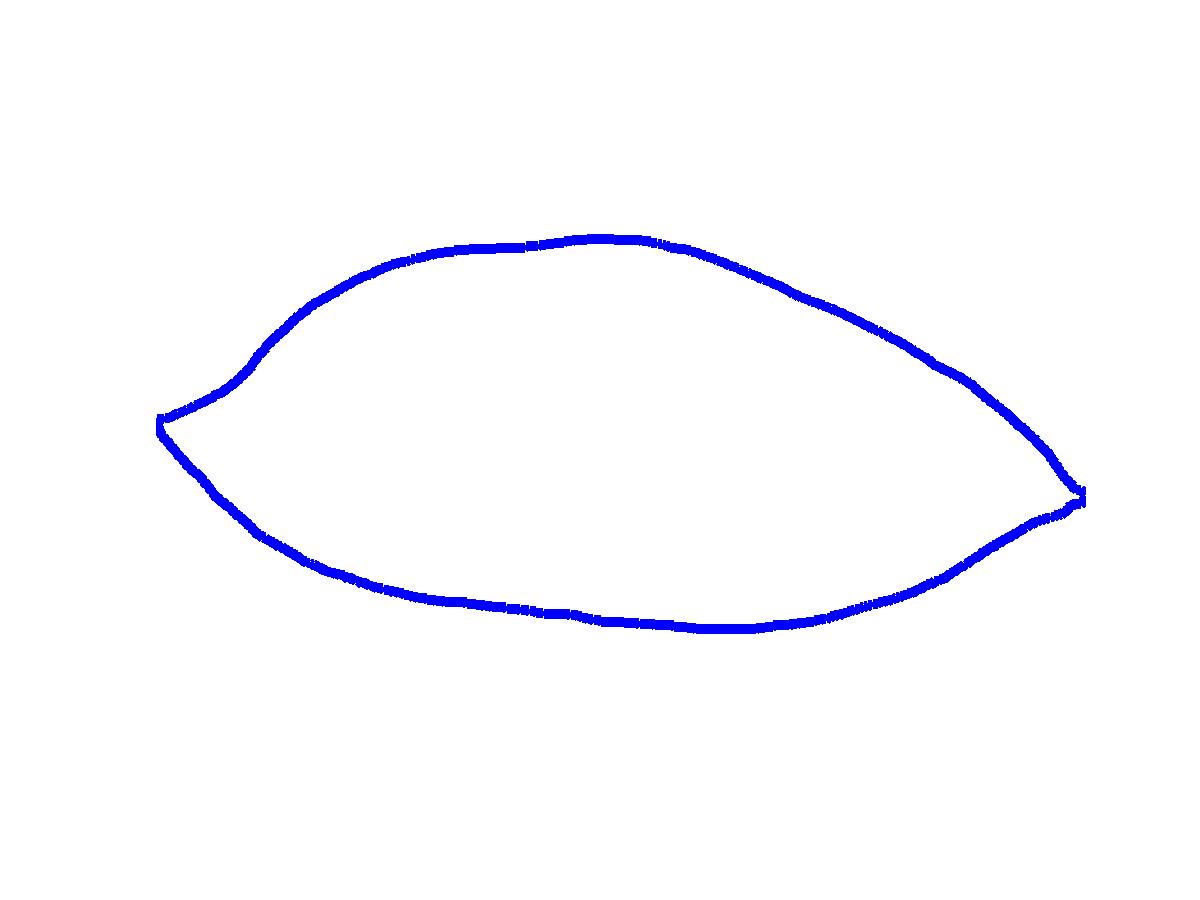}
\includegraphics[scale=0.03]{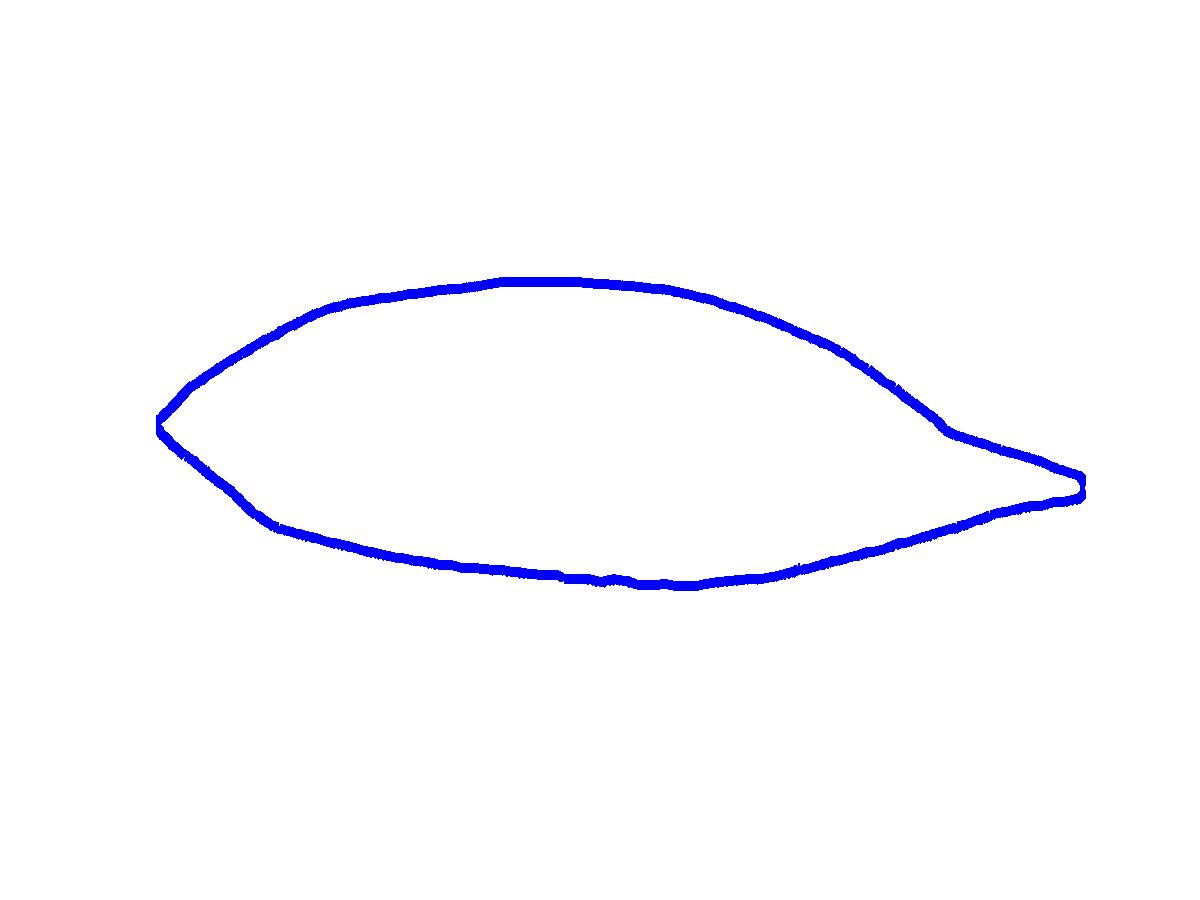}
\includegraphics[scale=0.03]{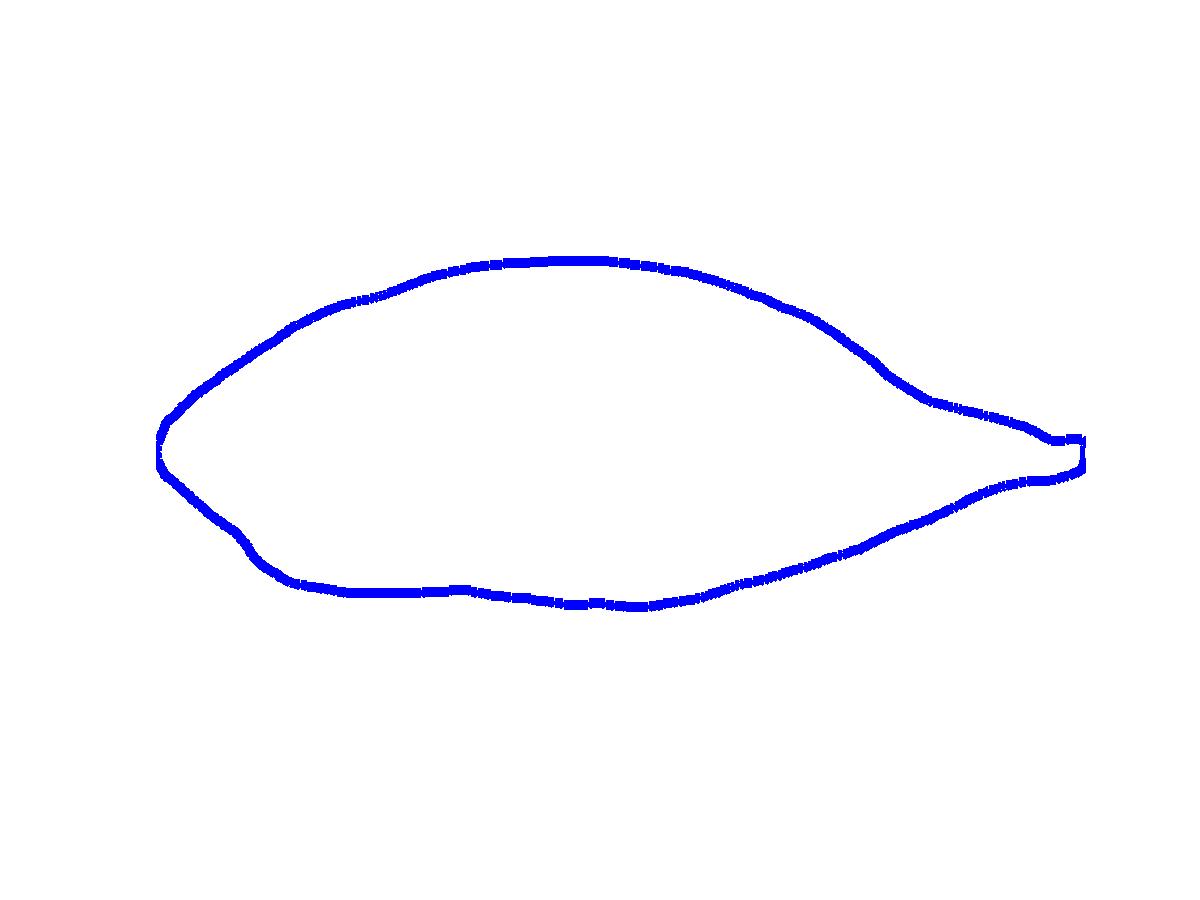}
\includegraphics[scale=0.03]{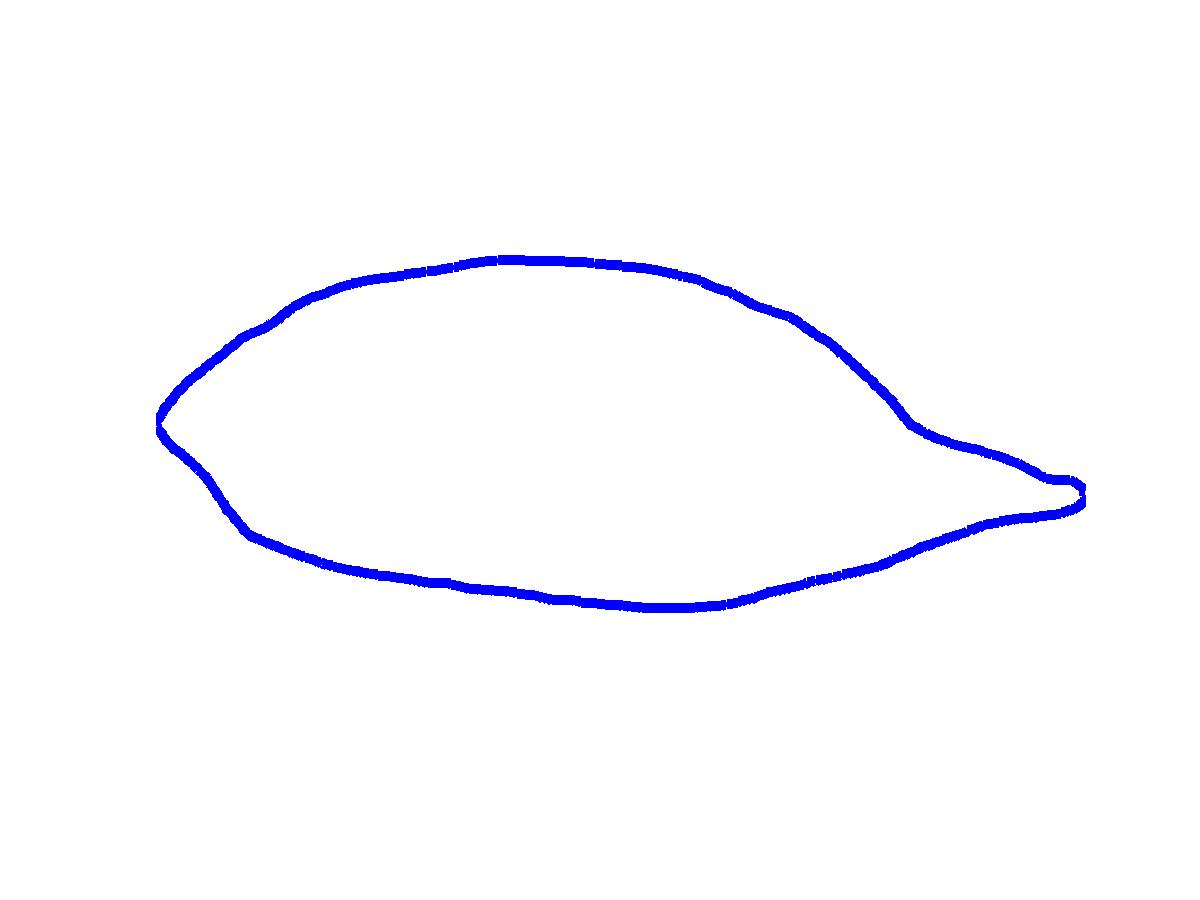}
\includegraphics[scale=0.03]{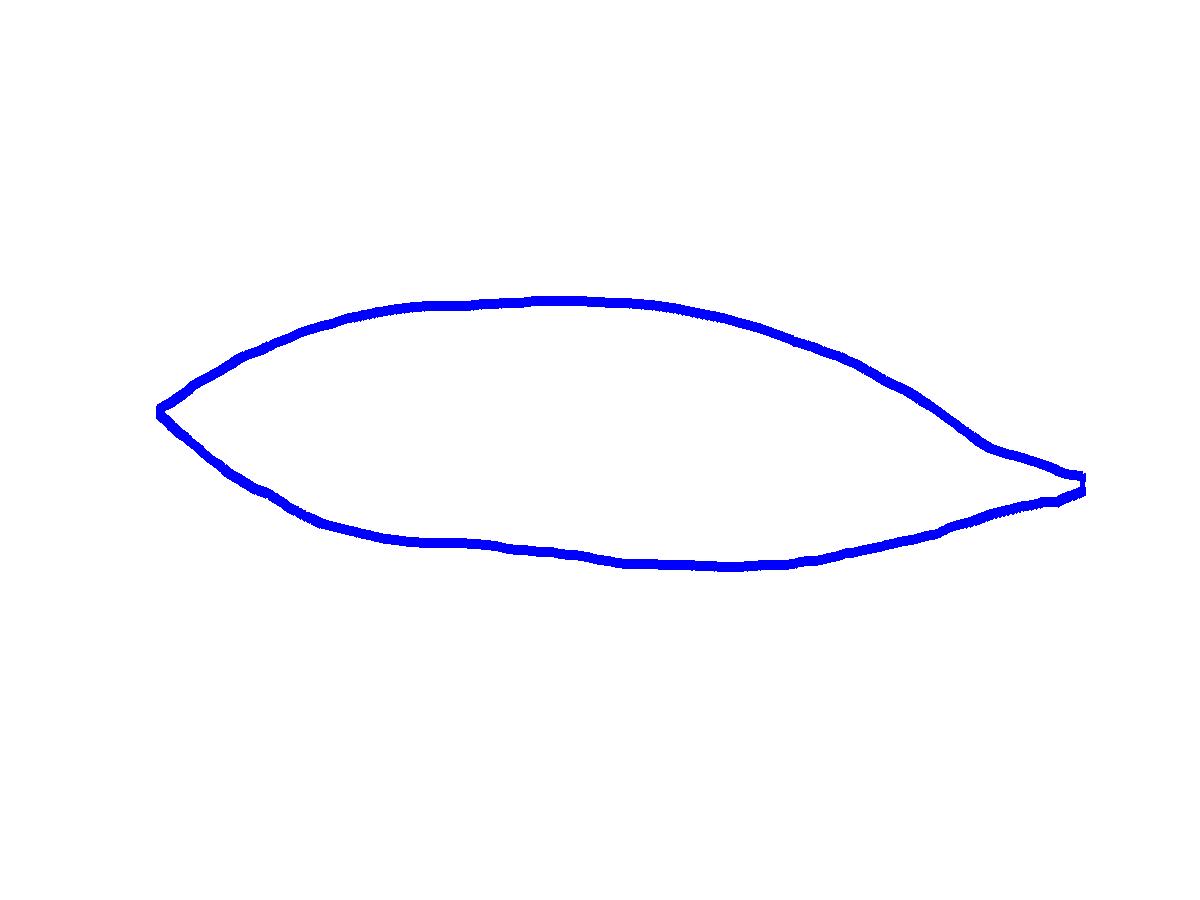}
\includegraphics[scale=0.03]{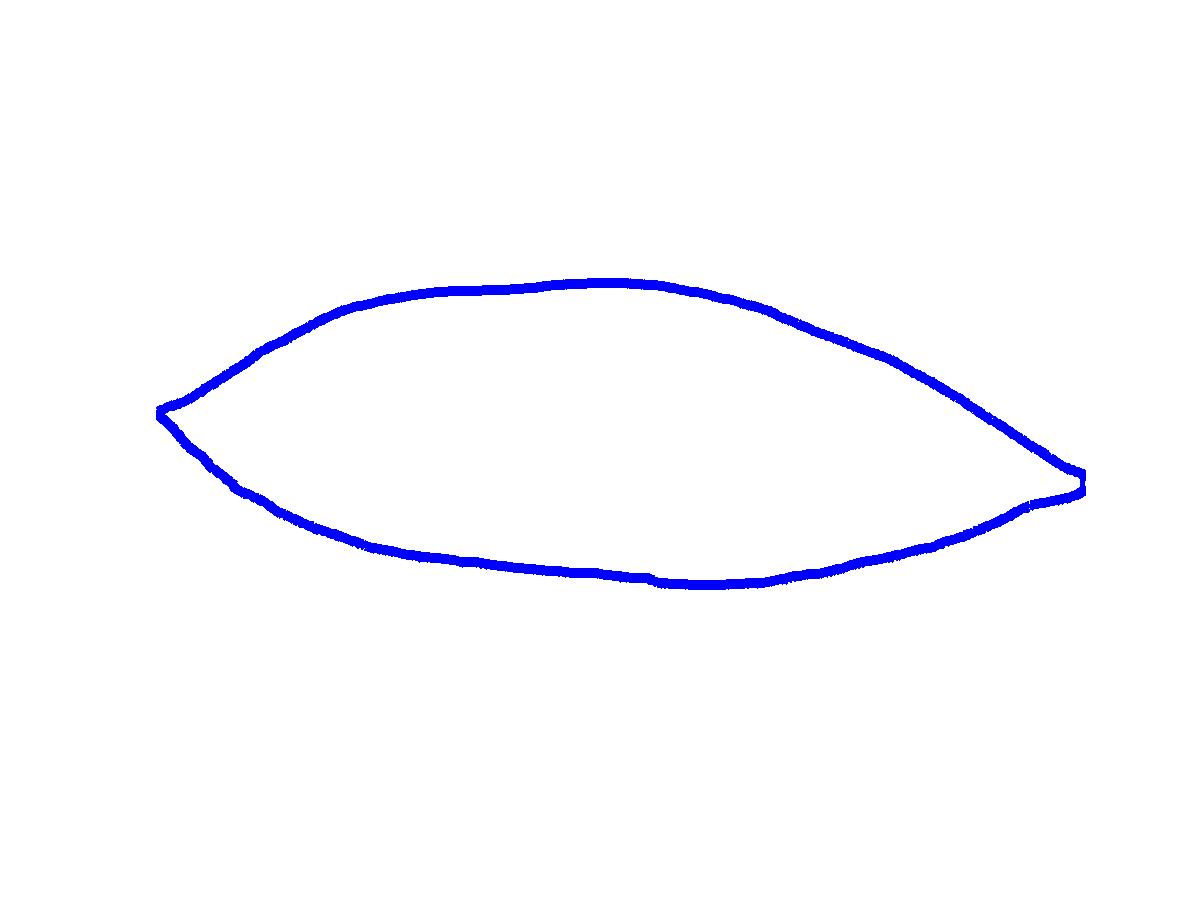}
\includegraphics[scale=0.03]{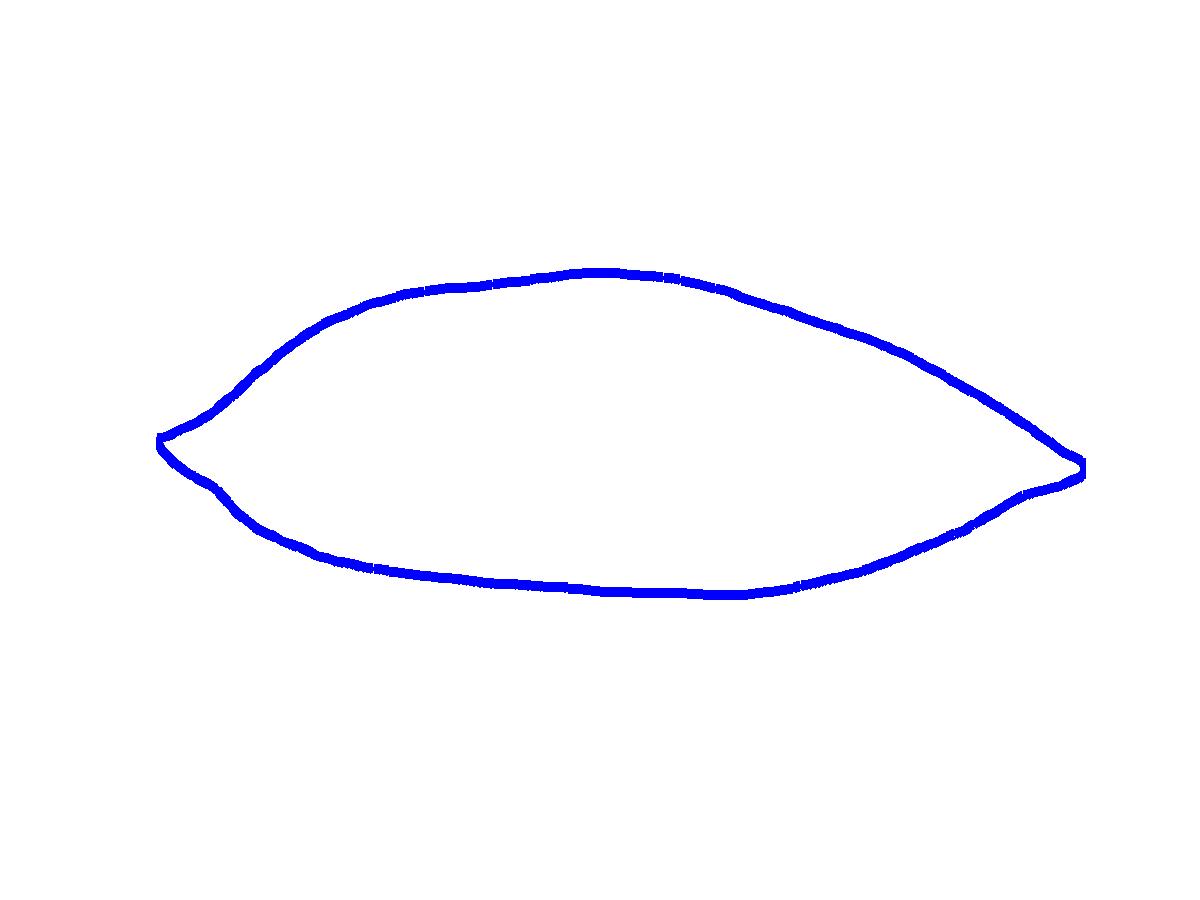}\newline%
\smallskip
\includegraphics[scale=0.03]{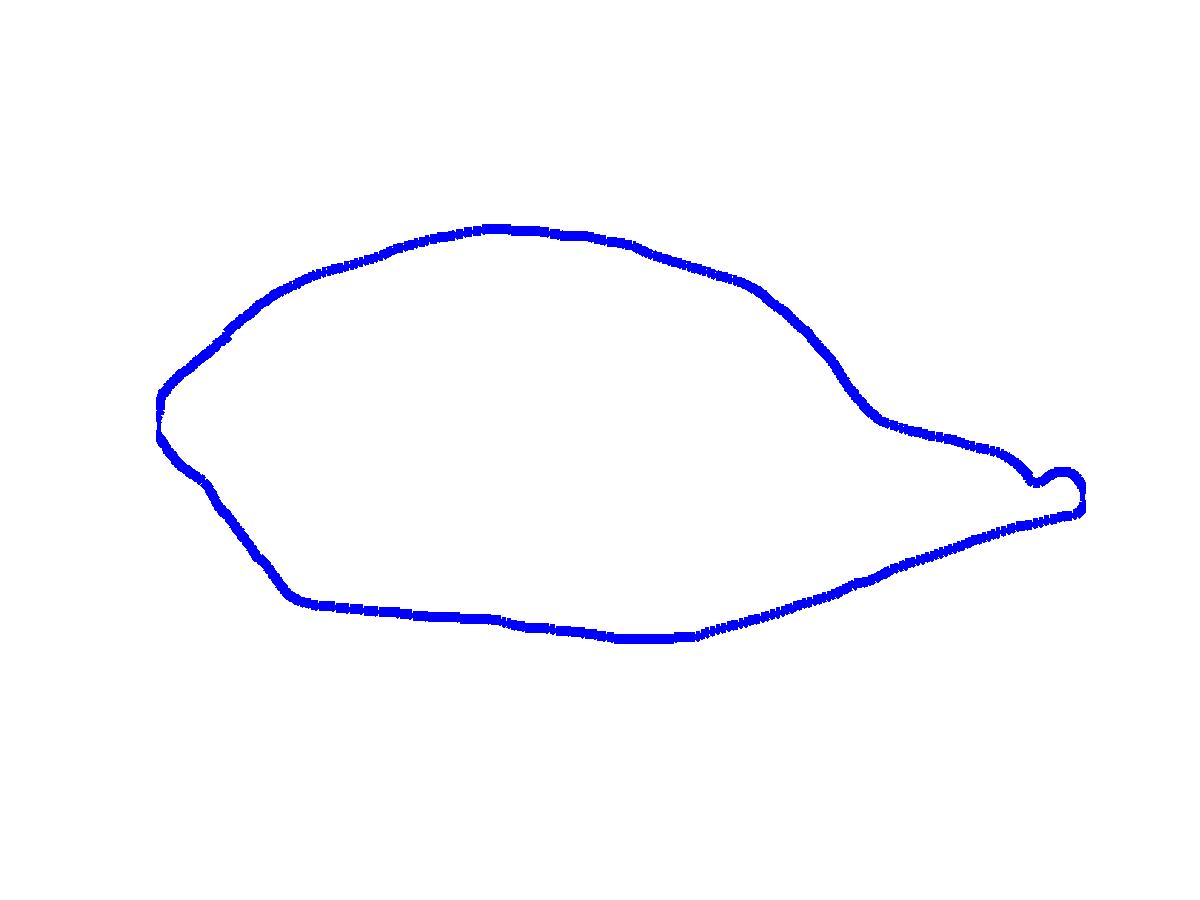}
\includegraphics[scale=0.03]{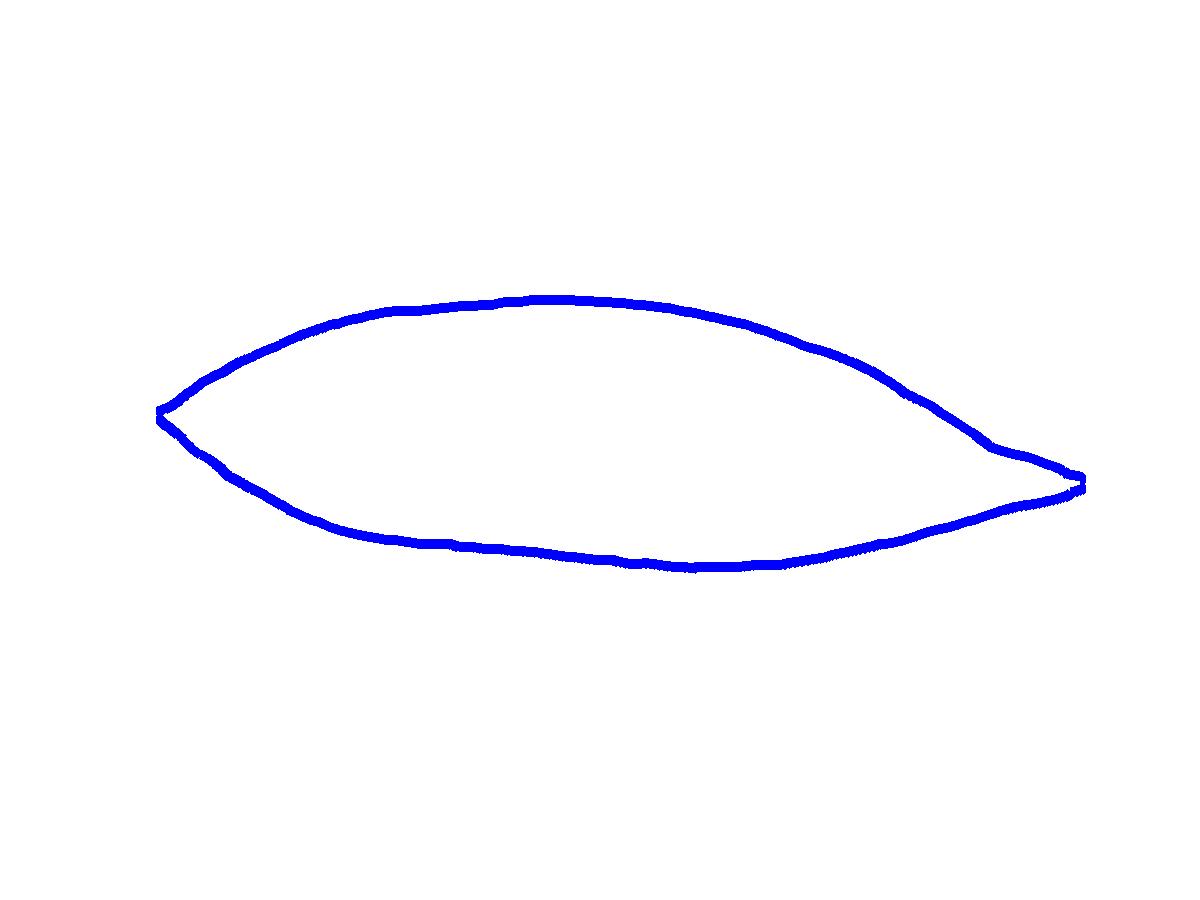}
\includegraphics[scale=0.03]{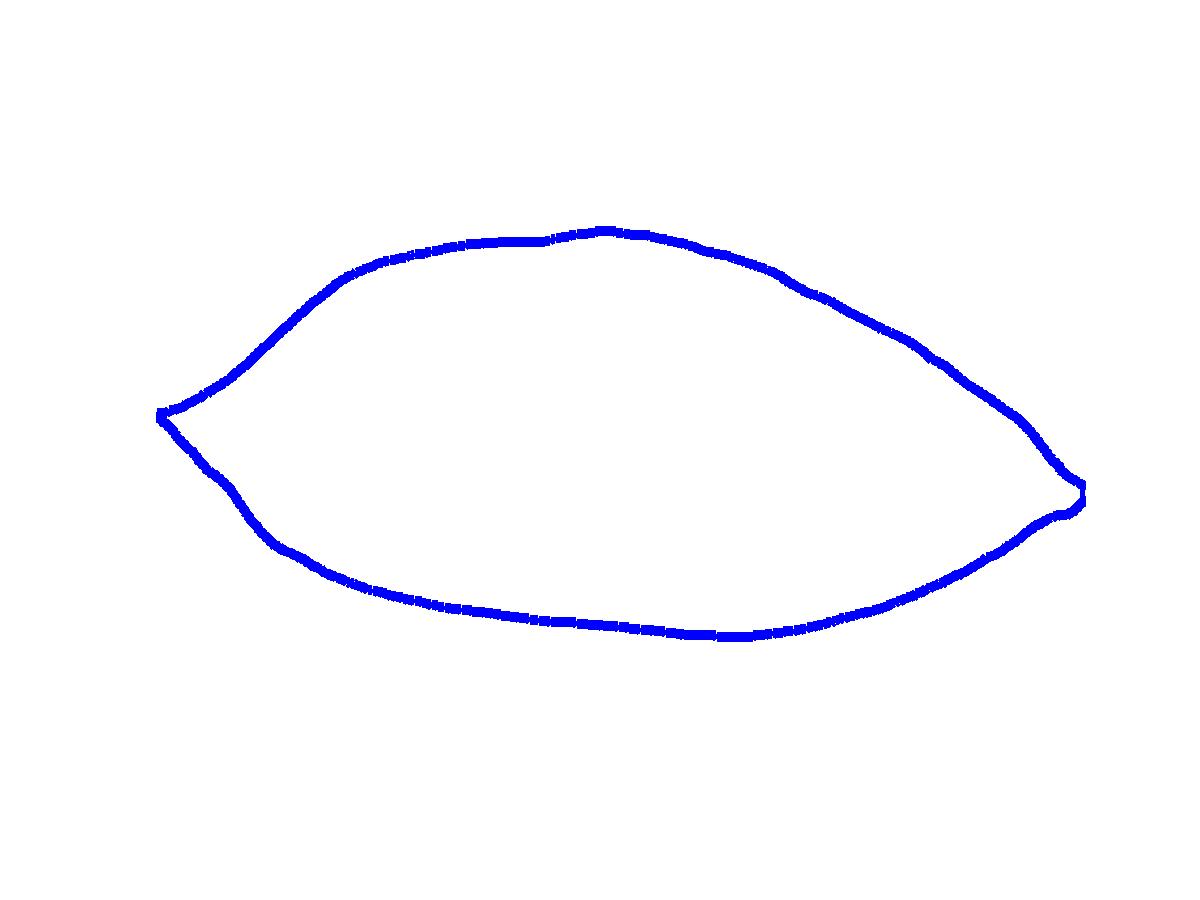}
\includegraphics[scale=0.03]{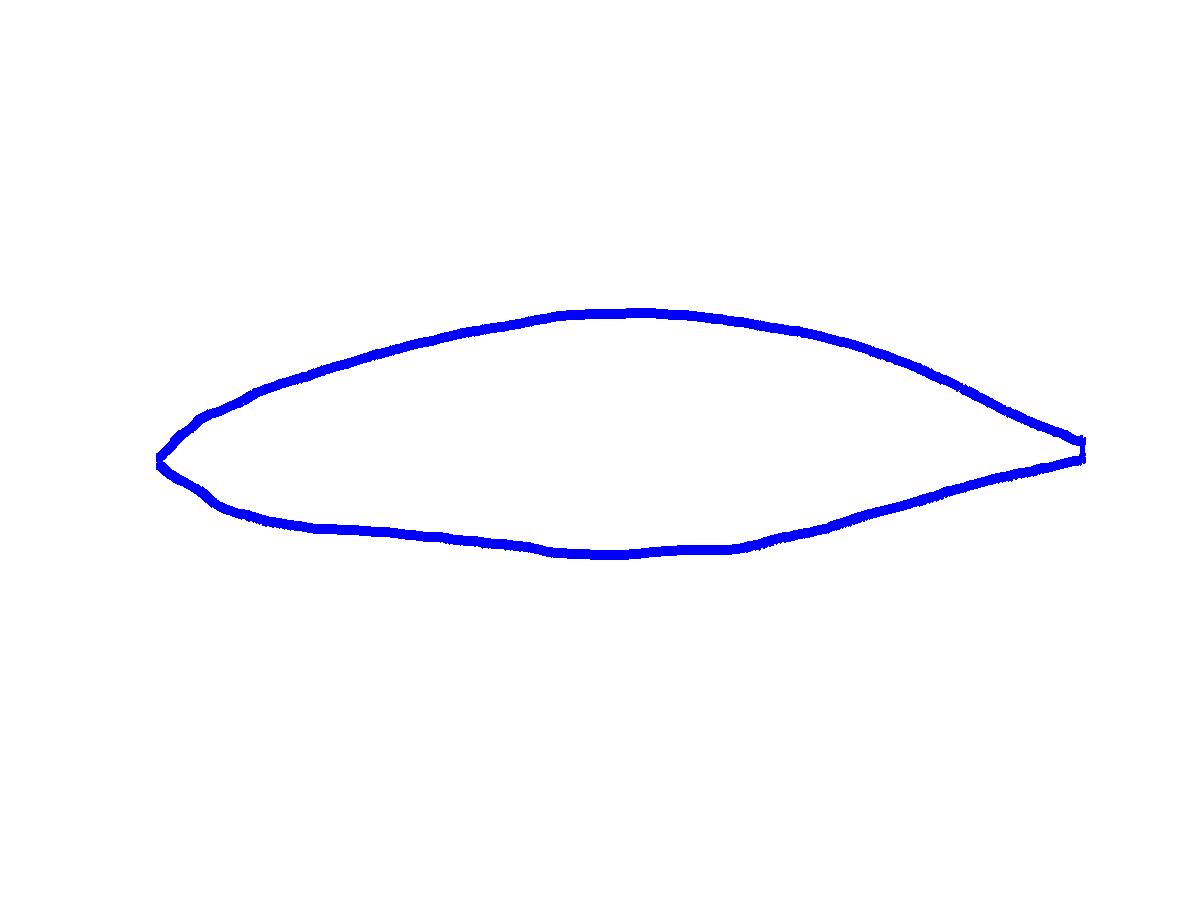}
\includegraphics[scale=0.03]{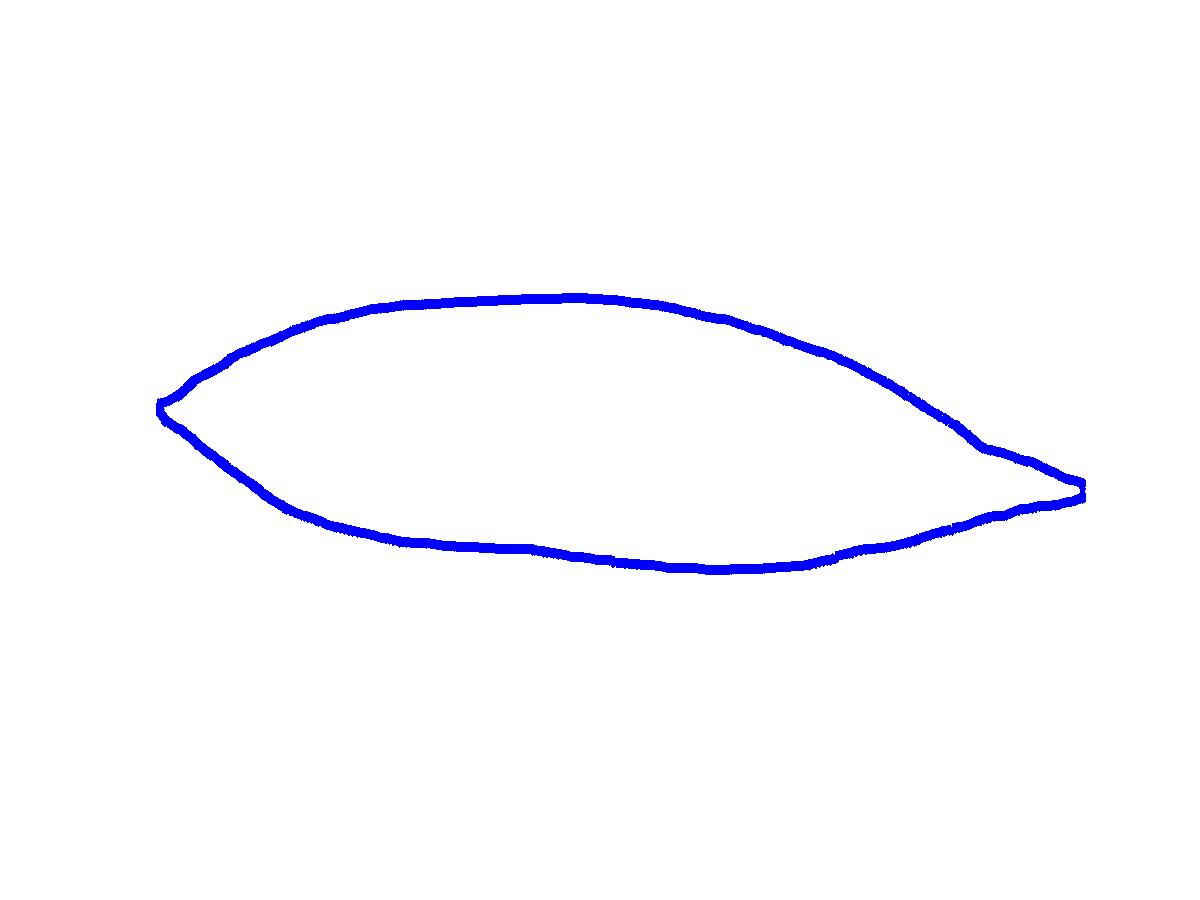}
\includegraphics[scale=0.03]{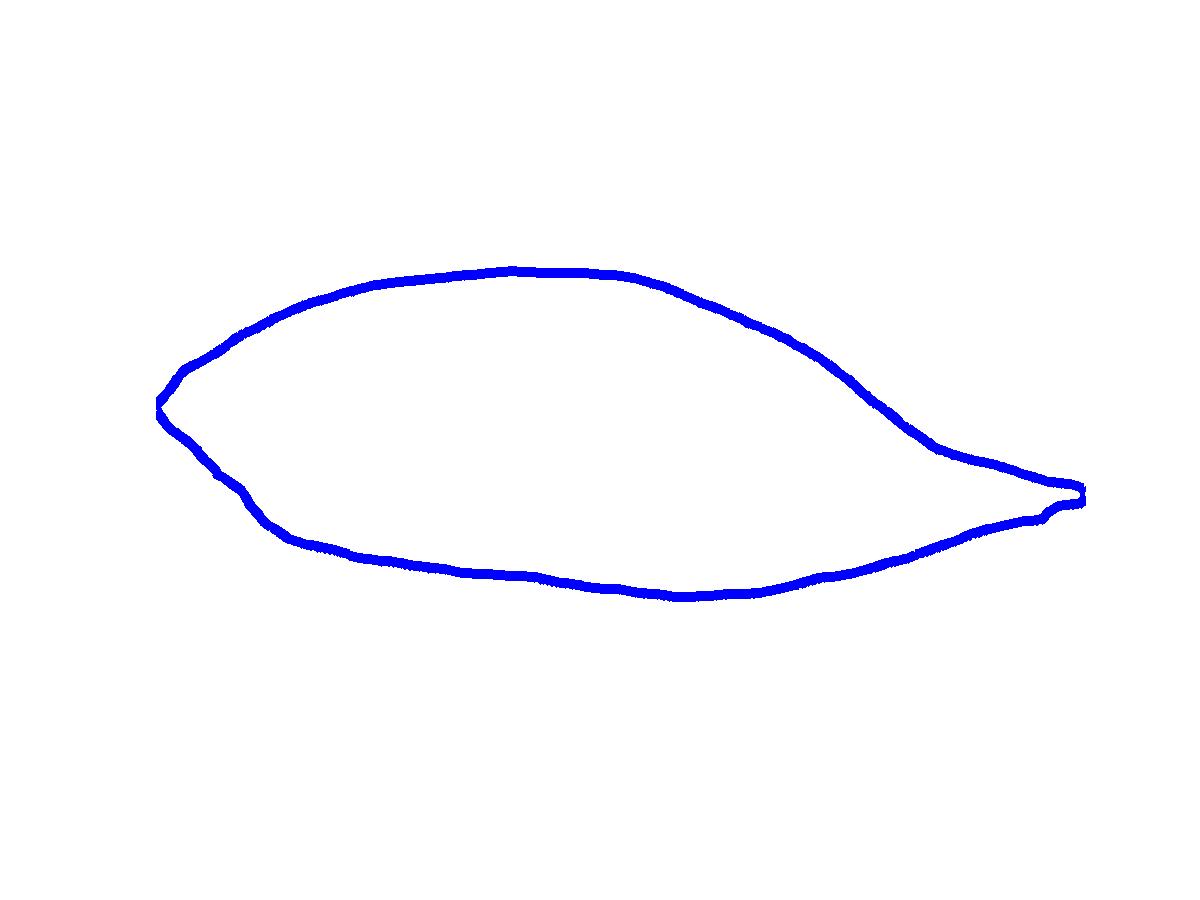}
\includegraphics[scale=0.03]{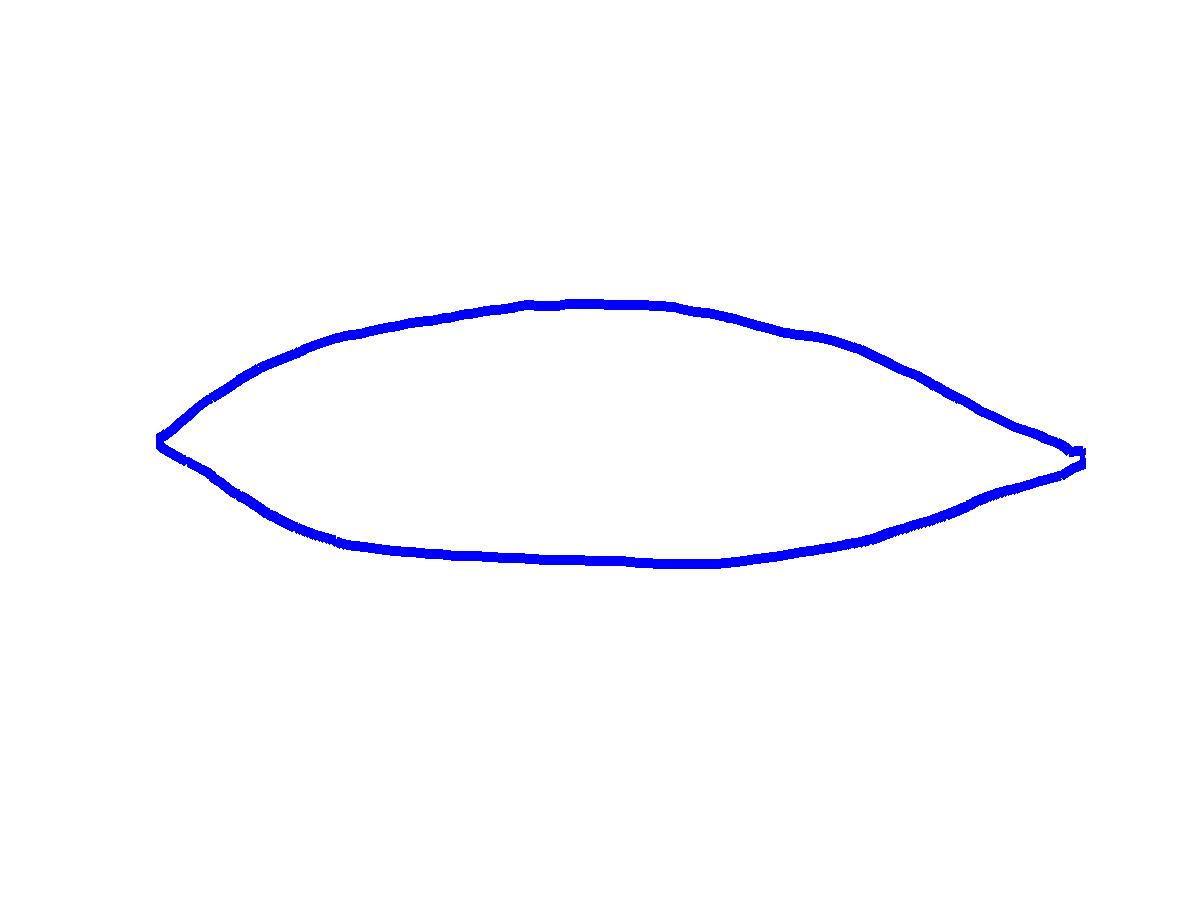}
\includegraphics[scale=0.03]{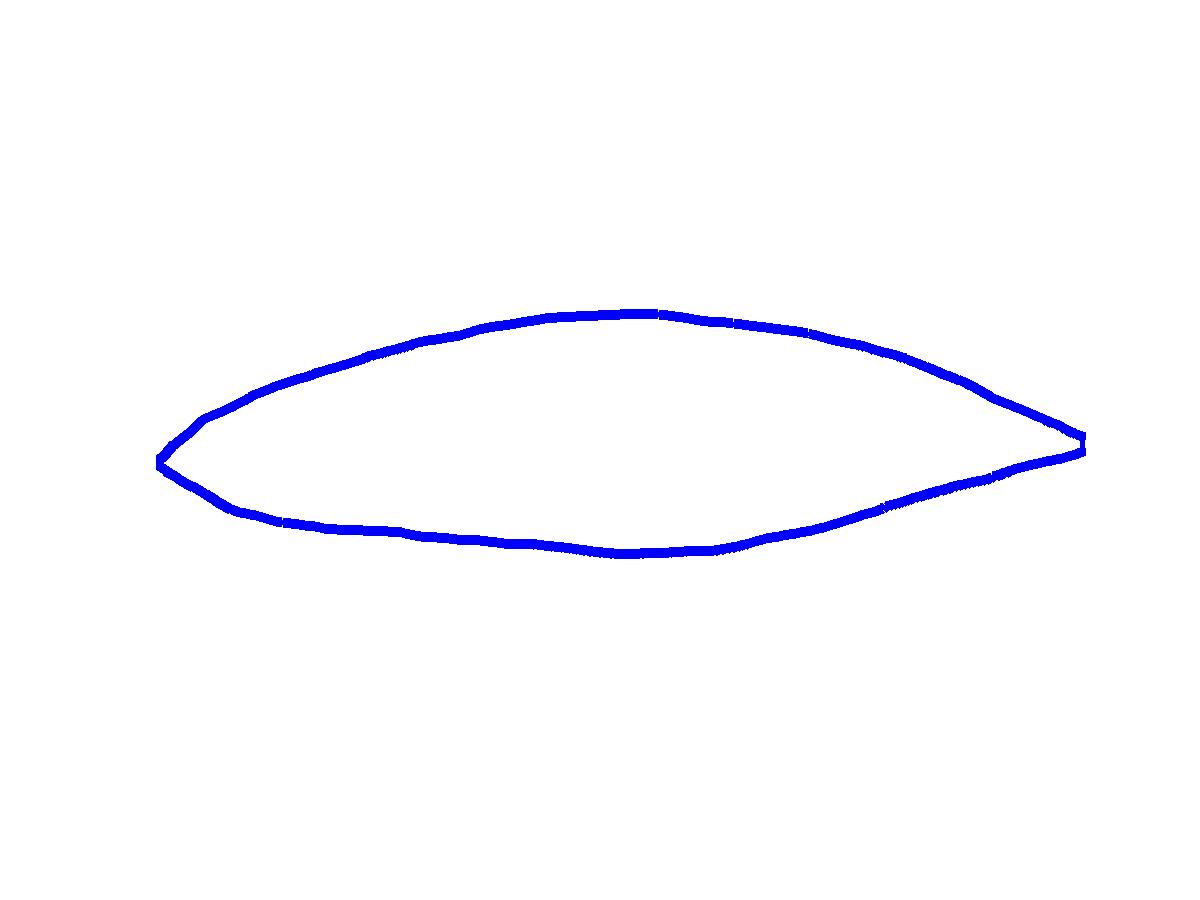}
\includegraphics[scale=0.03]{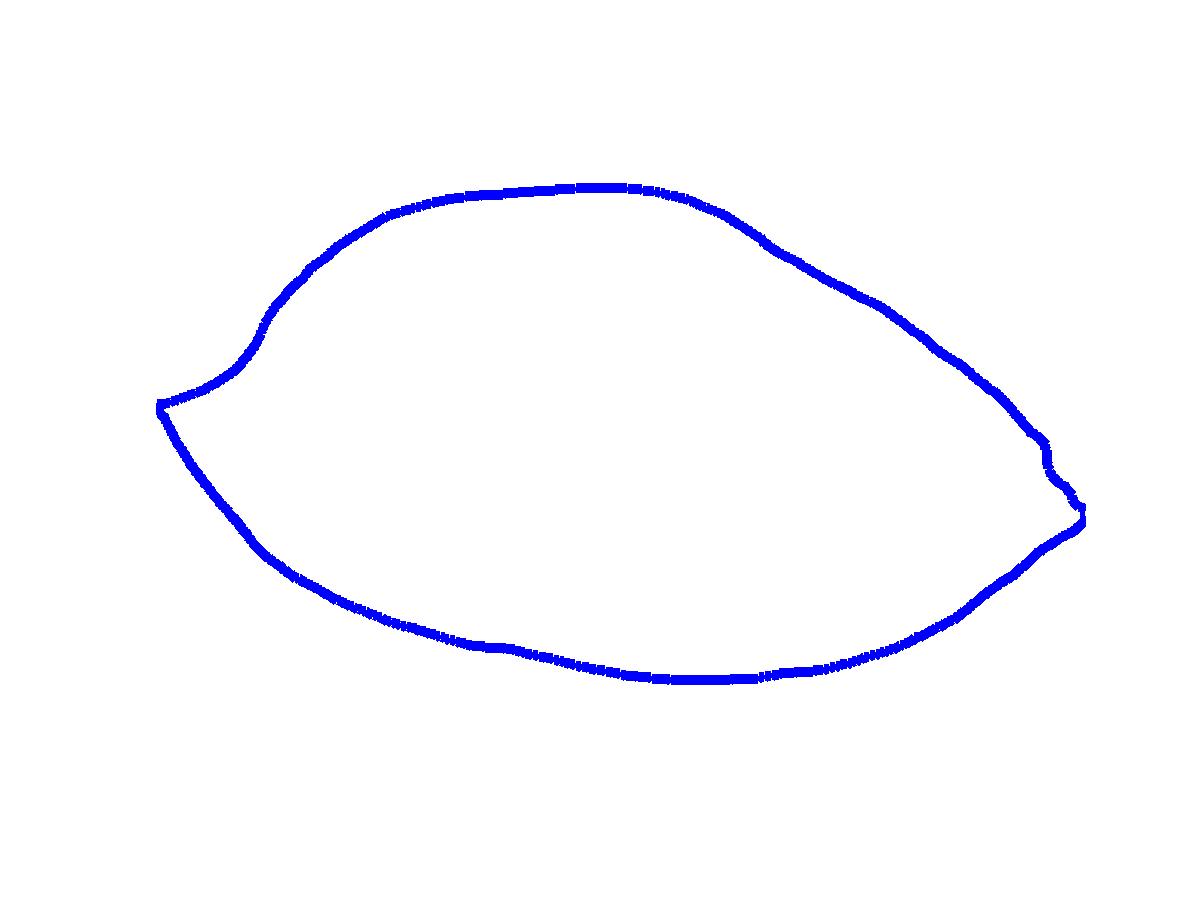}
\includegraphics[scale=0.03]{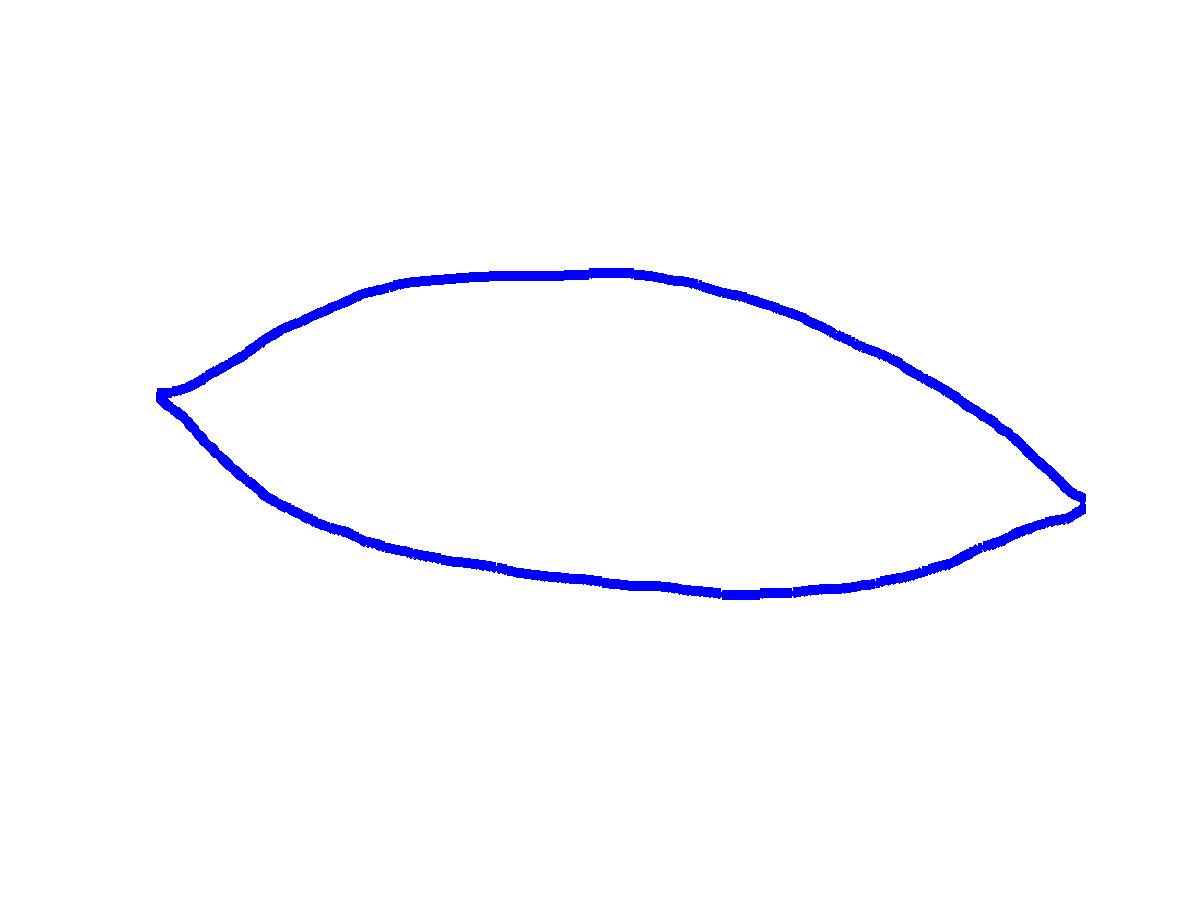}\newline%
\smallskip
\includegraphics[scale=0.03]{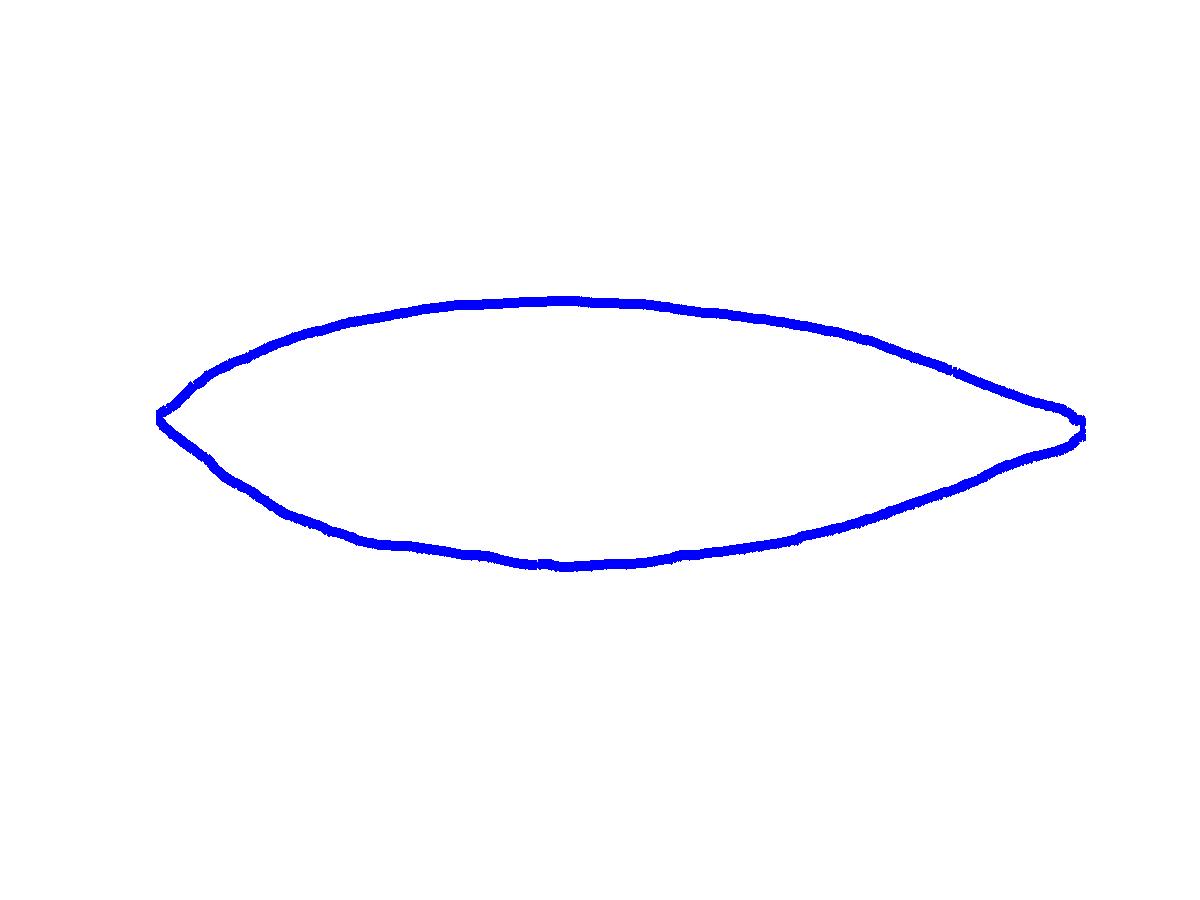}
\includegraphics[scale=0.03]{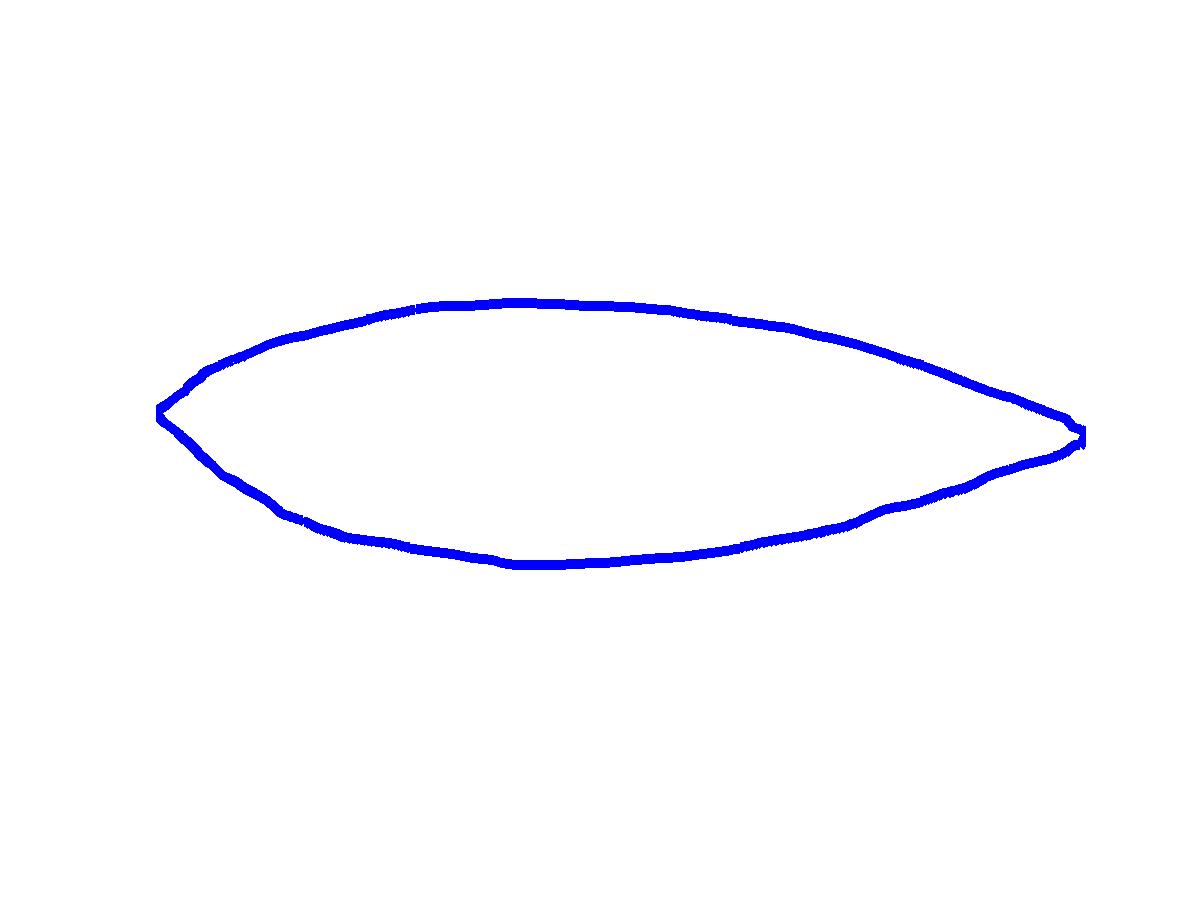}
\includegraphics[scale=0.03]{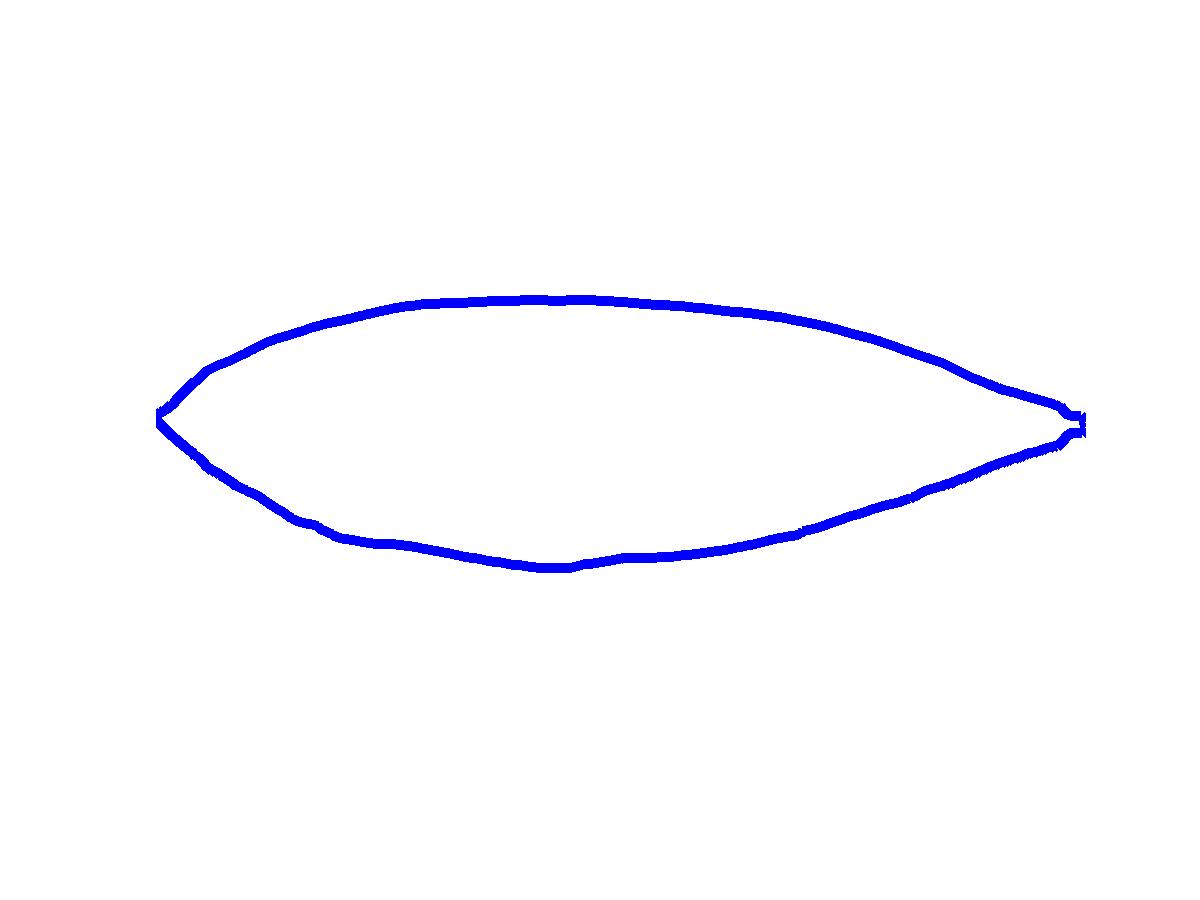}
\includegraphics[scale=0.03]{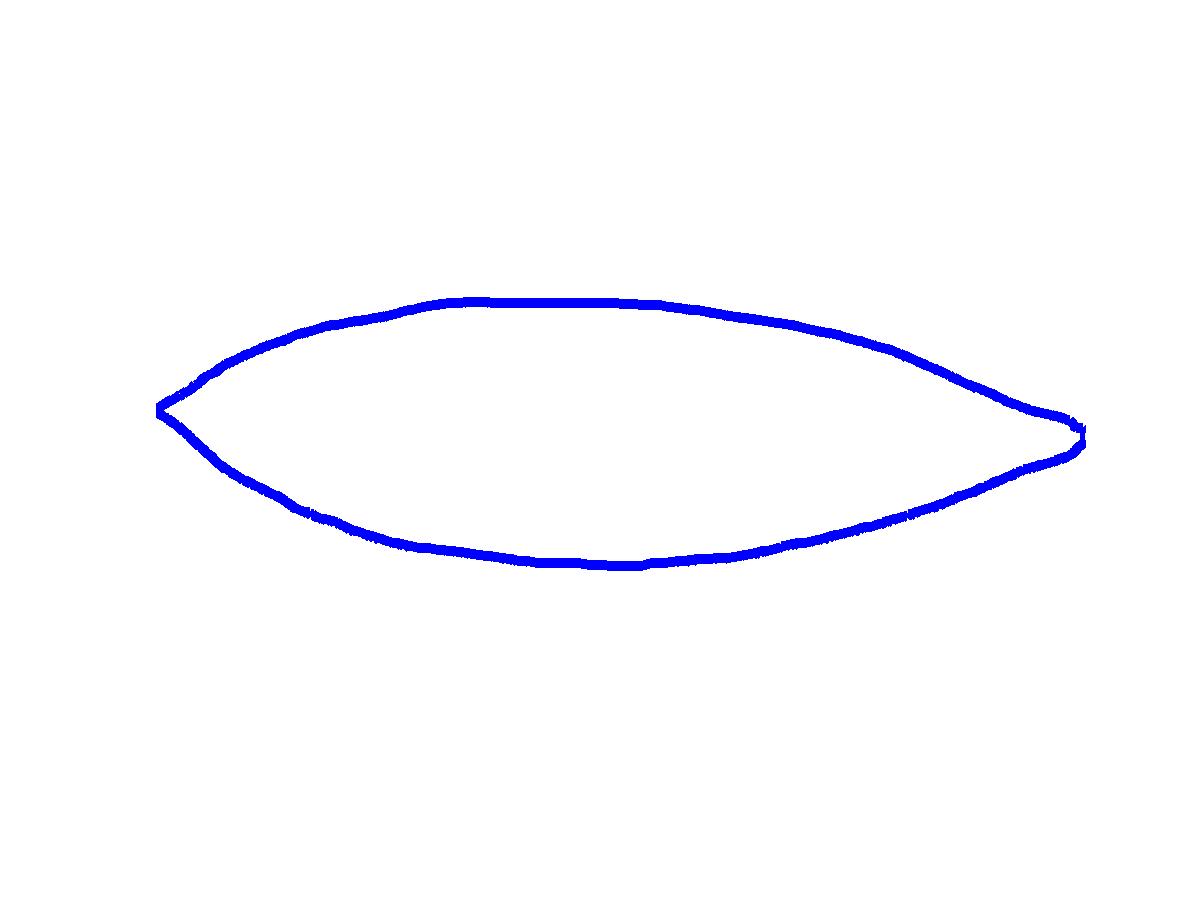}
\includegraphics[scale=0.03]{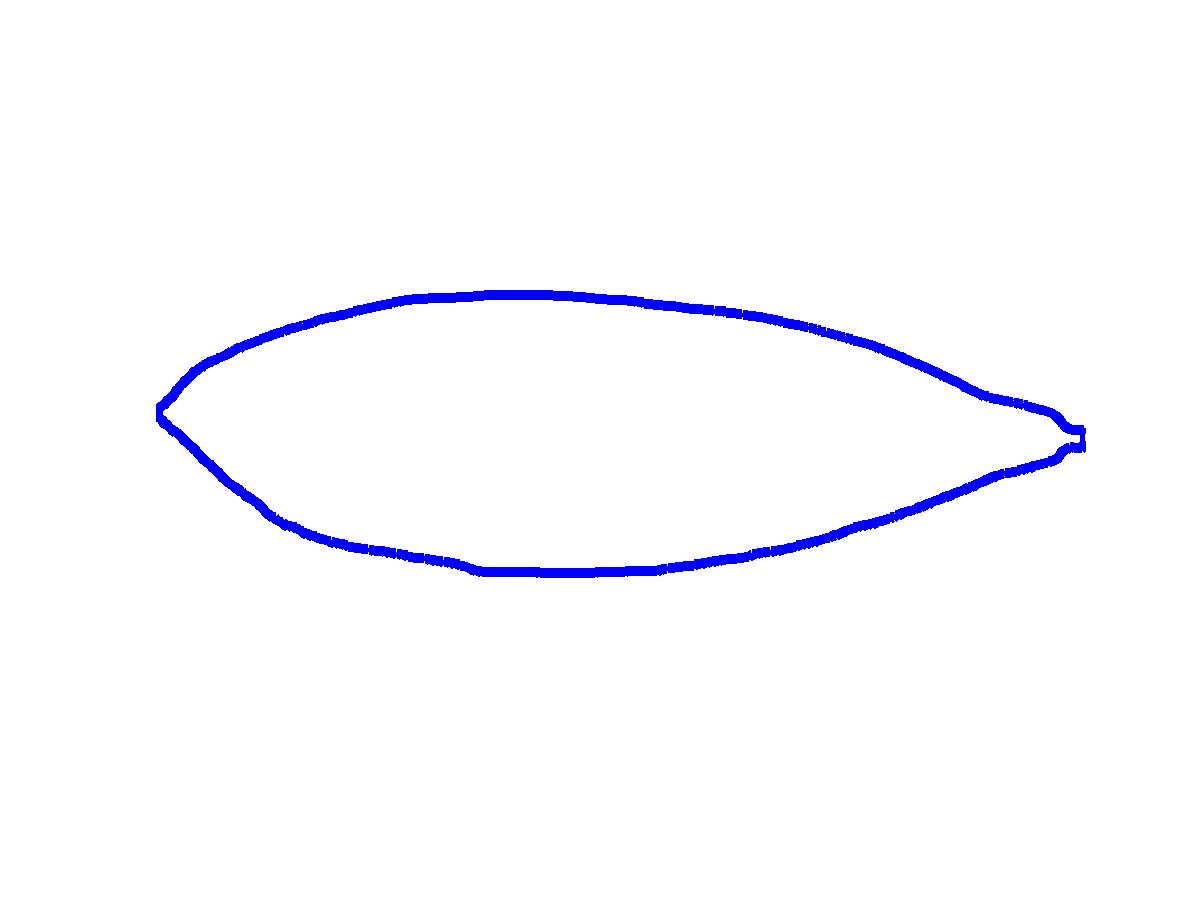}
\includegraphics[scale=0.03]{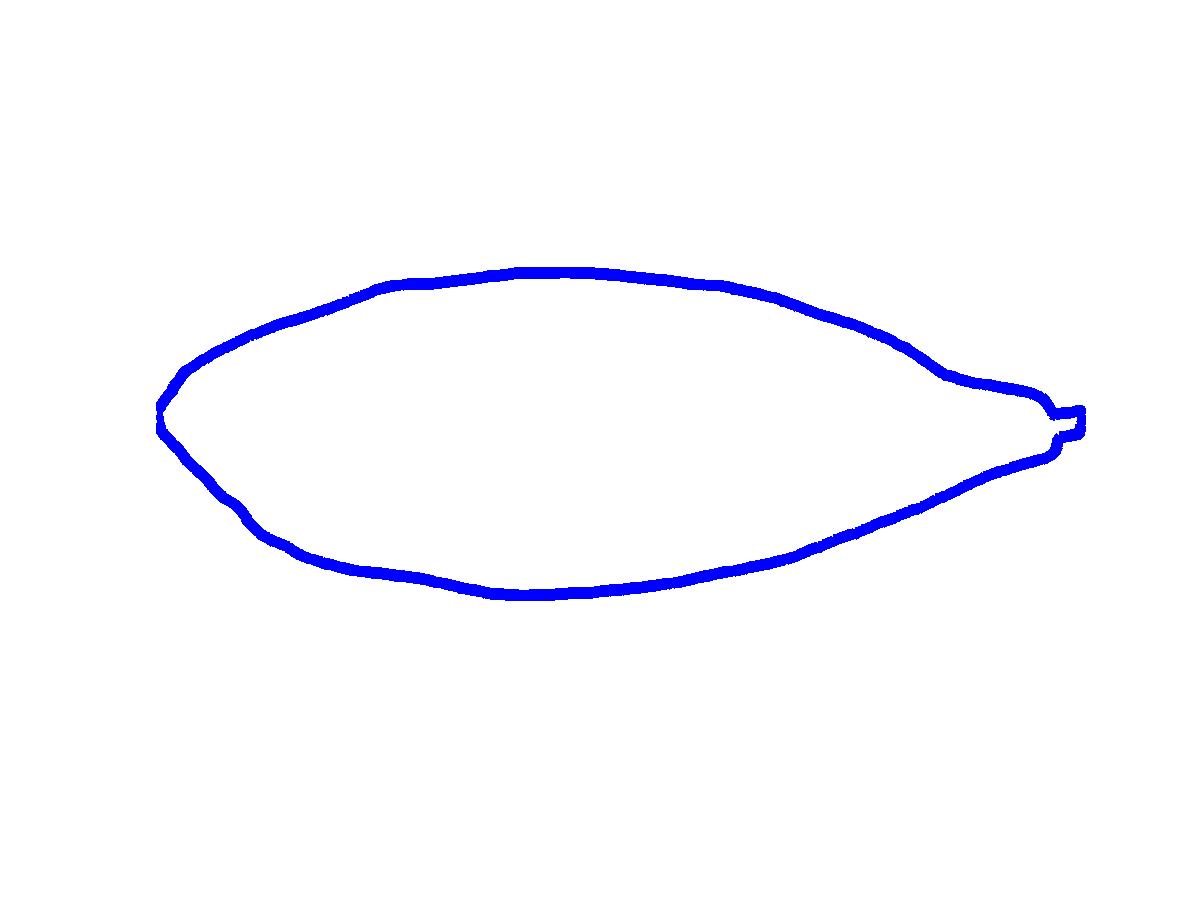}
\includegraphics[scale=0.03]{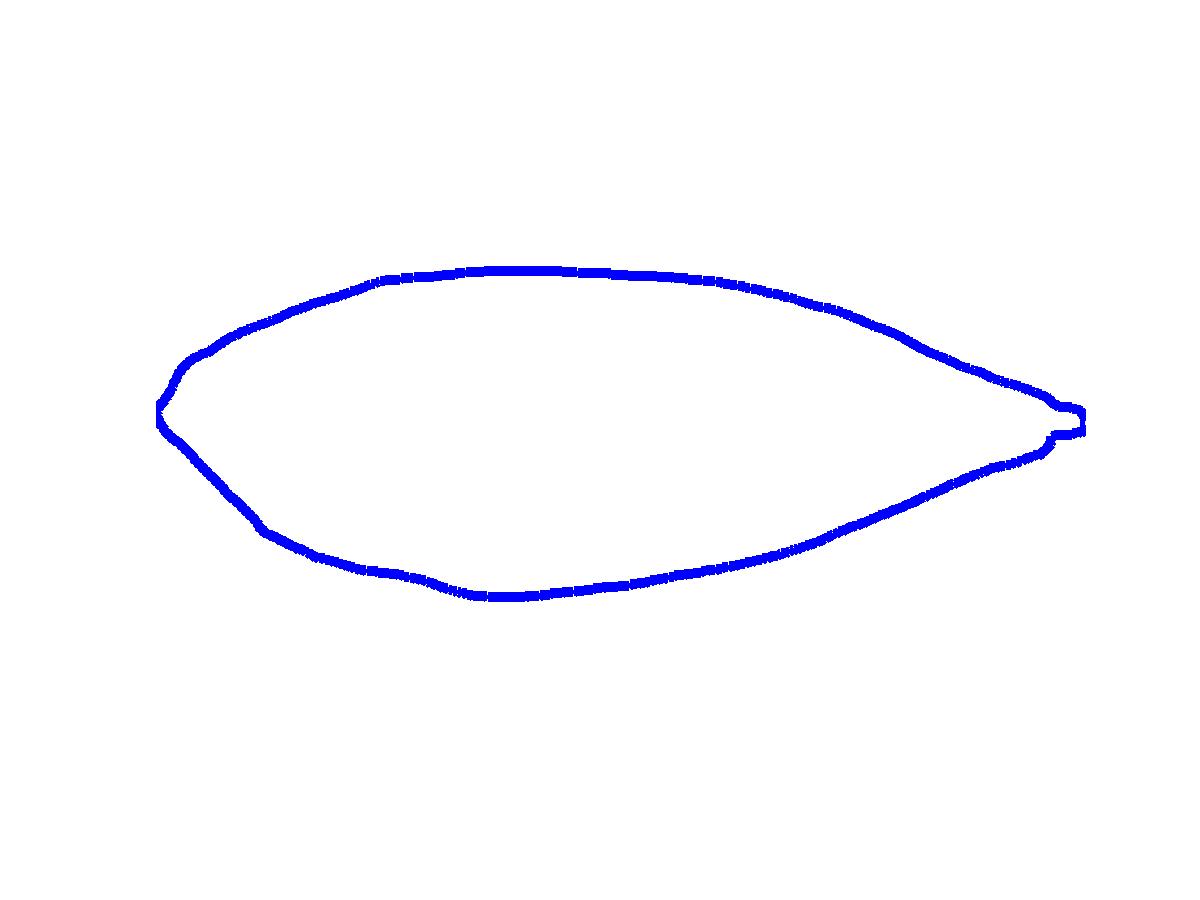}
\includegraphics[scale=0.03]{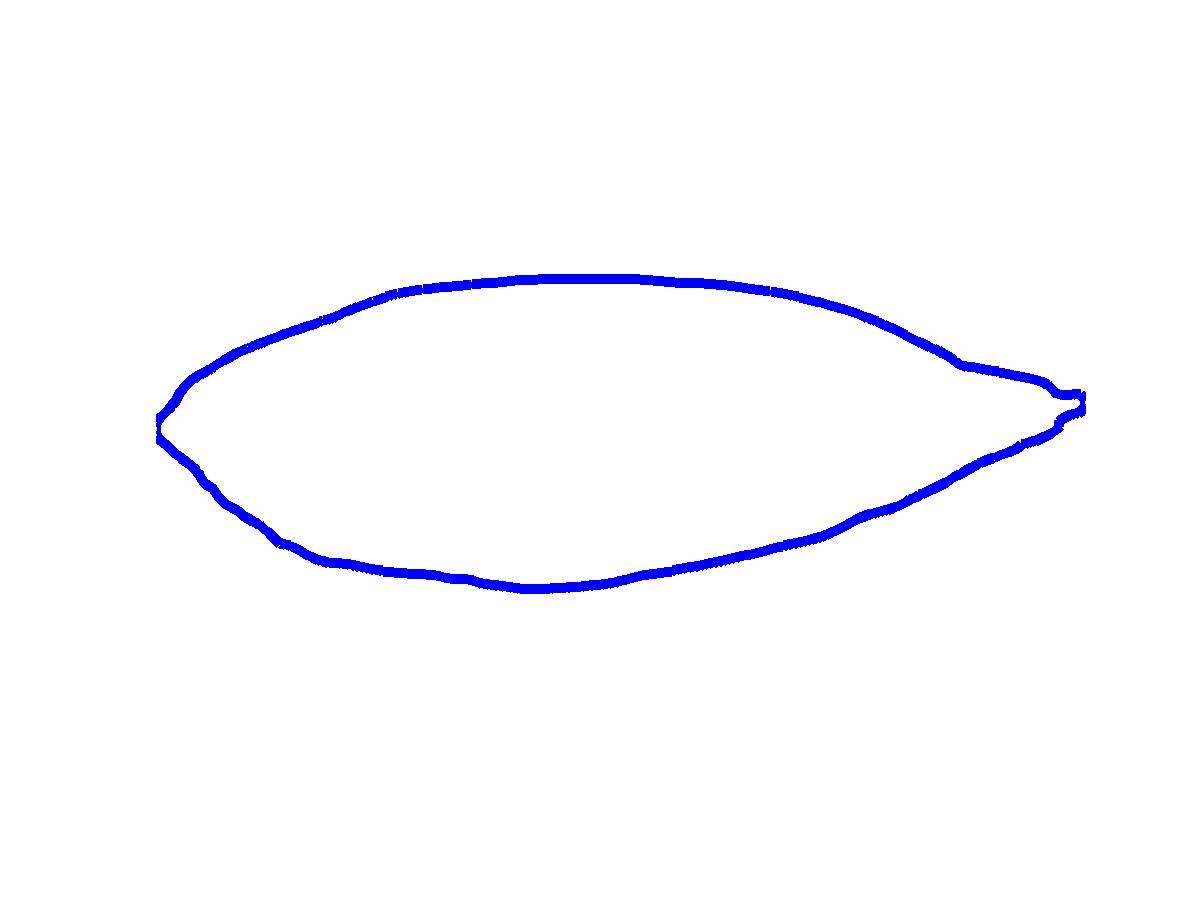}
\includegraphics[scale=0.03]{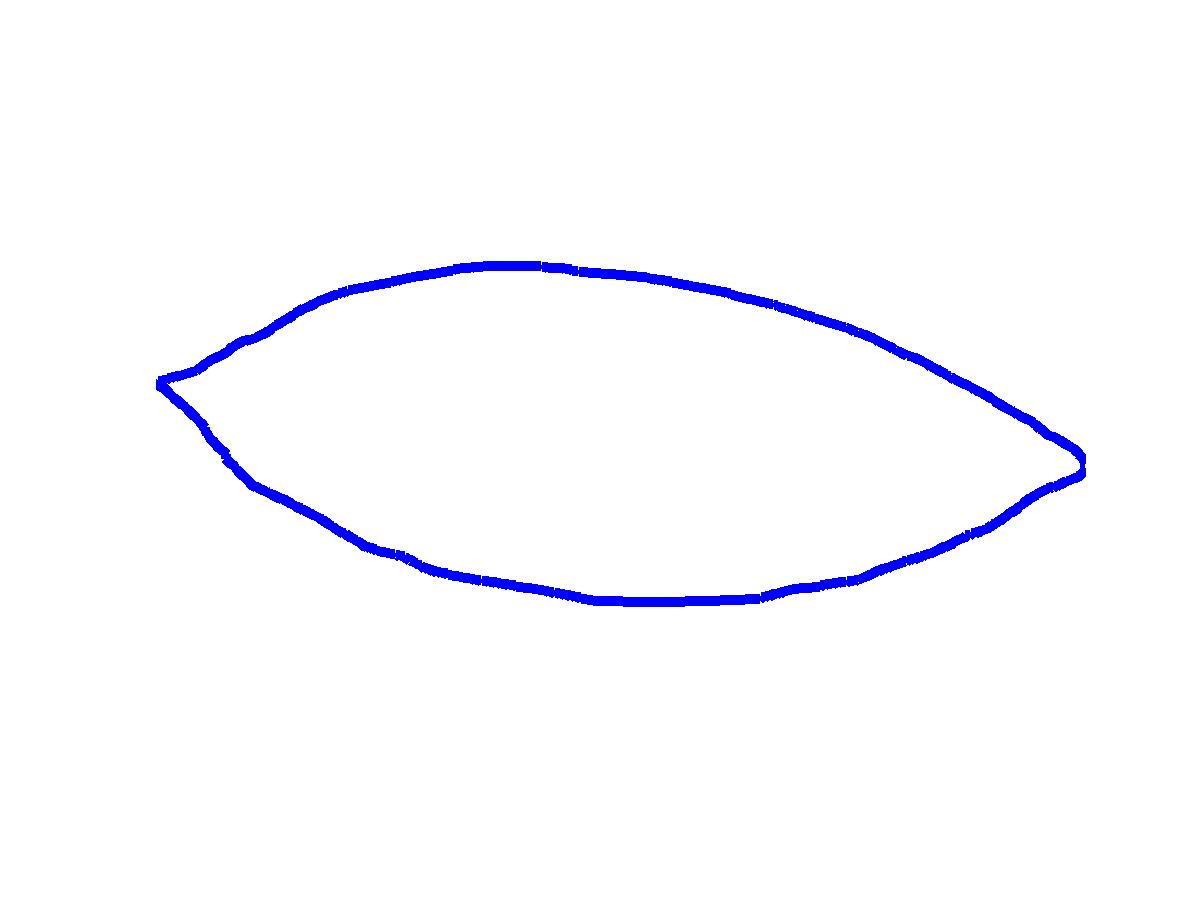}
\includegraphics[scale=0.03]{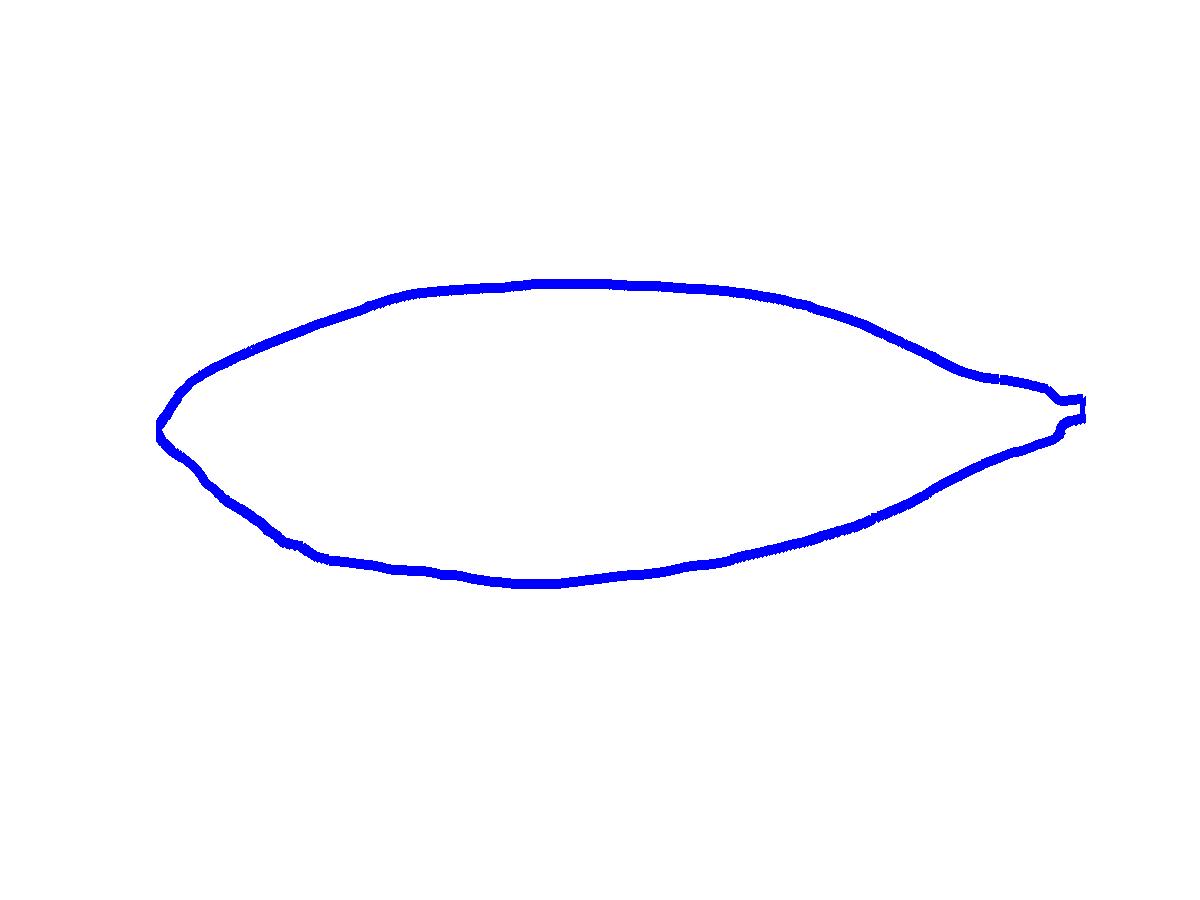}\newline%
\smallskip
\includegraphics[scale=0.03]{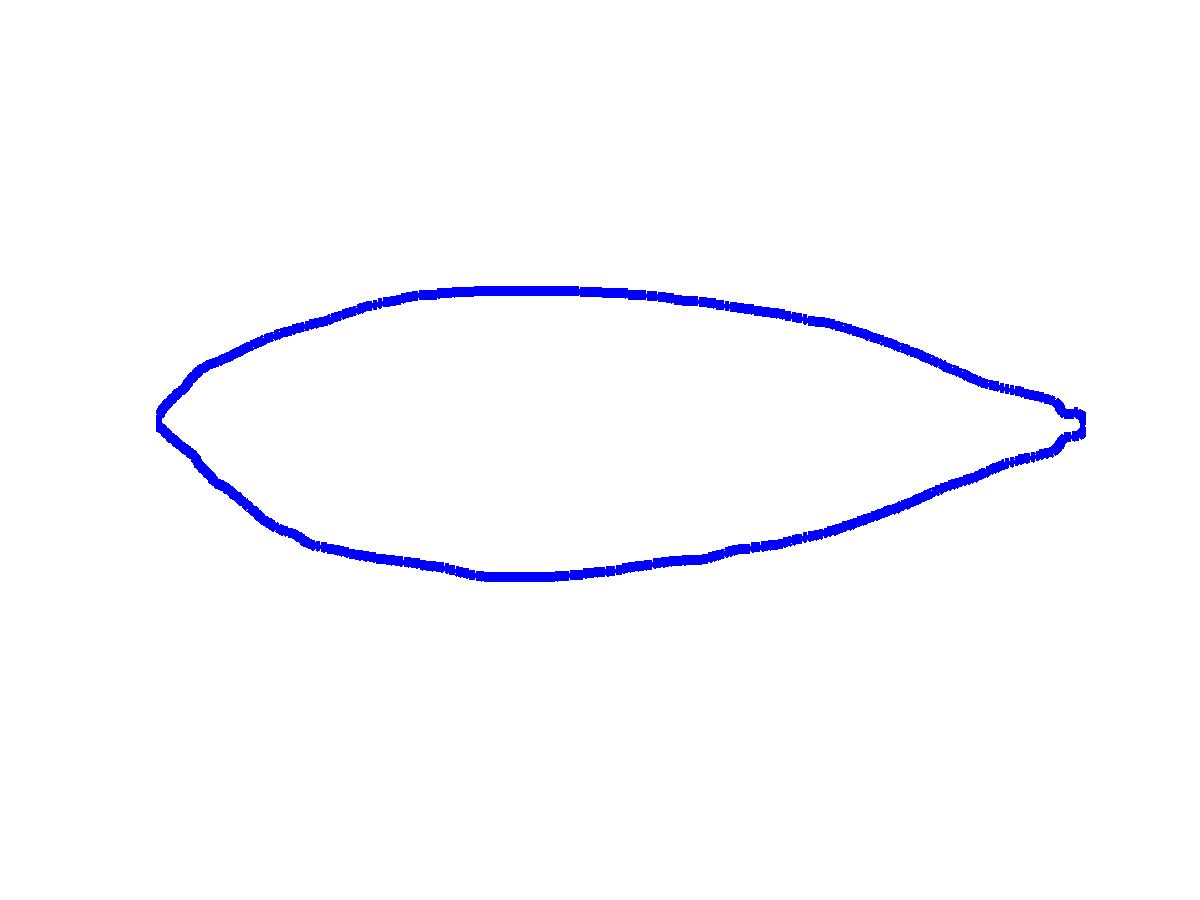}
\includegraphics[scale=0.03]{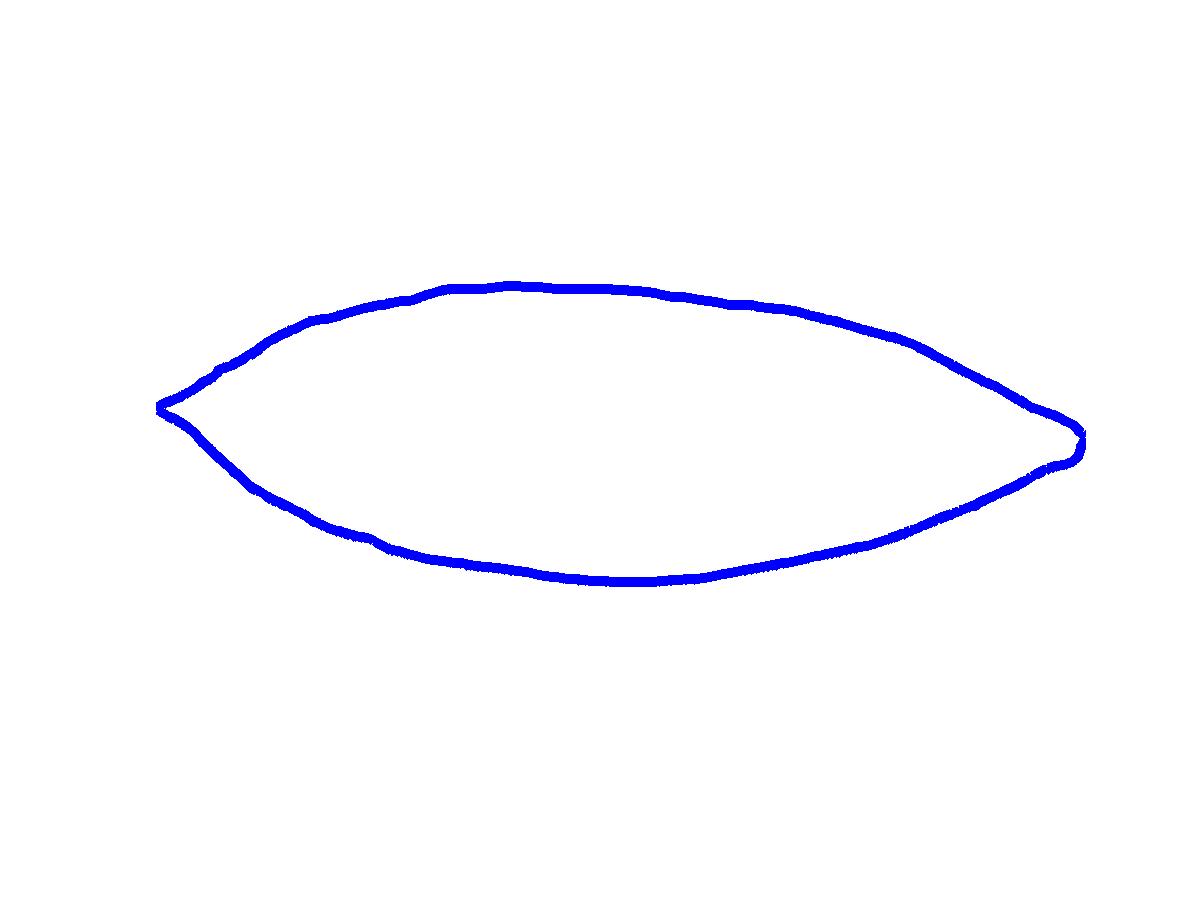}
\includegraphics[scale=0.03]{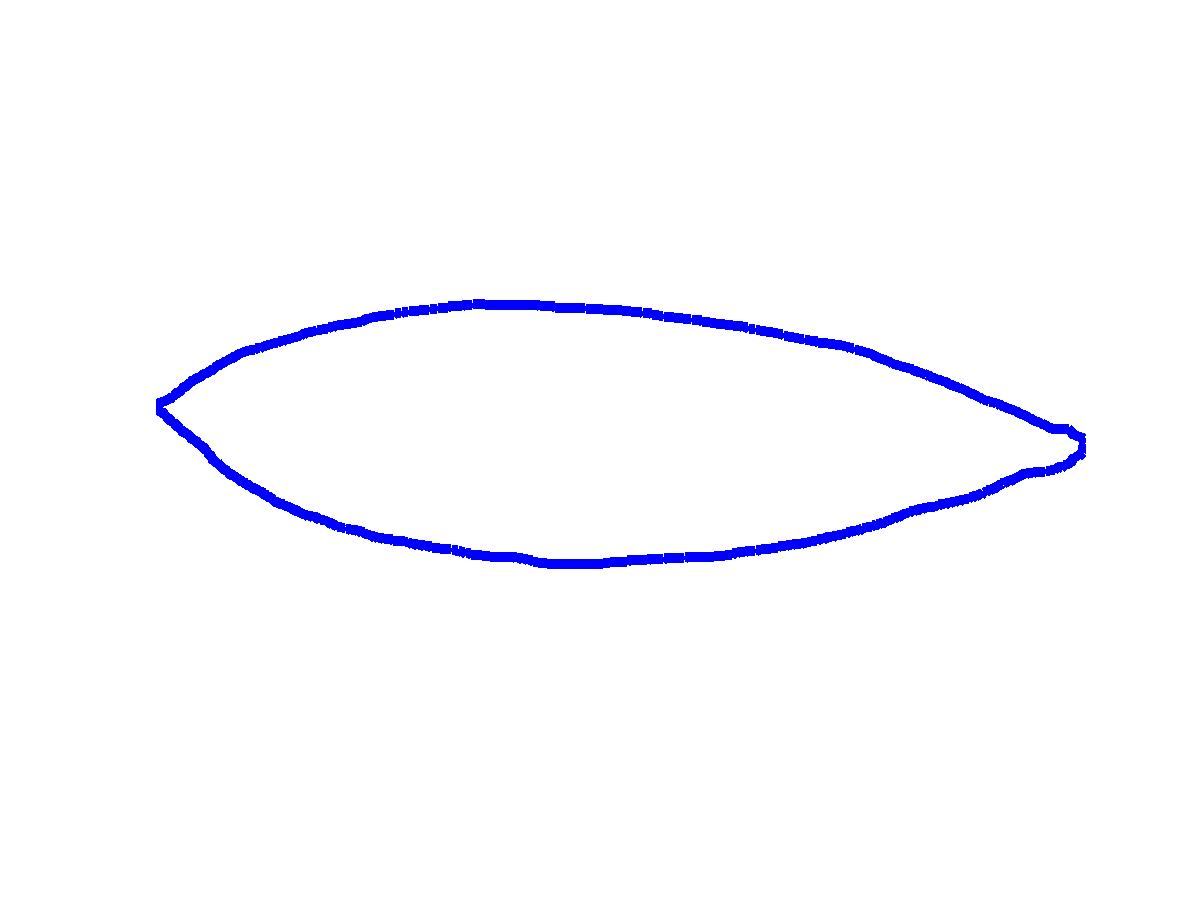}
\includegraphics[scale=0.03]{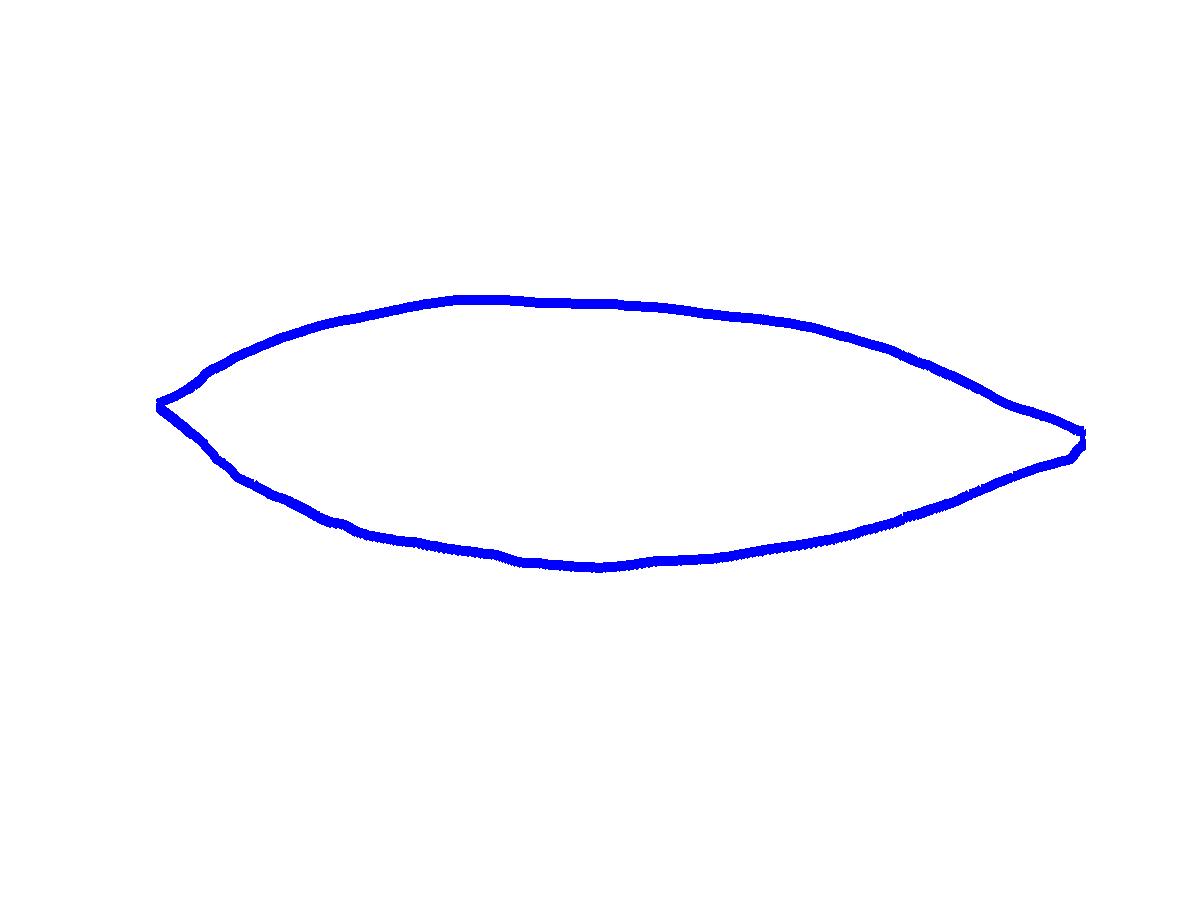}
\includegraphics[scale=0.03]{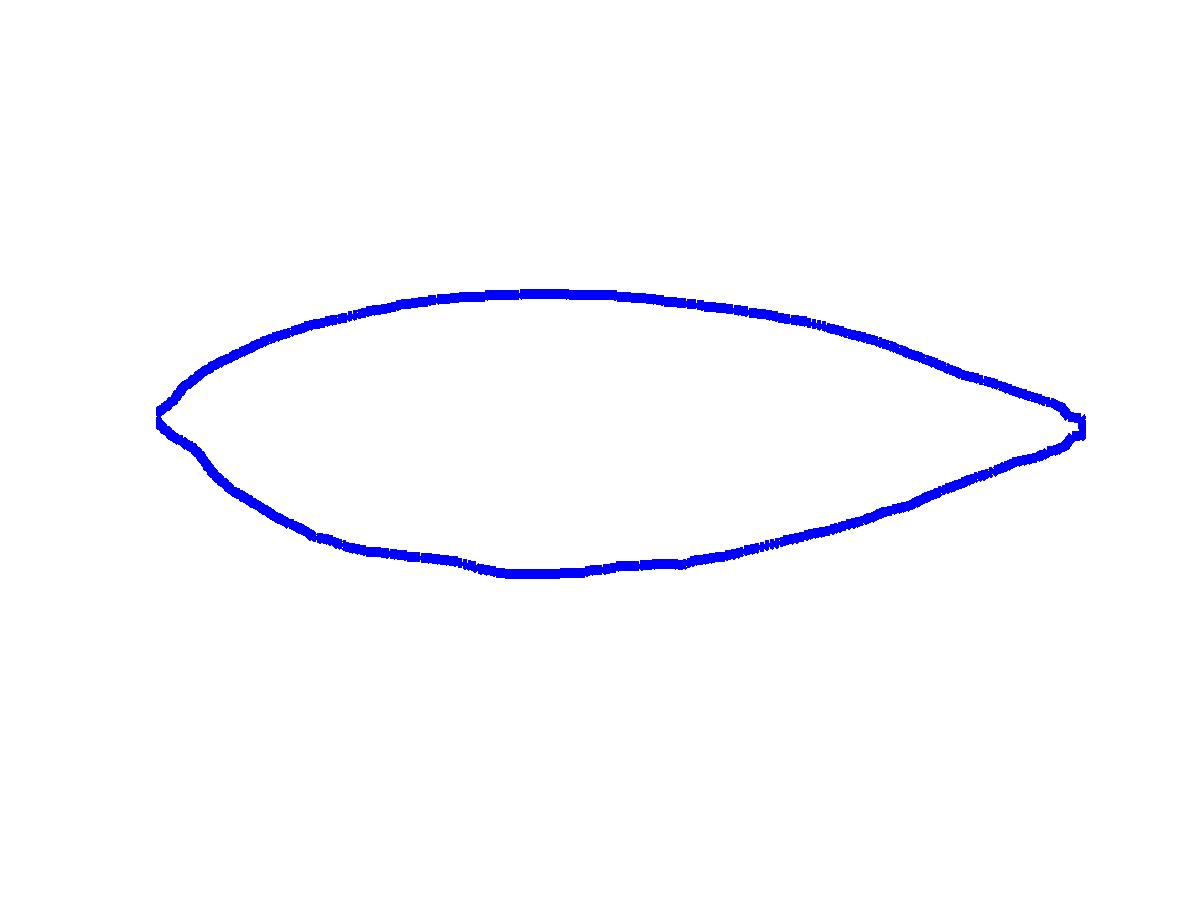}
\includegraphics[scale=0.03]{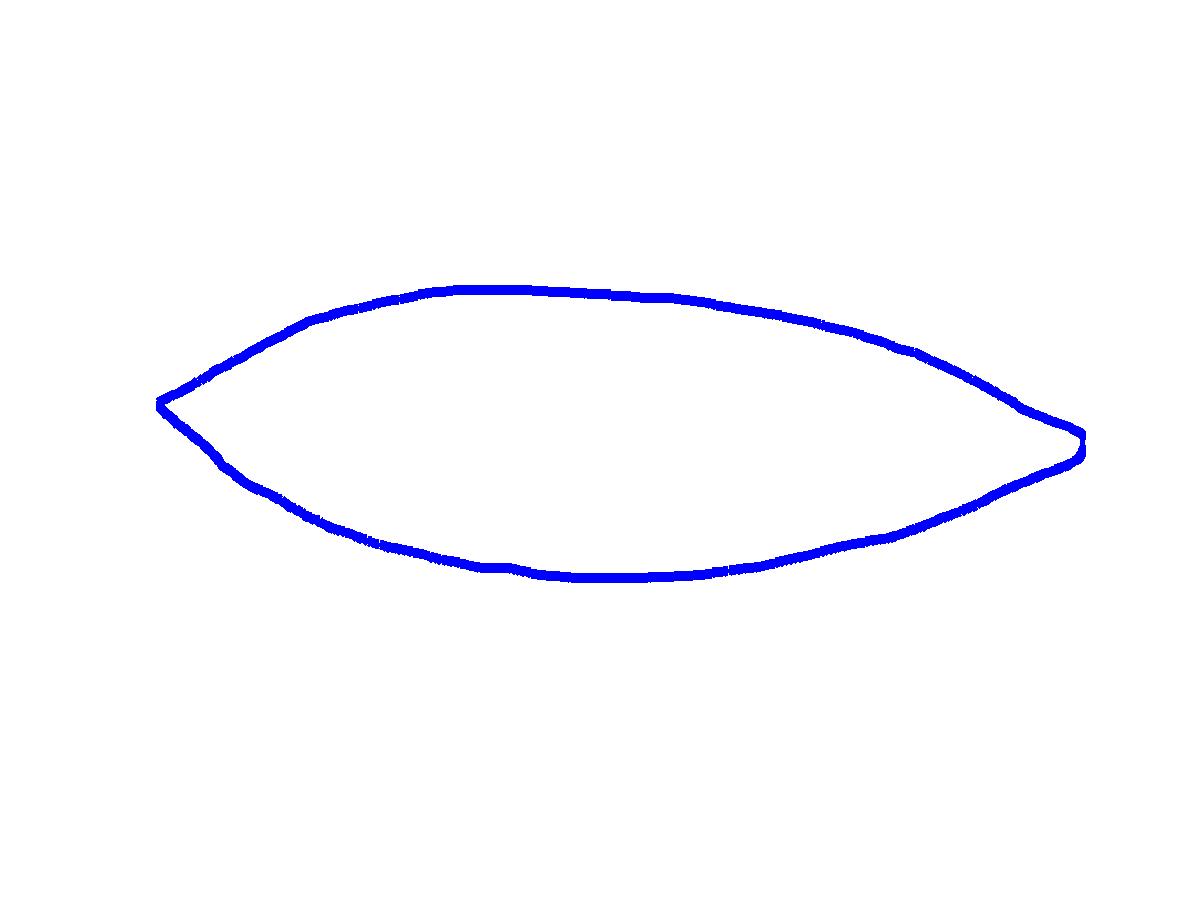}
\includegraphics[scale=0.03]{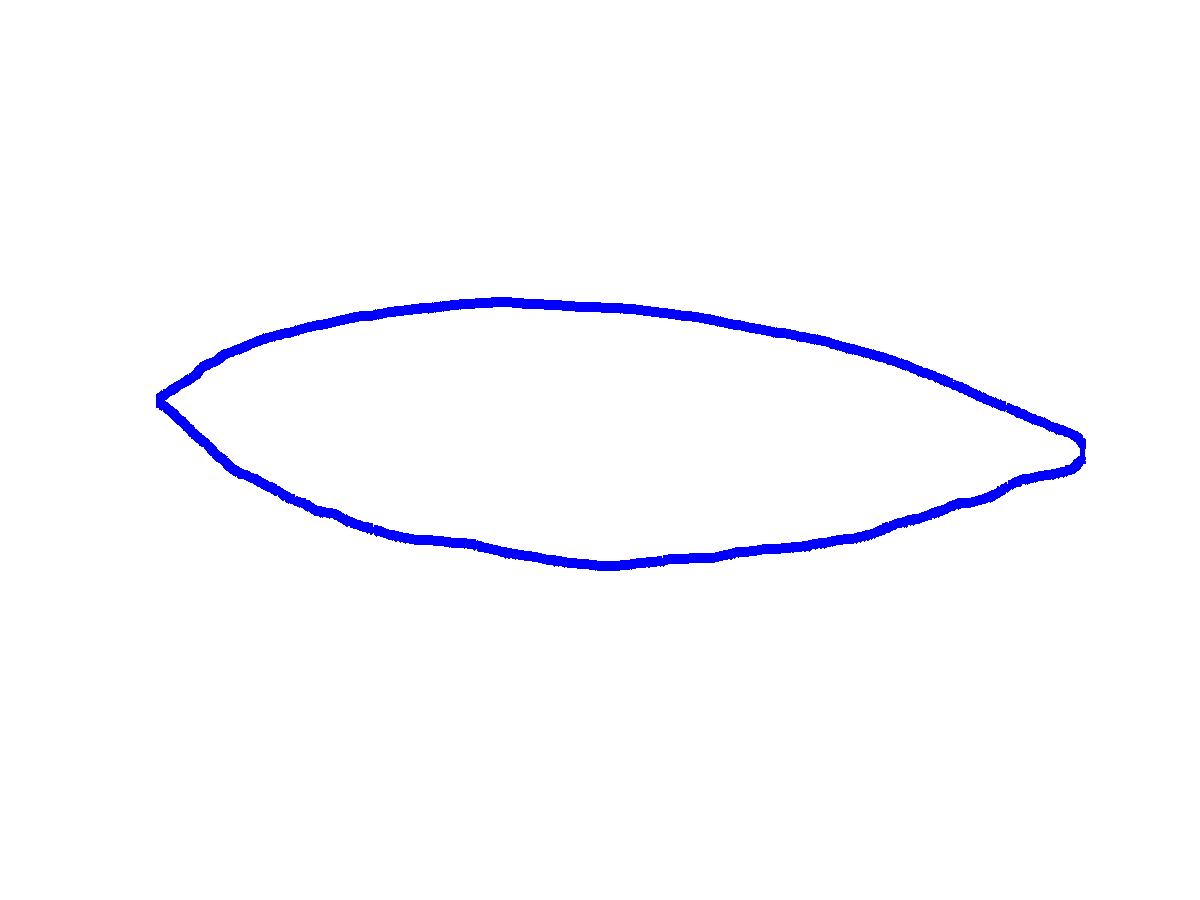}
\includegraphics[scale=0.03]{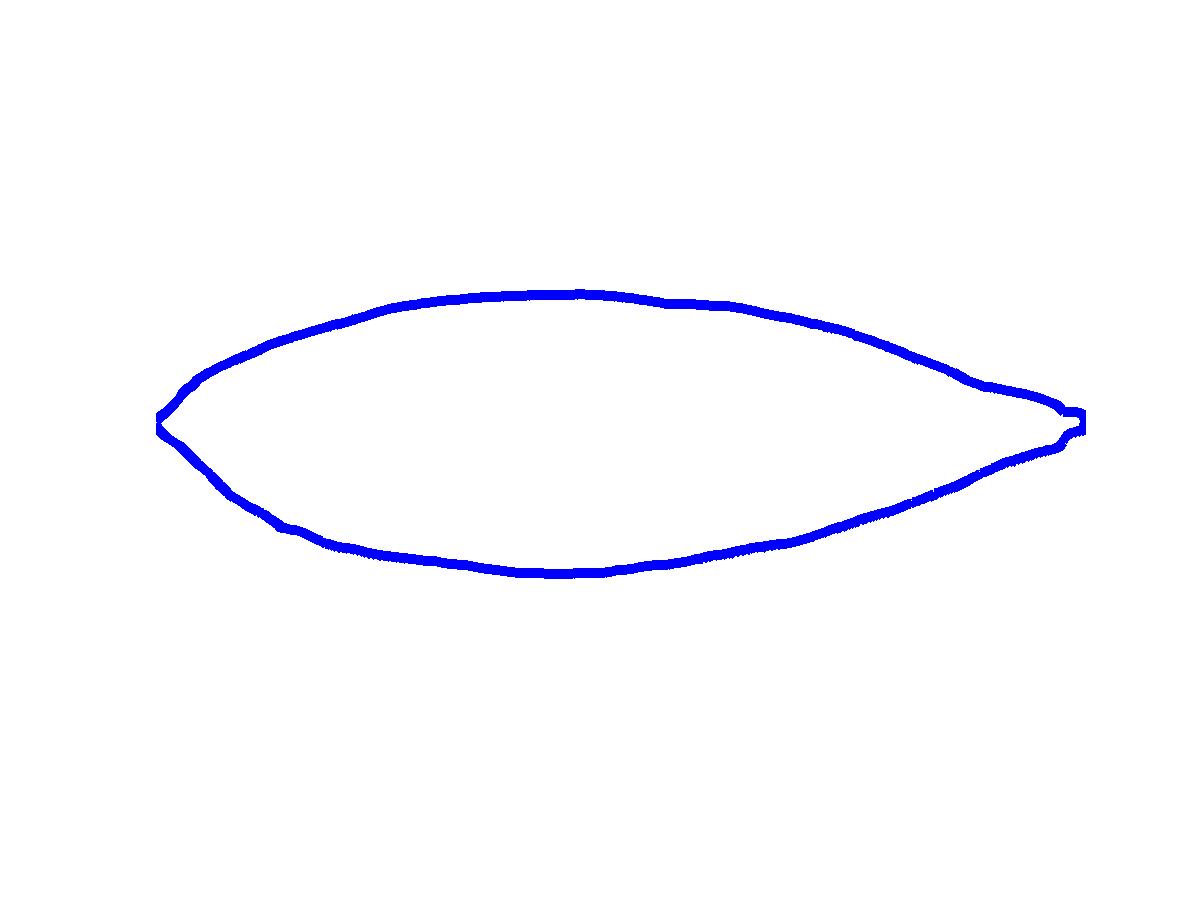}
\includegraphics[scale=0.03]{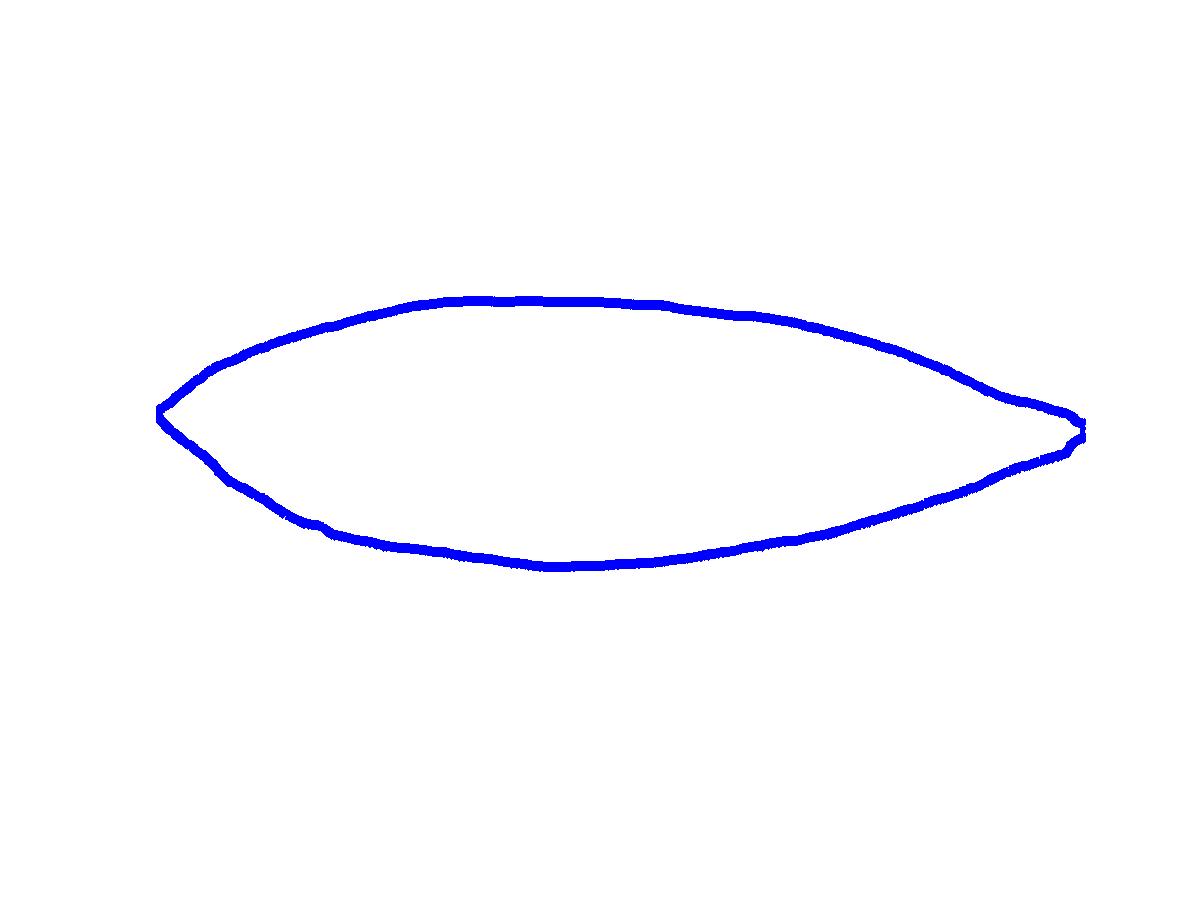}
\includegraphics[scale=0.03]{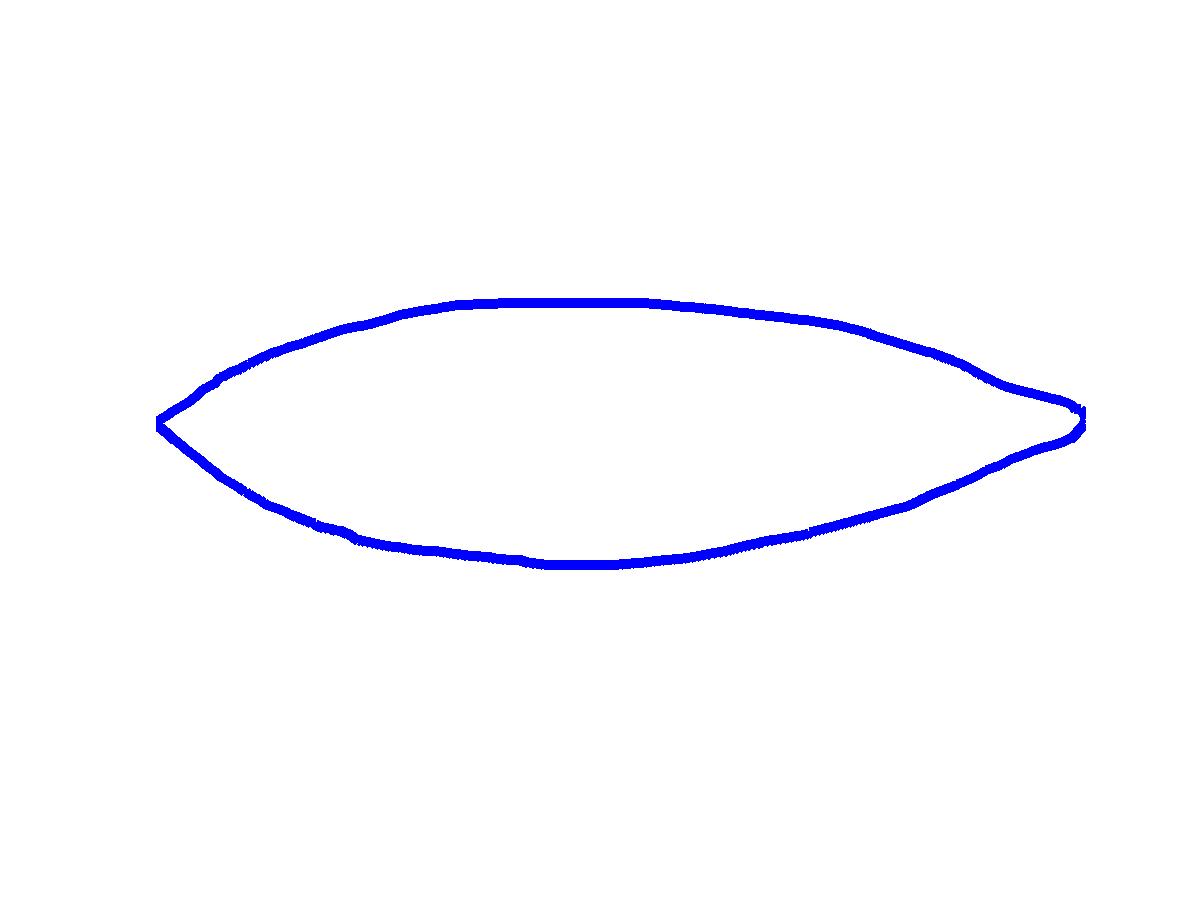}\newline%
\caption{Contours extracted from the original images.}%
\label{fig:original_contours}%
\end{figure}

Next a contour each of the 40 pictures was extracted, in MATLAB with using an
edge map, and then pairs of 2-dimensional contours were matched in resulting
in ten matched pairs of contours for each leaf using the method of Ellingson
et al. (2013)\cite{ElPaRu:2013}. After this a 3-dimensional contour was
reconstructed using the classical eight point algorithm (see for instance Ma
et al. (2006){\cite{MSKS2006}) }from each pair of 2-dimensional contours to
yield a total of ten reconstructed 3-dimensional contour for each leaf. The
two samples of 3-dimensional (reconstructed) contours for leaves A (left) and
B (right) as shown in Figure \ref{fig:original_contours}. 

Finally, a neighborhood hypothesis test for a difference in the mean
projective shape of the random 3-dimensional contours for leaves A and B was
performed and a no significant difference was found. Our TDA analysis which
follows outperforms that hypothesis testing procedure developed in  Patrangenaru et al. (2016)~\cite{PaPaYaQiLe:2016} in terms of (i) being computationally
much easier to implement, (ii) yielding statistically more powerful tests
which find a significant difference in the leaf A and leaf B images as one
would expect and (iii) providing much more information about the topological
differences in the leaf image point clouds. Our computations were performed in
MATLAB and Image Processing Toolbox Release 2013a, in R-3.4.1 with Pawel
Dlotko's Plot Of Landscapes package \cite{Dl:2018} and in C++ with Ulrich
Bauer's Ripser code~\cite{Ba:2017}. First edge detection was performed in
MATLAB with Sobel, Canny, Prewitt, Roberts, and Log (zero-crossing) methods in the Image Processing toolbox. After an inspection of the
results from all five edge detection methods for all 40 point clouds it was
found the Log method was essentially the best method in terms of detecting
points on the edge or contour and the veins of a leaf and filtering out points
not on contour or a vein.
The totality of 40 leaf edges, from the Log edge detection method, are shown
below in Figure \ref{fig:Leaf_edges}.

\begin{figure}[h]
\centering
\includegraphics[scale=0.09]{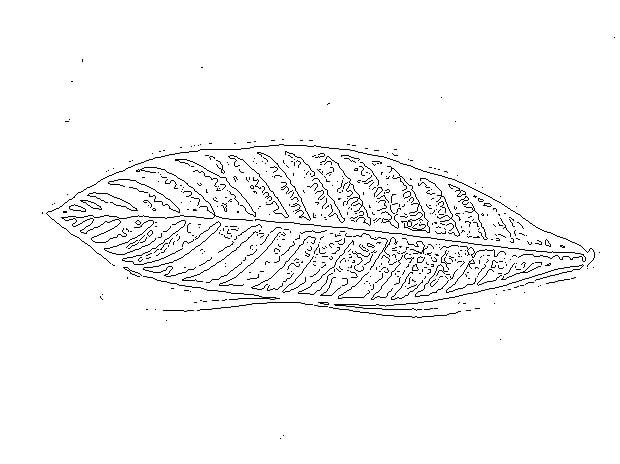}
\includegraphics[scale=0.09]{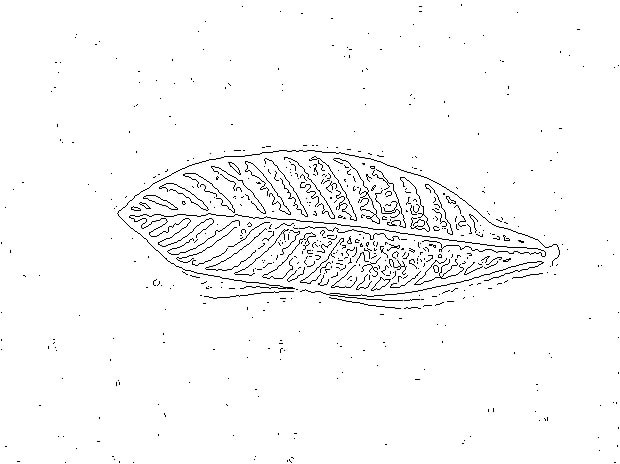}
\includegraphics[scale=0.09]{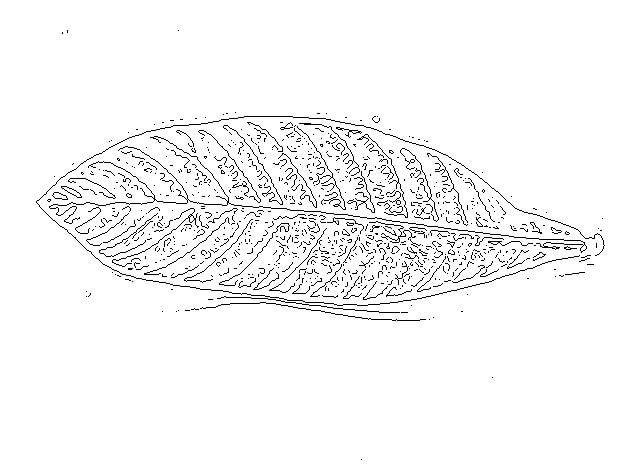}
\includegraphics[scale=0.09]{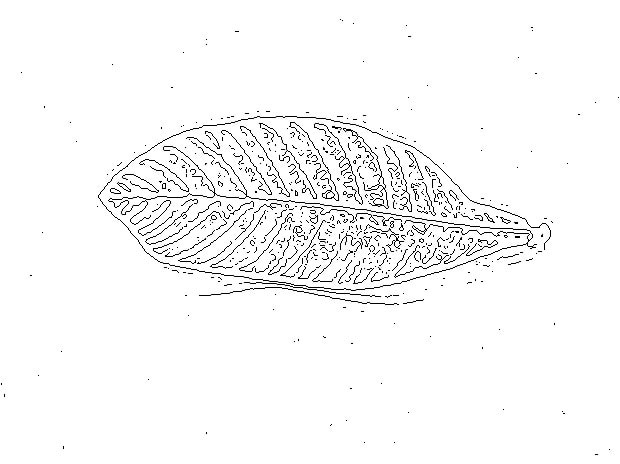}
\includegraphics[scale=0.09]{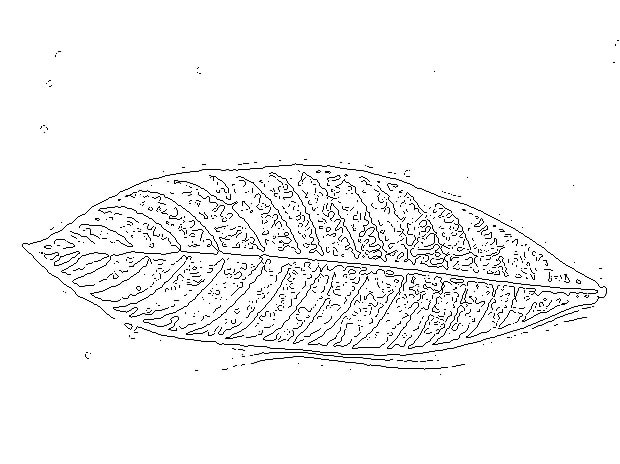}
\includegraphics[scale=0.09]{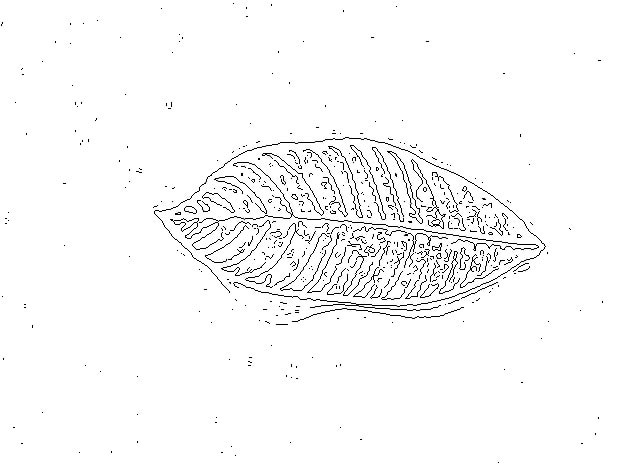}
\includegraphics[scale=0.09]{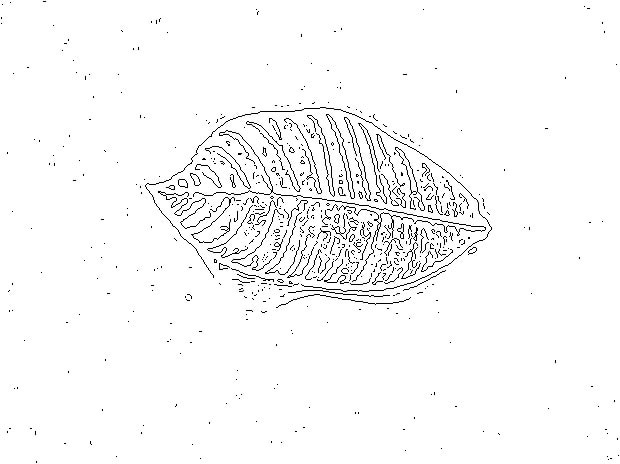}
\includegraphics[scale=0.09]{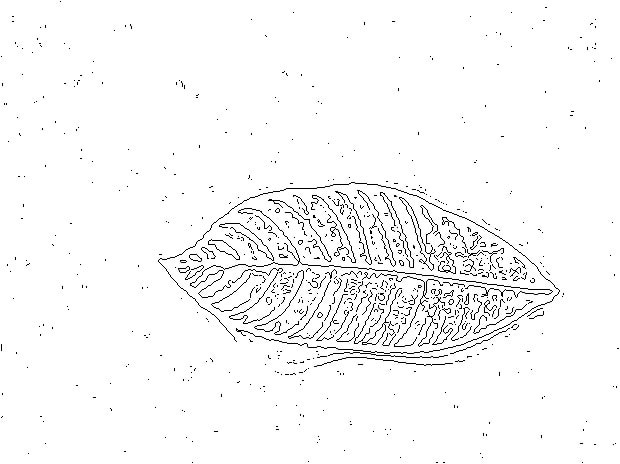}
\includegraphics[scale=0.09]{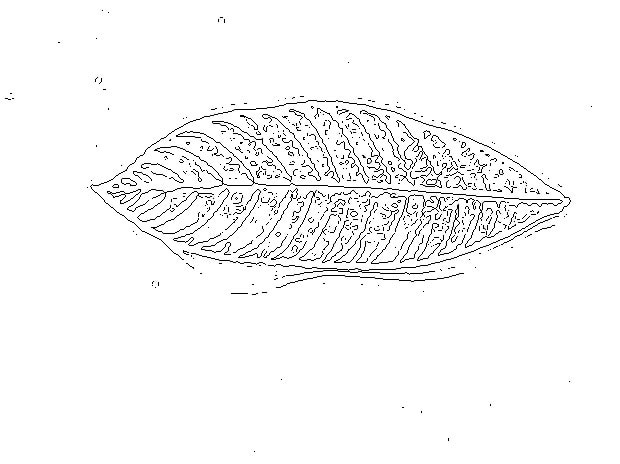}
\includegraphics[scale=0.09]{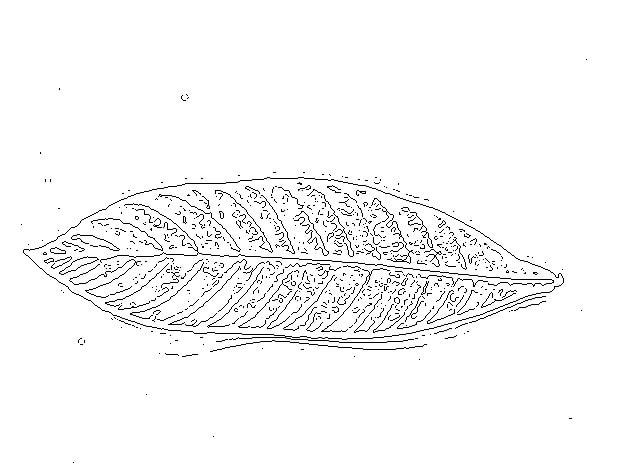}\newline%
\smallskip\includegraphics[scale=0.09]{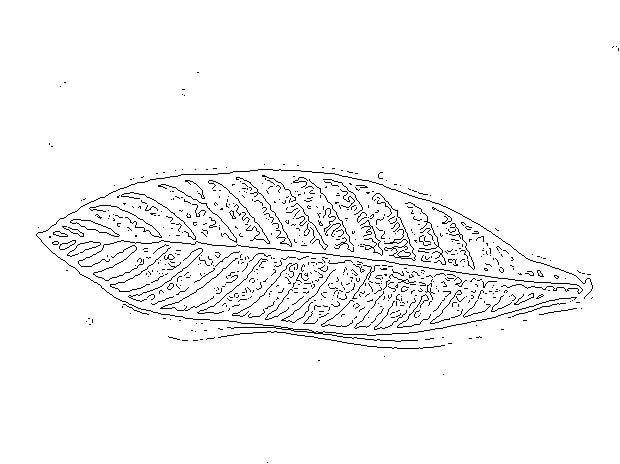}
\includegraphics[scale=0.09]{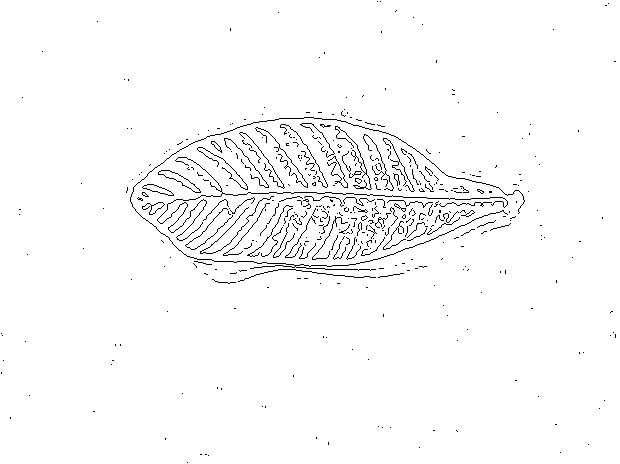}
\includegraphics[scale=0.09]{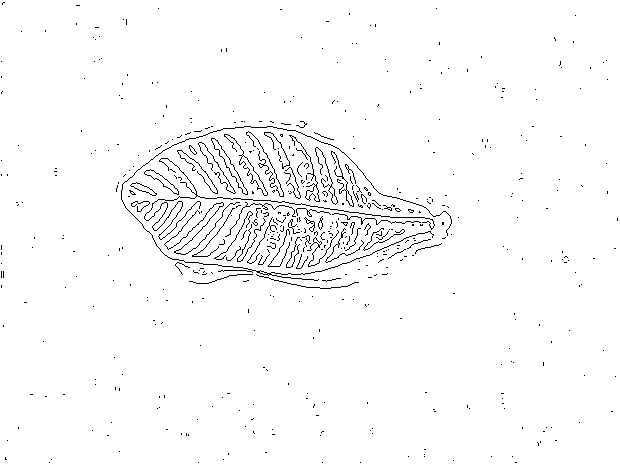}
\includegraphics[scale=0.09]{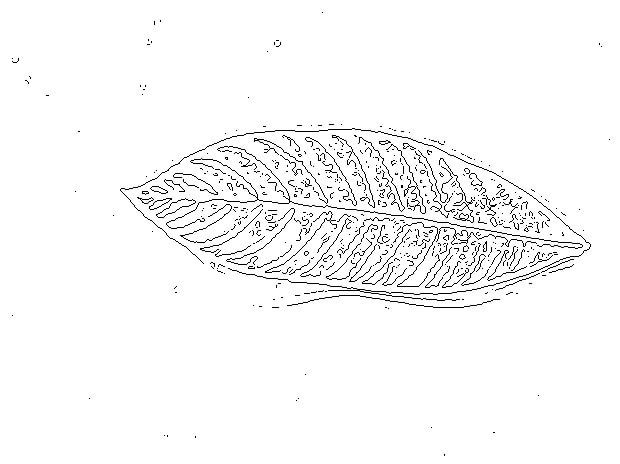}
\includegraphics[scale=0.09]{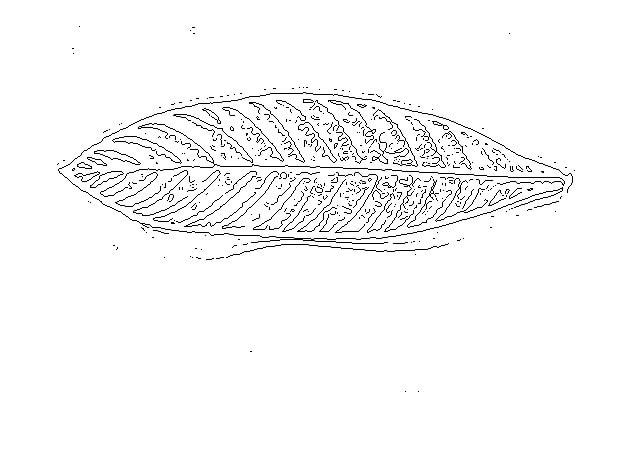}
\includegraphics[scale=0.09]{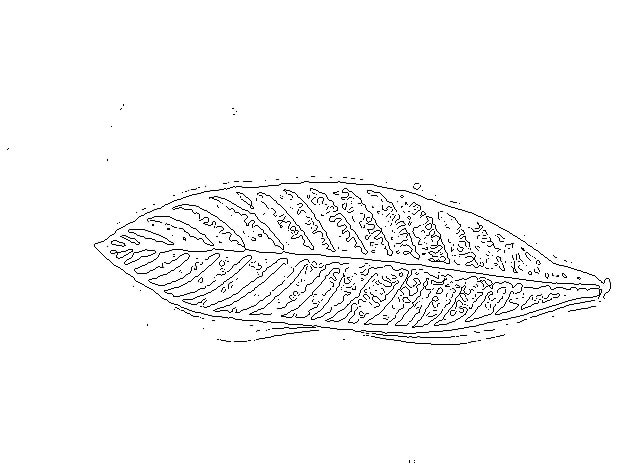}
\includegraphics[scale=0.09]{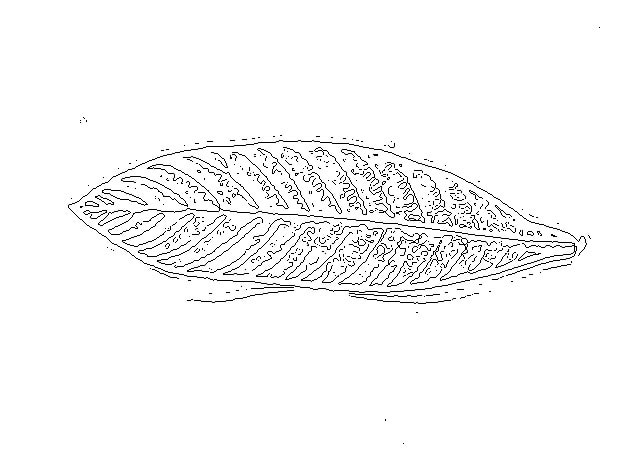}
\includegraphics[scale=0.09]{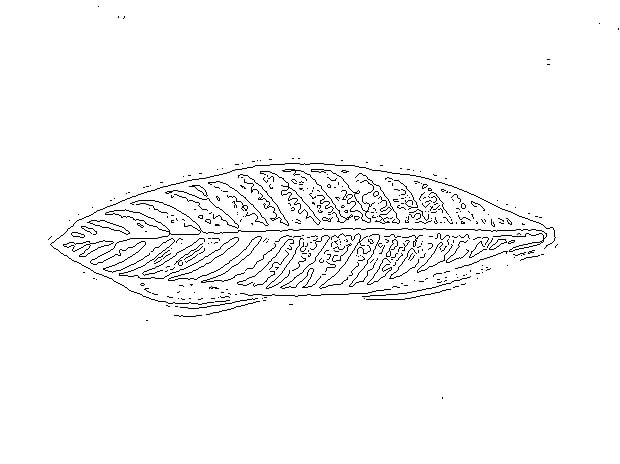}
\includegraphics[scale=0.09]{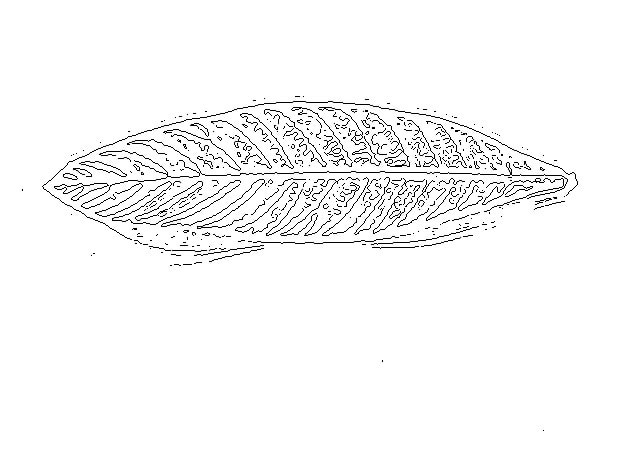}
\includegraphics[scale=0.09]{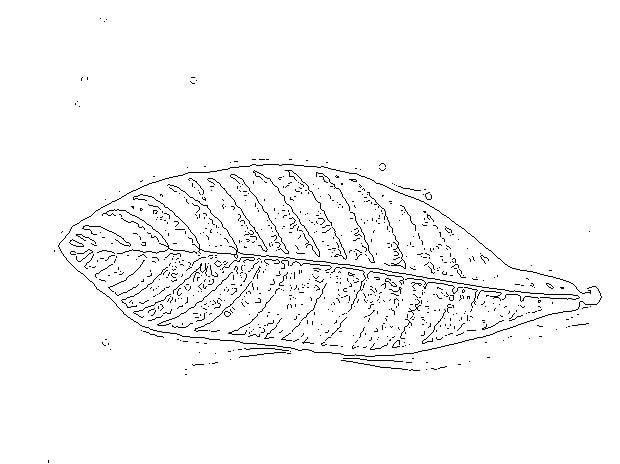}\newline%
\smallskip\includegraphics[scale=0.09]{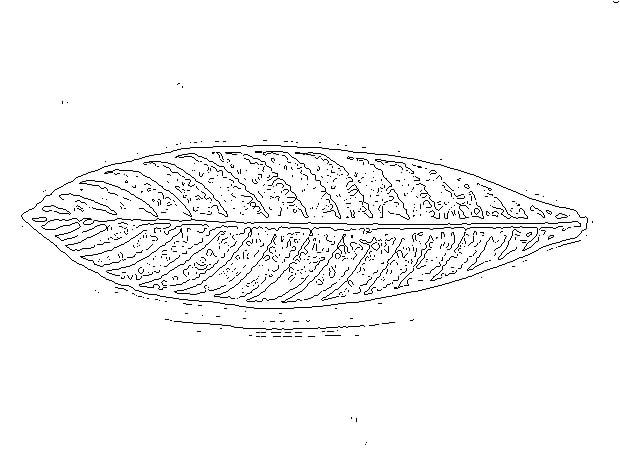}
\includegraphics[scale=0.09]{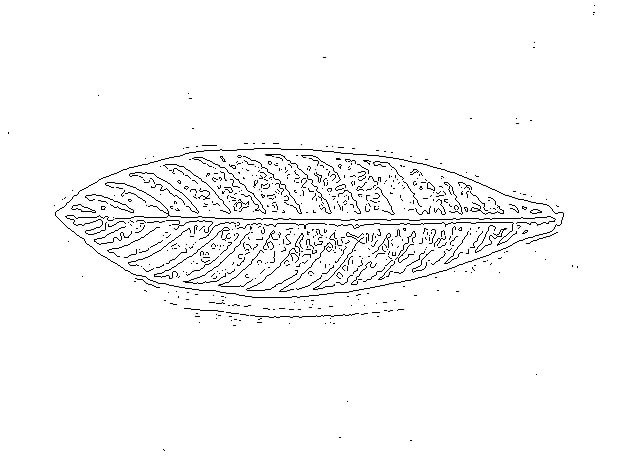}
\includegraphics[scale=0.09]{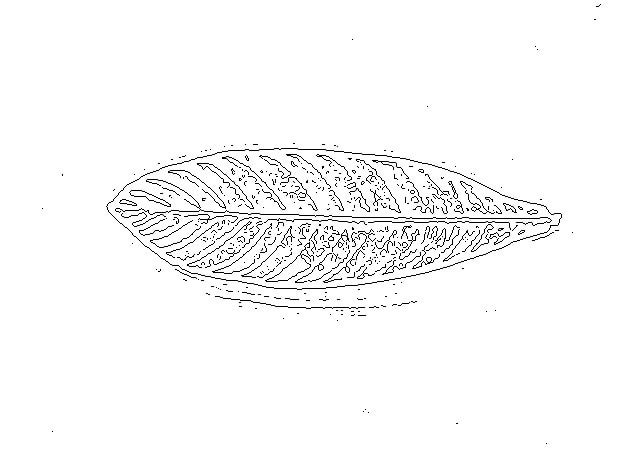}
\includegraphics[scale=0.09]{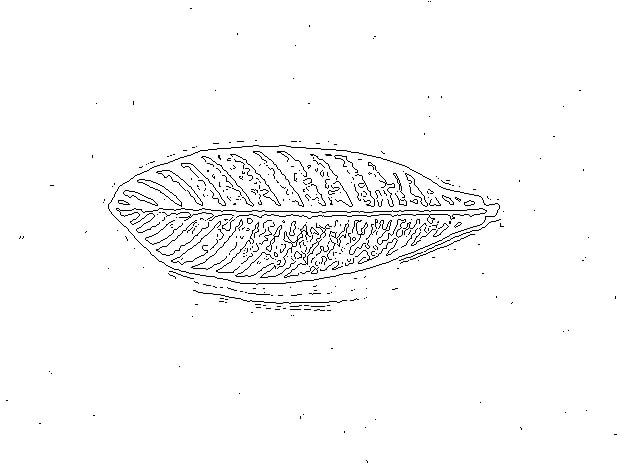}
\includegraphics[scale=0.09]{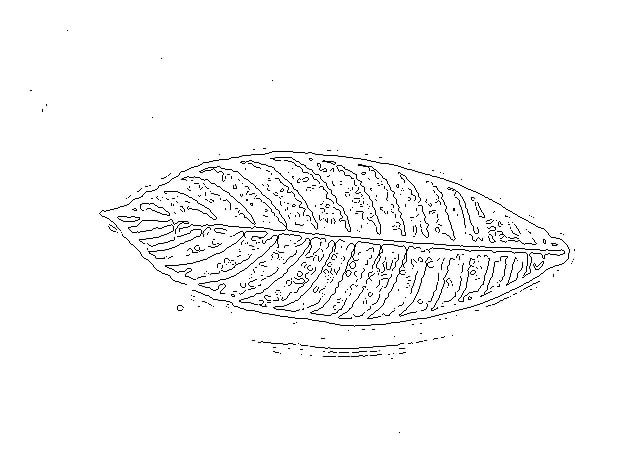}
\includegraphics[scale=0.09]{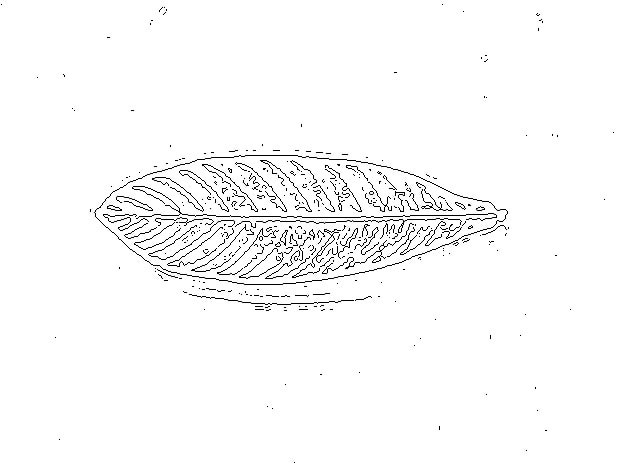}
\includegraphics[scale=0.09]{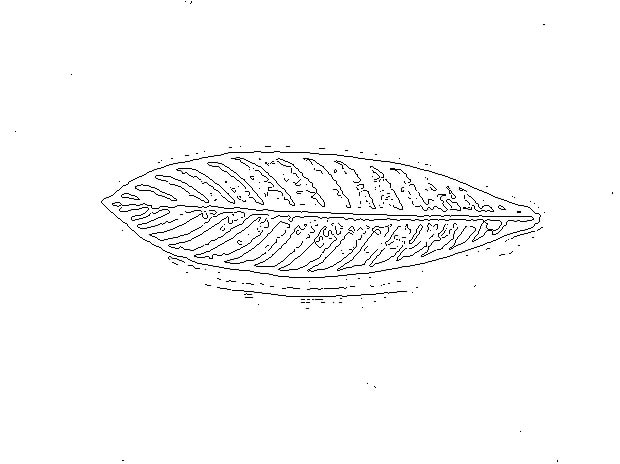}
\includegraphics[scale=0.09]{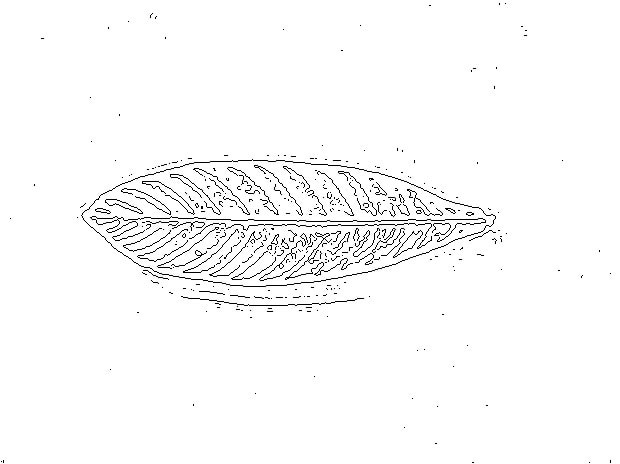}
\includegraphics[scale=0.09]{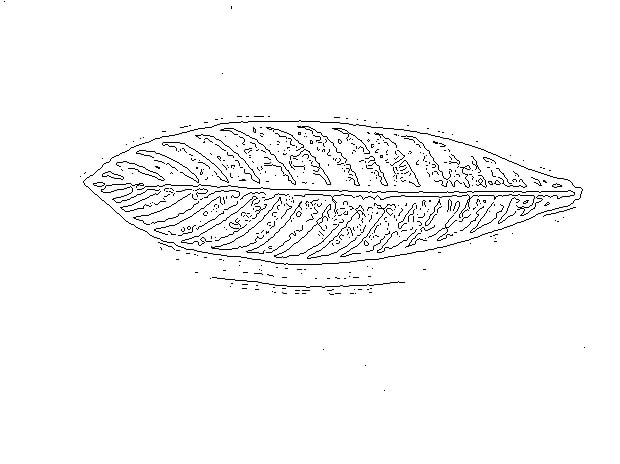}
\includegraphics[scale=0.09]{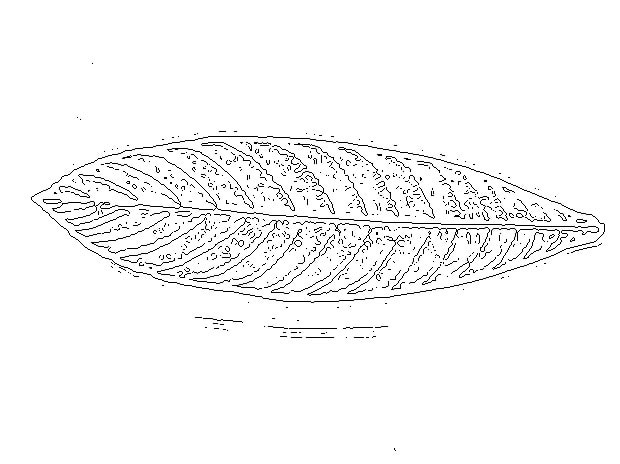}\newline%
\smallskip\includegraphics[scale=0.09]{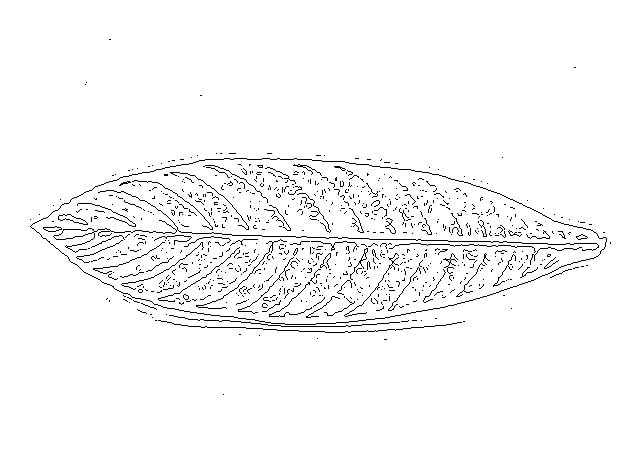}
\includegraphics[scale=0.09]{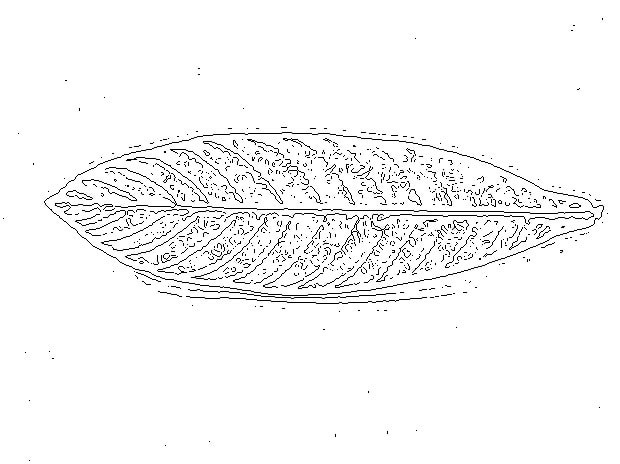}
\includegraphics[scale=0.09]{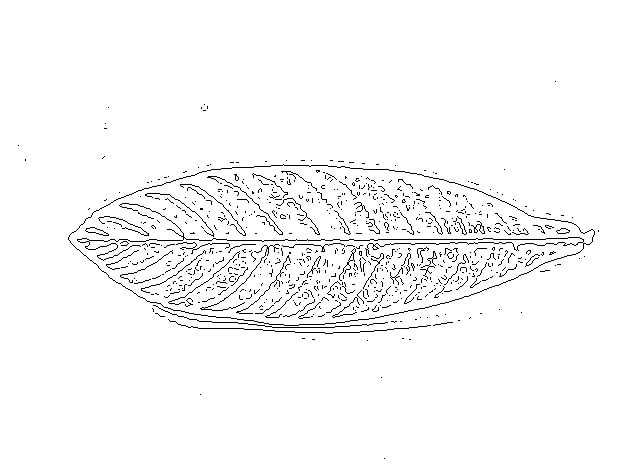}
\includegraphics[scale=0.09]{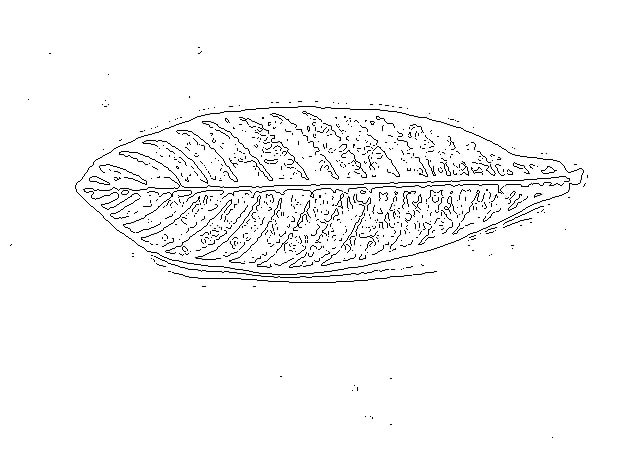}
\includegraphics[scale=0.09]{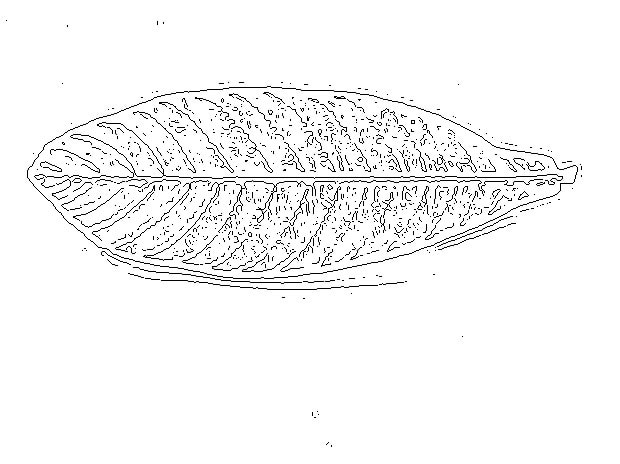}
\includegraphics[scale=0.09]{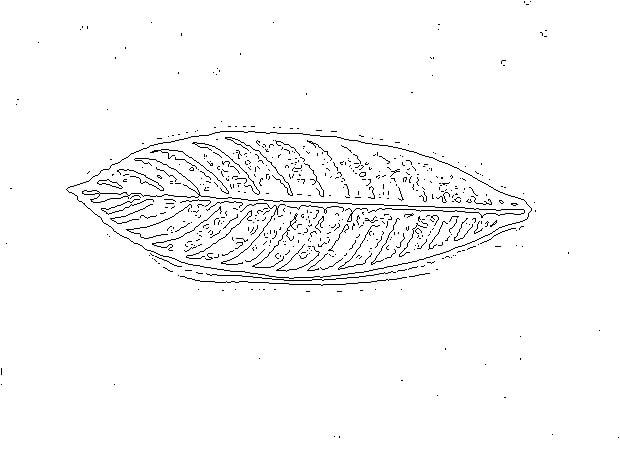}
\includegraphics[scale=0.09]{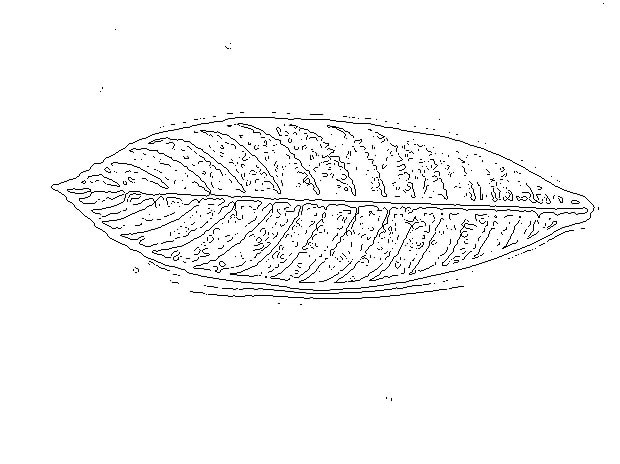}
\includegraphics[scale=0.09]{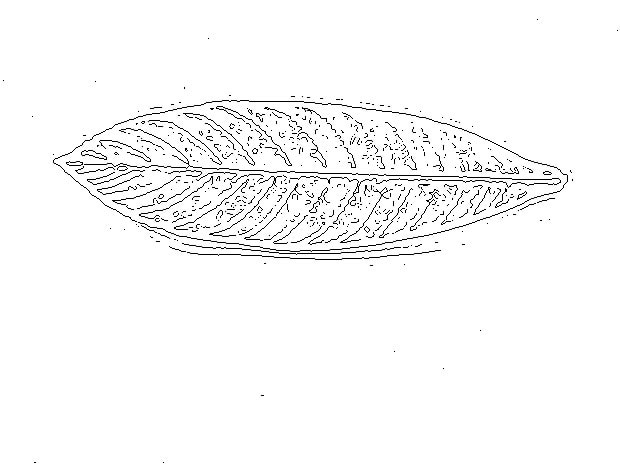}
\includegraphics[scale=0.09]{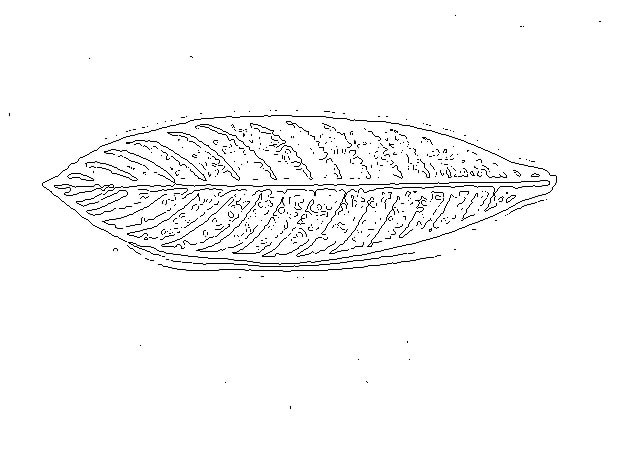}
\includegraphics[scale=0.09]{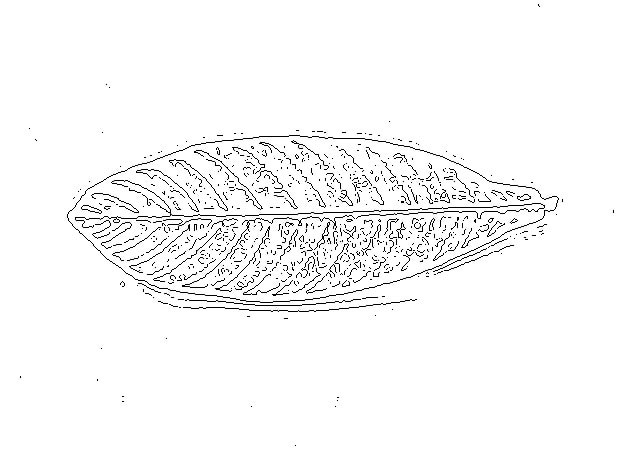}\newline%
\caption{Leaf edges from the original images.}%
\label{fig:Leaf_edges}%
\end{figure}

\begin{figure}[h]
\centering
\includegraphics[width=50mm]{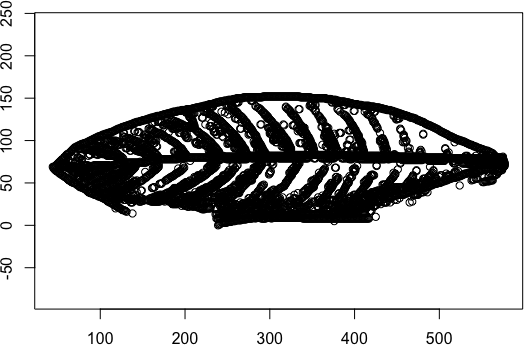} \quad
\includegraphics[width=50mm]{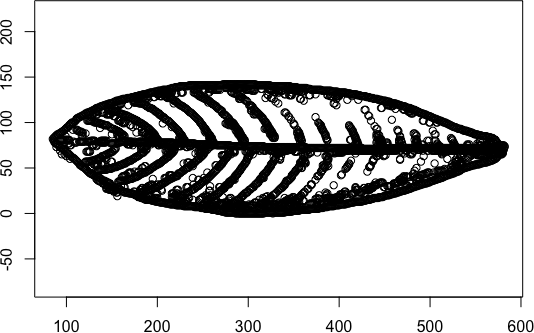}
\par
\includegraphics[width=50mm]{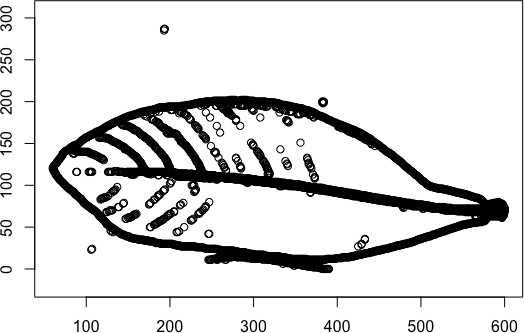} \quad
\includegraphics[width=50mm]{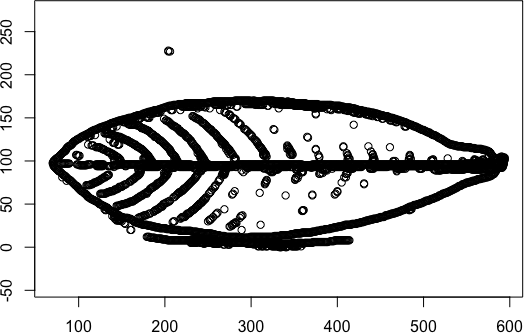}\caption{Sample point clouds
from leaf A (top left and bottom left) and leaf B (top right and bottom
right).}%
\label{fig:sample-leaf-points}%
\end{figure}

From these leaf edges point clouds consisting of approximately 4300 points
were sampled. See
Figure~\ref{fig:sample-leaf-points}.

Next, the persistence diagrams for the Vietoris-Rips complexes of all the
point cloud data sets were computed using Ripser~\cite{Ba:2017} (see Sections
\ref{sec:sc} and \ref{sec:ph}).
These persistence diagrams were then converted into vectors to facilitate
statistical analysis. Specifically, the persistence diagrams for homology in
degree 0 were converted into death vector (see Section~\ref{sec:pl}) and the
persistence diagrams for homology in degree 1 were converted into persistence
landscapes (see Section~\ref{sec:ph}) using the Persistence Landscapes
Toolbox~\cite{BuDl:2017}.

The persistence landscapes were converted into vectors by evaluating on a grid as follows. Specifically, we evaluated the persistence landscape functions $\lambda_1, \ldots, \lambda_{60}$ (all further landscape functions were identically zero) at the values $0,0.1,0.2,0.3,\ldots,39.9,40$. The resulting $60 \times 401 = 24060$ values were concatenated to obtain vectors in $\mathbb{R}^{24,060}$. For an example of the death vector and the persistence landscape, see Figure~\ref{fig:dv-pl}.

\begin{figure}[h]
  \centering
  \includegraphics[width=50mm]{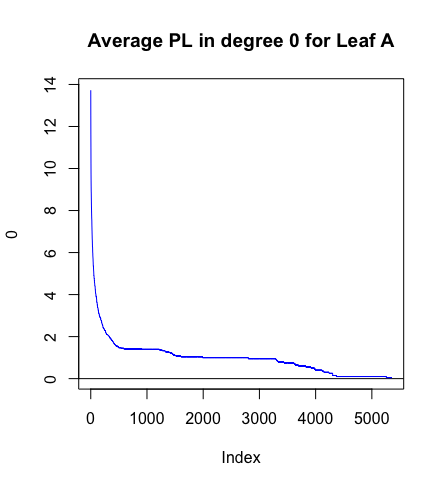} \quad
  \includegraphics[width=50mm]{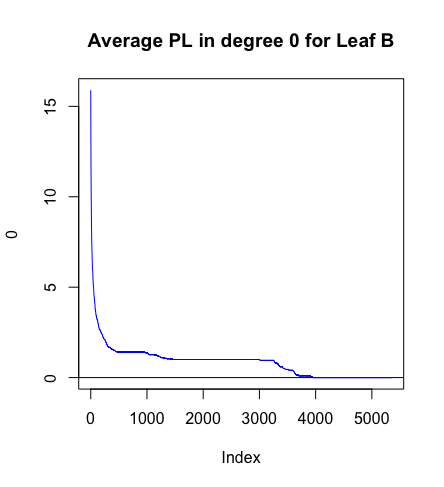} \quad
  \includegraphics[width=50mm]{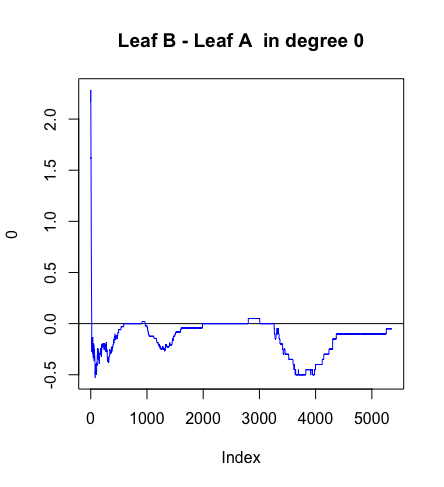}
  \par
  \includegraphics[width=50mm]{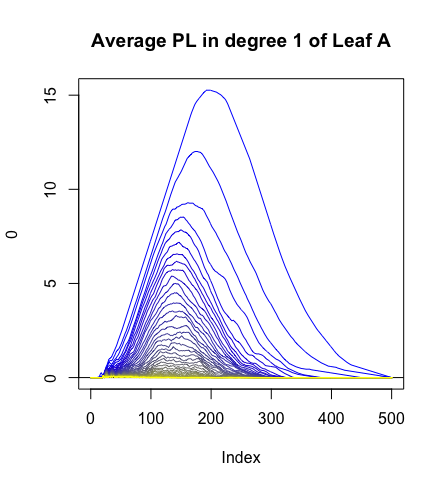} \quad
  \includegraphics[width=50mm]{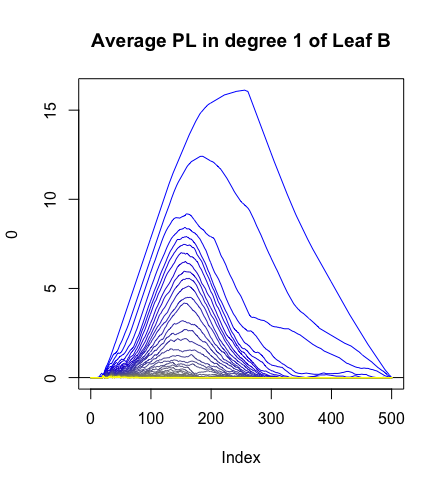} \quad
  \includegraphics[width=50mm]{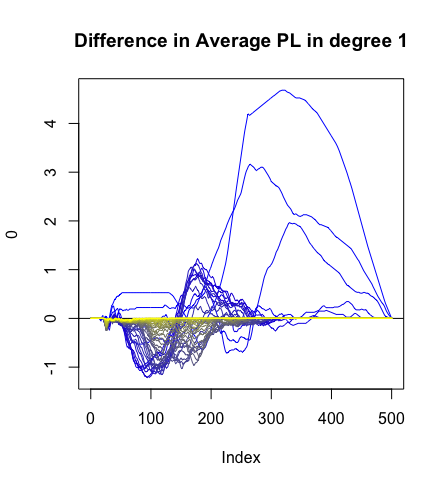}
  \caption{Average death vectors for leaf A (top left) and leaf B (top middle) and their difference (top right). Average persistence landscapes for leaf A (bottom left) and leaf B (bottom middle) and their difference (bottom right).}
  \label{fig:avg-diff}
\end{figure}

Next, we consider the average death vectors and average persistence landscapes for the two leaves and the differences between these averages. See Figure~\ref{fig:avg-diff}.

\begin{figure}[h]
\begin{center}
\includegraphics[scale = 0.5]{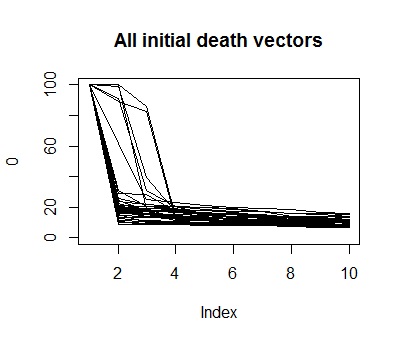}
\end{center}
\caption{{\small All initial death vectors}}%
\label{fig:initial-dv}%
\end{figure}

Upon inspection, of the initial terms of the death vectors, it was observed
that the first three coordinates are very noisy (see Figure~\ref{fig:initial-dv}). As a result, these
coordinates were excluded from further statistical analyses (with an eye
toward removing ``topological noise'' from our data to better detect the
``geometric signal''), and the resulting permutation test p-value for a
difference in the death vectors, of leaves and A and B, was found to be a
highly significant 0.0007.

\begin{figure}[h]
\begin{center}
\includegraphics[scale = 0.5]{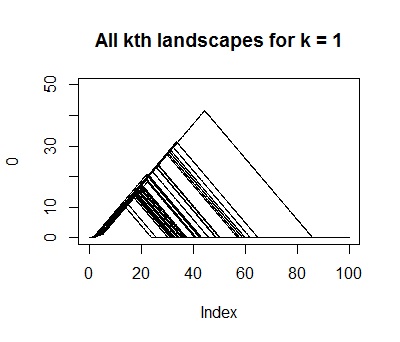}
\includegraphics[scale = 0.5]{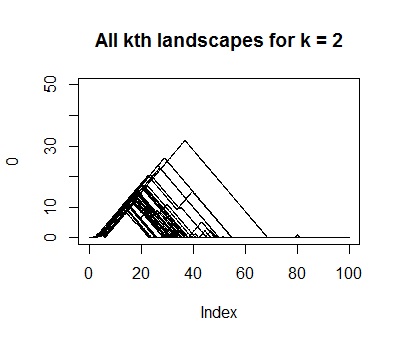}
\end{center}
\caption{{\small The first (left) and second (right) persistence landscape functions  for all of the leaves.}}%
\label{fig:leaf-land}%
\end{figure}

The degree one persistence landscape functions were plotted to look for
outliers and none were found. However, that the first 20 or so degree one
persistence landscape functions contained large variability and hence large
amounts of ``topological noise''. In Figure~\ref{fig:leaf-land},
the plots of all of the first two degree one persistence landscape functions are displayed.

Subsequently, the first 20 degree one persistence landscape functions were
excluded (again, to filter out ``topological noise''), leaving the remaining degree one
persistence landscape functions, and the permutation test p-value for the
difference in the lower frequency degree one persistence landscape functions,
for leaves and A and B, was found to be highly significant at 0.0019. This p-value is not very sensitive to changing the number of excluded persistence landscape functions.
When we did not exclude the first 20 degree one persistence
landscape functions then the permutation test p-value was marginally
insignificant at 0.0821.


After this we considered classification with support vector machines (SVMs).
Here our feature vector was taken to be the death vector concatenated with all
degree one persistence landscape vectors. Unlike the statistical analysis, we did not remove any ``topological noise''. Using 10-fold cross validation we obtain a fitted
classifier with 90\% classification accuracy. In fact, we
did just as well in terms of classification accuracy when the death vector was
removed from the feature vector.

In summary, we see that the points in the persistence diagram that are closest to the diagonal, (that is, the points of high frequency, low variance in our topological statistics)
best capture local geometry of the leaves and
are better able to distinguish between the two leaves.

\bigskip

\section{New Directions in Object Data Analysis}
\label{sec:oda}

To date, Object Data Analysis (ODA) is the most inclusive type of statistical analysis as far as the complexity of the objects under investigation is concerned. In particular, ODA includes Linear Data Analysis (LDA), shape analysis (see Dryden and Mardia(2016)\cite{DrMa:2016}, Patrangenaru and Ellingson(2015)\cite{PaEl:2015}), directional and axial data analysis (see Mardia and Jupp(2000)\cite{MaJu:2000}), data analysis on spaces of phylogenetic trees (see Billera et al(2001)\cite{BiHoVo:2001}), to mention just a few. Mathematically, ODA is data analysis on an {\em object space}, which is a complete separable metric space $(\mathcal M, \rho)$ and typically has a {\em manifold stratification} (see Bhattacharya et al.(2013)\cite{BhBuDrElGrHeHuLeLiMaOsPaScThWo:2013}, Patrangenaru and Ellingson(2015,p.475)\cite{PaEl:2015} and the references therein). A {\em random object} is a function $X: \Omega \to \mathcal M$ defined on a probability space $(\Omega, \mathcal A, \mathbb P)$, such that $X^{-1}(B)\in \mathcal A,$ for any Borel set $B \in \mathcal B_\rho$. Let $\mathcal S $ be the support of the probability measure (distribution) $Q = \mathbb P_X$ on $\mathcal M, Q(B) = \mathbb P(X^{-1}(B)), \forall B \in \mathcal B_\rho.$

{The main techniques for a nonparametric analysis of object data are Fr\'echet function based (see e.g. Patrangenaru and Ellingson (2015)\cite{PaEl:2015}). Since in general there is no group structure on the object space, the {\em total variance} of $X$ is defined as the minimum of the expected square distance from $X$ to an arbitrary fixed object $x$ on $\mathcal M,$ and the minimizers of this {\em Fr\'echet function} (see Fr\'echet \cite{Fr:1948}) form the {\em mean set} of $X.$ The fastest and theoretically sound quantitative methods for analyzing object data, are {\em extrinsic}, based on an object space embedding in a numerical space, with the induced ``chord" distance. The embedding also allows using certain linear techniques when testing for equality of two or more distributions on an object spaces, via an {\em extrinsic energy methodology} (see e.g. Guo and Patrangenaru(2017)\cite{GuPa:2017}).}

\subsection{Fr\'echet Object Data Analysis}\label{ssec:foda}

Fr\'echet object data analysis (FODA) is a ODA on a metrizable object space,
based on a preferred distance. There are two key types of
distances used in FODA: a ``chord distance" induced by the Euclidean distance on the numerical space where the object space is embedded, and a geodesic distance associated with a Riemannian structure on the nonsingular part
of the object space. Extrinsic FODA (based on the chord distance) has multiple advantages over intrinsic analysis (using a geodesic distance), including computational and methodological advantages (see Bhattacharya et al(2012)\cite{BhElLiPaCr:2012}, Patrangenaru et al(2015, p.168)\cite{PaEl:2015}).

More general location parameters, reflecting the topological structure of both the support of the random object $X$ and of the underlying object manifold $\mathcal M,$ extending those in Patrangenaru(2016a, 2016b)\cite{PaGuYa:2016,PaYaGu:2016} are introduced below.
\begin{definition} \label{d:mean-r}Assume the Fr\'echet function associated with a random object $X$ on $\mathcal{M}$ is a Morse function (for a definition see eg Bubenik et al \cite{BuCaKiLu:2010}). The set of nondegenerate critical points
of the Fr\'echet Morse function $\mathcal F_X$, with fixed index $r$ is the {Fr\'echet mean set of index $r$} of $X.$
\end{definition}

If $\mathcal M$ has dimension $m,$ the Fr\'echet mean sets of index $0$ and $m$ are, respectively, the Fr\'echet antimean set (see Patrangenaru and Ellingson (2015), p.139), and the Fr\'echet mean set. In case the Fr\'echet mean set of index $r$ of $X$ has one point, that point is called {\em Fr\'echet mean of index $r$} of $X.$ Given a random sample of size $n$ from the distribution $Q$ associated with an random object on $\mathcal M,$ we define the {\em Fr\'echet sample mean (set) of index $r$} to be the Fr\'echet mean (set) of index $r$ of the empirical distribution $\hat Q_n.$
If the manifold $(\mathcal{M}, \rho_0)$ is compact, where $\rho_0$ is the chord distance associated with the embedding $j$ of $\mathcal M$ in  $\mathbb R^N,$ we define the {\em extrinsic mean (set) of index $r$ of $Q$} to be the Fr\'echet mean (set) of index $r$ of $Q$ associated with the distance $\rho_0,$ and given a sample of size $n$ from $Q,$ its {\em extrinsic sample mean (set) of index} $r$ is the extrinsic mean (set) of index $r$ of the empirical distribution $\hat Q_n.$

\begin{remark} Using similar, {somewhat more sophisticated} techniques as for antimeans (see Patrangenaru et al(2016a)\cite{PaGuYa:2016}), one may prove the consistency of extrinsic sample means of index $r$ as estimators of the for extrinsic means of index $r$.
Along these lines one may also derive the asymptotic distributions for extrinsic sample means of index $r$ that help estimate their population counterparts.
To prove consistency, the key new ingredient involves the fact that the class of Morse functions is {\em generic} (open and dense).
\end{remark}
The new location parameters introduced above allow for an extension of nonparametric regression to extrinsic regression and extrinsic anti-regression, when the response variable is a random object on a compact set. For more details and an application of time dependent anti-regression in projective shape analysis for biological growth of a clam shells species, see Deng et al.(2018)\cite{DePaBa:2018}

\subsection{Statistical challenges of ODA}

Unlike linear data analysis, including functional data analysis, ODA was designed to mainly analyze imaging data, since these days a datum is often an electronic image of some form. Arguably, the most widespread type of imaging data are digital camera images. Among the many challenges arising in with camera images, one of the most difficult is the 3D scene retrieval from its digital camera images (see eg. Ma et al.(2006)\cite{MSKS2006} or Chapter 22 in Patrangenaru and Ellingson (2015)\cite{PaEl:2015}). If the dimensionality of the scene is of secondary interest, as opposed to its fine structure, as is the case with the TDA leaf data in Section \ref{sec:leafs}, Fr\'echet function based methods become computationally costly and methodologically challenging. In addition, a plethora of usual methods for random vectors, raise difficulties with random objects on a nonlinear object space. Starting with a proper definition of location and spread parameters as shown in subsection \ref{ssec:foda}, dimension reduction, regression, MANOVA and other inference problems, all the way to designing appropriate distributional models on object spaces and designing nonparametric tests for their goodness of fit, one encounters many unanswered, potentially difficult questions. Moreover new challenging questions arising from images of the Universe, on recognising dark matter and singularities as voids in the 3D continuum, raise qualitative questions that can be formulated in more in a TDA setting, rather than a ``classical" FODA way. Note that in the multivariate case, Chen et al(2017)\cite{ChGeWa:2017} already developed TDA techniques for qualitative aspects of data analysis. The challenge is to find similar methods for nonlinear ODA.

\subsection{A homology approach to qualitative analysis of distributions on object spaces}

From the object data analysis perspective, a closed manifold can be regarded as the support of the distribution on an object space. A basic example is provided by the Fisher von-Mises distribution on $\mathbb S^1$ \cite{Fi:1983}, whose support is the entire circle. In general any compact manifold $\mathcal M$, endowed with a Riemannian structure $g$ may be regarded as the support of a uniform distribution on it, whose density w.r.t. the volume measure is $\frac{1}{Vol_g(\mathcal M)}.$

\begin{remark}A well known result, the
topological classification of compact orientable surfaces, shows that an algebraic homology invariant, {\em the rank of the first homology group}, which is twice the genus of such a surface,  is a
classifier for the homeomorphism class of such a surface. In dimensions three, the problem of classifying homeomorphism classes of closed
manifolds based on their homology, was advanced only in the eighties and nineties, especially by
Thurston-Perelman's theorem (see Thurston(1982)\cite{Th:1982}, Perlman
\cite{Pe:2003}, Scott(2003)\cite{Sc:1983},
Patrangenaru(1996)\cite{Pa:1996}), and in dimension four by M.H.Freedman, S. Donaldson and their collaborators (see eg.
\cite{DoKr:1990}, \cite{FrQu:1990}).
\end{remark}

Unfortunately, object data is high dimensional, and in dimensions five or higher, it is way more difficult to
classify homeomorphism classes of closed manifolds, even in the smooth case,
due to the so called \emph{moduli spaces}. So, 
despite the preferred equivalence via homeomorphisms, one has to accept the
idea of a weaker form equivalence relation for topological spaces, leading to
the notion of \emph{homotopy type}, which is often used. Intuitively, two topological
spaces have the same homotopy type if one can be continuously deformed, but
not necessarily in one-to-one correspondence, into the other. The basic
definitions are as follows:

\begin{definition}
Given two topological spaces $X,Y,$ we say that two continuous functions
$F_{0}, F_{1}: X \to Y$ are homotopic, and we write $F_{0} \cong F_{1},$ if
there is a continuous function $F:[0,1] \times X \to Y,$ such that $\forall x
\in X, F(0,x) = F_{0}(x), F(1,x) = F_{1}(x).$ $X$ and $Y$ have the same
homotopy type, if there are continuous functions $F:X \to Y, G:Y \to X,$ such
that $G\circ F \cong Id_{X}, F\circ G\cong Id_{Y}.$
\end{definition}

One view of TDA is that it aims to consider the homotopy type of an object or of the support of a distribution, using a random sample of its points in a numerical space. More precisely, it really
computes certain invariants associated with the homotopy type, that persist, while gradually inflating this sample by balls of growing radii around them, which is somewhat similar with the recovery of the CW homotopy type of a submanifold $M$ in the Euclidean space, via a filtration by sublevel sets of a Morse function (see Bubenik et al.(2010)\cite{BuCaKiLu:2010}). Persistent homology measures these invariants
associated with the homotopy type of the ``telescoping" limit via this filtration $M_n, n \in \mathbb N$ of $M$. Essentially if $\mathcal S$ is regarded as a union of sub-level level sets $\mathcal S_n$ of the support of a probability distribution on $\mathbb{R}^{p},$ one may consider the homotopy type of $(\mathbb{R}^{p},\mathcal S_n),$ as an ``estimate" of the homotopy type of the pair $(\mathbb{R}^{p},\mathcal S).$ Note that two pairs $(\mathbb{R}^{p},A_{1}),(\mathbb{R}^{p},A_{2}),$ have the same homotopy type, if there are continuous functions $h,k:\mathbb{R}^p \circlearrowleft,$ with $h(A_1)\subseteq A_2, k(A_2)\subseteq A_1,$ such that $h\circ k$ and $k\circ f$ are homotopic to the identity of $\mathbb R^p.$ Thus within the same homotopy type, the continuity relation between the contiguous
regions and the number of holes or voids of the subspaces $A_{1},A_{2}$ of
$\mathbb{R}^{p}$ remains unchanged. Furthermore, when it comes to the support $\mathcal{S}$ of a
distribution on an object space $\mathcal{M},$ this continuity relation of the pair $(\mathcal{M}%
,\mathcal{S})$ is reflected by the {\em homology of the pair}
$(\mathcal{M},\mathcal{S}),$ more precisely, its algebraic consequence, the {\em exact homology sequence of this
pair} (see Patrangenaru and Ellingson
(2015)\cite{PaEl:2015}, p.131-132). Homology is our approach to studying the support of a distribution, unlike homotopic trees that used for understanding connections (or continuity) between the contiguous connected components in machine vision (eg Sonka et al.(2015)\cite{SoHlBo:2015}, p.699, and an illustration in the Figure \ref{fig:h-tree} below).
\begin{figure}[h]
\begin{center}
\includegraphics[scale = 0.35]{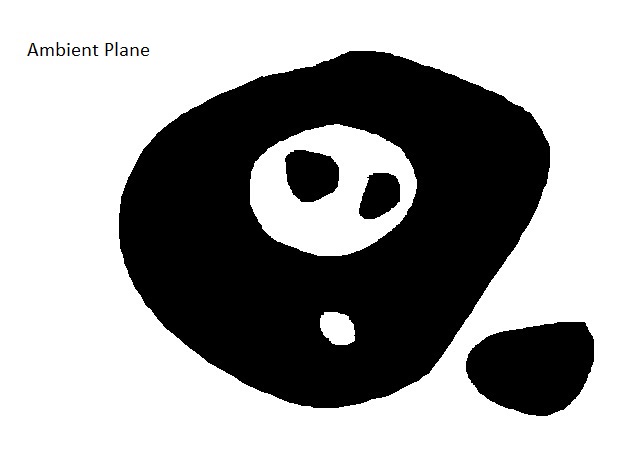}
\includegraphics[scale = 0.55]{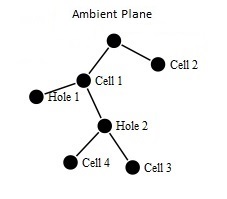}
\end{center}
\caption{\small Contiguity of connected components explained via a homotopic tree.}%
\label{fig:h-tree}%
\end{figure}

\subsection{Topological Object Data Analysis}

Given the pair $(\mathcal{M}, \mathcal{S}),$ where $\mathcal{M}$ is the ambient {\em object space} and
$\mathcal{S}$ is the support of a distribution on $\mathcal{M},$ there is a long exact sequence in homology, via the inclusions $i:\mathcal{S} \to \mathcal{M}$ and $j:\mathcal{M}\to (\mathcal{M}, \mathcal{S}),$ where $\mathcal{M})$ is the shorthand for $(\mathcal{M},\emptyset):$
\begin{equation}\label{eq:exact-pair}
        \cdots \to H_{k}(\mathcal{S})\to ^{{\!\!\!\!\!\!i_{k}}}H_{k}(\mathcal{M})\to ^{{\!\!\!\!\!\!j_{k}}}H_{k}(\mathcal{M},\mathcal{S})\to ^{{\!\!\!\!\!\!\partial _{k}}}H_{{k-1}}(\mathcal{S})\to \cdots .
        \end{equation}
Topological Object Data Analysis (TODA) is a data driven homology based statistical analysis of the relative homology spaces $H_{k}(\mathcal{M},\mathcal{S}).$

Note that since the homology groups of the Euclidean space are all trivial,
except for $H_0(\mathbb R^m)=\mathbb Z,$ from (1) it follows that in case of a random vector
$X$, $H_k(\mathcal M, \mathcal S) \backsimeq H_{k-1}(\mathcal S),$ therefore TODA aims at estimating the homology of $\mathcal S,$
for example via a persistent homology. \\

Some stratified spaces of interest like spaces $T_k$ of phylogenetic trees with $k$ leafs (see Billera et al(2001)\cite{BiHoVo:2001}) are contractible, therefore they
have trivial reduced homology as well, thus for any random phylogenetic tree $X$ with $k$ leafs, whose distributional support is $\mathcal S,$
the relative homology $H_k(T_k,S) \backsimeq  H^{k-1}(S),$ similar with the
case of a random vector. From this perspective, TODA applies to phylogenetic tree spaces, via persistence homology techniques. This opens a new venue to qualitative analysis for certain types of Big Data.

\section*{Acknowledgment}
We  are  most  grateful  to  an anonymous referee for comments which have led to substantial improvements of the initial manuscript.

\end{document}